\newif\ifarXiv
\arXivtrue
\ifarXiv
  \documentclass[10pt,twocolumn]{article}
  \usepackage[noobjstm,merged]{agnostic}
  \renewcommand{\agappendixgeometry}{textwidth=17.8cm,top=0.5in,bottom=0.5in}
\else
  \documentclass[9pt,twocolumn,twoside]{pnas-new}
\fi
\usepackage{svg}
\usepackage{bm}

\articletype{Physical Sciences: Applied Mathematics and Medical Sciences}%%%% Article topic/classification

\templatetype{pnasresearcharticle} % Choose template
\usepackage{mathtools, dsfont}
\usepackage{subcaption}
\newcommand*{\Reals}{\mathbb{R} }

\newcommand*{\Expectation}{\mathbb{E} }
\newcommand*{\Prob}{\mathbb{P} }
\newcommand*{\defeq}{\coloneqq}

\newcommand*{\arXiv}[1]{\bgroup\href{https://arxiv.org/abs/#1}{arXiv:#1}\egroup}

\newcommand*{\apd}{\text{APD}_{90} }
\newcommand*{\diag}{\text{diag}}

\begin{document}
\title{Multi-Fidelity Gaussian Processes for Translational Modelling of Clinical Outcomes}

% Use letters for affiliations, numbers to show equal authorship (if applicable) and to indicate the corresponding author
\author[a,1]{Isaac S. Hayden}
\author[b]{Alicia D’Souza}
\author[b,c]{Steven Niederer}
\author[a]{Sarah Filippi}

\affil[a]{Department of Mathematics, Imperial College London, London, UK, SW7 2RH}
\affil[b]{National Heart and Lung Institute, Imperial College London, London, UK, SW3 6LY}
\affil[c]{Division of Cardiovascular Medicine, Stanford Cardiovascular Institute, Stanford University, CA 94305, USA}

% Please give the surname of the lead author for the running footer
\leadauthor{Hayden}

% Please add a significance statement to explain the relevance of your work
\significancestatement{New medical treatments must be tested extensively in animals and humans before regulatory approval. Not only does this raise both costs and ethical concerns, but there is no formal quantitative framework to integrate all this data. By treating animal experiments as informative, low-fidelity approximations of human ones, we use statistical machine learning to jointly learn nonlinear exposure–response relationships and cross-species similarity, while providing uncertainty quantification for decision making. This enables transfer learning across both species and experiments, allowing efficient use of data and potentially reducing the number of human and animal tests required. We demonstrate improved prediction of drug-induced changes to two important indicators of cardiac function, one using simulated data and the other serving as validation on clinical data.}

% Please include corresponding author, author contribution and author declaration information
\authorcontributions{I.S.H., S.N. and S.F. designed research; I.S.H. performed research; A.D.S. provided data; I.S.H. and A.D.S. analysed data; I.S.H. 
wrote the paper; and A.D.S., S.N. and S.F. reviewed drafts of the paper.}
\authordeclaration{The authors declare no competing interests.}
\equalauthors{\textsuperscript{1}To whom correspondence should be addressed. E-mail: i.hayden24@imperial.ac.uk}

% At least three keywords are required at submission. Please provide three to five keywords, separated by the pipe symbol.

\begin{abstract}
Bridging the gap between animal and human experiments remains a major challenge in translational medicine, particularly in early drug development. Progress is constrained by financial cost, the difficulty of integrating heterogeneous \textit{in vitro} and \textit{in vivo} data, and the desire to reduce the use of animal testing balanced against minimising the risk to human participants. 

We present a statistical machine learning framework using multi-fidelity Gaussian processes, in which animal studies are considered as lower fidelity but informative approximations to human experiments. This allows cross-species similarities and nonlinear exposure-response relationships to be learned simultaneously, enabling principled extrapolation between species while quantifying uncertainty. By leveraging information from multiple experimental fidelities, our method improves estimation of clinically relevant quantities of interest and supports the replacement, reduction, and refinement of \textit{in vivo} testing.

We first illustrate this approach in a simulated scenario, before validating it on real clinical data. We simulate data for drug-induced QT-interval prolongation, a key cardiac safety assessment required for regulatory approval. This framework provides a probabilistic surrogate capable of integrating \textit{in vitro} pharmacology, animal experiments and human data within a unified statistical model. Crucially, it achieves this at no additional experimental cost while also enabling transfer learning across compounds. As a result, predictions and uncertainty quantification for new drugs can be generated from \textit{in vitro} findings alone, providing additional efficiency gains and accelerating decision making. For validation, we use a clinical dataset measuring change in heart rate under autonomic blockade, which represents some of the challenges commonly found in multi-species datasets.
\end{abstract}

\dates{This manuscript was compiled on \today}
\doi{\url{www.pnas.org/cgi/doi/10.1073/pnas.XXXXXXXXXX}}

\maketitle
\thispagestyle{firststyle}
\ifthenelse{\boolean{shortarticle}}{\ifthenelse{\boolean{singlecolumn}}{\abscontentformatted}{\abscontent}}{}

\Firstpage
%The \Firstpage command is used to format the first page text column size. The same size will be maintained for subsequent paragraph until the \Endparasplit or \Parasplit command is encountered.
Animal models are used ubiquitously in medical research as proxies for human experiments, particularly in drug development \cite{voisin1990extrapolation}. Despite this, translating results from animal studies to human outcomes remains challenging \cite{mittal2025exploring}, and regulatory bodies worldwide have adopted the ``3Rs'' principles of reduction, refinement and replacement of animals in biomedical research \cite{madden2012strategies}. For these reasons, there is growing focus on ``new approach methodologies,'' including those which make better use of computational methods to integrate this data more efficiently \cite{mehta2025modernizing, Sewell2025NAMs}. Developing such methods is an ongoing challenge: existing constraints include financial cost, the integration of multiple sources of heterogeneous data, and physiological differences between animals and human beings \cite{mehta2025modernizing}.

Existing approaches in diverse biomedical applications include developing physiologically-based pharmacokinetic models to describe the shared mechanistic pathways between species \cite{bloomingdale2022pbpk, shah2012towards}, and using generative artificial intelligence to emulate animal experiments \textit{in silico} \cite{bordukova2024generative, chen2022tox, chen2023generative}. Meanwhile, neural networks and other machine learning architectures have shown good predictive accuracy in extrapolating from animal or cell experiments to human outcomes \cite{sharifi2020aitl, mourragui2021predicting, fang2026prediction}, or predicting drug properties directly \Parasplit from chemical structure \cite{myung2024deep, rao2019novel, pillai2024machine}. While these models have the benefit of handling complex input data such as genetic or molecular, this must be balanced against their lack of interpretability or inbuilt uncertainty quantification, which hinders their use for decision-making. They also do not provide a framework in which \textit{in vitro} data, animal and human experimental data can be combined into a single model.
% \Endparasplit

We propose a method inspired by multi-fidelity modelling methods common in the field of engineering \cite{fleck2025physics, folch2023combining, cutajar2019deep}, in which we consider animal experiments as cheaper, low-fidelity approximations of experiments on humans. This approach uses Gaussian processes (GPs) to simultaneously learn both the relationship between continuous inputs and outputs as well as the similarities between animal and human response. This method has several advantages:

\begin{itemize}
    \item It allows the simultaneous integration of all animal and human experimental data, as well as continuous \textit{in vitro} data.
    \item It requires no assumption of linearity in the exposure-response relationship, and provides calibrated predictive uncertainty which can be used for subsequent decision making.
    \item In the context of drug development, it facilitates the transfer of learning across species and across drugs, allowing predictions for new chemical entities to be made using only \textit{in vitro} data, and ultimately permitting both the reduction of human experimentation and more efficient use of animal data.
    \item Again in the drug development context, it incurs no additional financial expense or data collection burden to the drug developer, and can reduce cost by accelerating early termination of drugs exhibiting excess risk, and by eventually reducing the total number of \textit{in vivo} samples required.
\end{itemize}
We demonstrate this approach first in an \textit{in silico} test bed with simulated data. We choose the example of drug-induced change in the corrected QT (QTc) interval of an electrocardiogram (ECG), since this is a well-studied case of drug toxicity where animal testing is crucial \cite{li2017drug, yap2003drug}, and where physiological simulators that can mimic drug effects are available in literature \cite{benson2008canine, pasek2008model, tomek2019development, tomek2020tor, li2010mathematical, gaur2021computational, mahajan2008rabbit}. Then, as validation on a clinical dataset, we use multi-species experimental measurements for drug-induced change in heart rate \cite{dsouza2017targeting, soattin2026endurance}, chosen based on data availability. Though we choose to focus on response to drug exposure, we emphasise that it is generalisable to any scenario where human and animal experiments share the same input and response variables.

\section*{Multi-Fidelity Gaussian Process Modelling}
We focus on problems of the following type: suppose we have some treatment or experiment that can be represented by $P$ continuous inputs $\bm{x} \in \Omega \subset \Reals^P$, which we are able to test in available species \( \mathcal{S} = \{1, \hdots, S\}\).  Let $s=S$ refer to the species of interest, or high-fidelity (HF) species (typically human beings), while $s<S$ denotes auxiliary or low-fidelity (LF) species. While we consider only animals as LF approximations, note that this approach extends to other surrogate models for human responses available \textit{in vitro} or \textit{in silico}. Suppose we have a univariate response variable of interest with true underlying response function \(f(s,\bm{x}):\mathcal{S} \times \Reals^P \mapsto \Reals\), and have access to noisy observations \(y(s,\bm{x}) = f(s,\bm{x}) +  \epsilon \cdot \sigma(s,\bm{x})\), where \(\epsilon \sim  \mathcal{N}(0,1) \) and the standard deviation \(\sigma(s,\bm{x}) > 0\) may be  species-dependent and heteroskedastic.

We propose to take a Bayesian non-parametric approach and suppose that \(f(\cdot,\cdot)\) is a random variable that can be modelled by a GP, allowing us to simultaneously describe the response and species similarity. This requires specifying prior mean and kernel functions:
\begin{align}
    \mu &: \mathcal{S} \times \Omega \mapsto \Reals \;, \nonumber \\
    k &:  (\mathcal{S} \times \Omega) \times (\mathcal{S} \times \Omega) \mapsto \Reals \;, \label{eq:gp_fns}
\end{align}
as well as a function for $\sigma$, respectively parametrised by hyperparameters $ \bm{\lambda}_\mu, \bm{\lambda}_k$ and $\bm{\lambda}_\sigma$. We then denote \(f \sim \mathcal{GP}(\mu, k)\) as our prior function. Given a set of training data, $\mathcal{X} = \left\{\left(s_j, \bm{x}_j, y_j\right) \right\}_{j=1}^J$, we can update this prior to obtain posterior: \(f | \mathcal{X} = f^\mathcal{X} \sim \mathcal{GP}(\mu^\mathcal{X}, k^\mathcal{X})\), where \(\mu^\mathcal{X}\) and \(k^\mathcal{X}\) can be written analytically once values for $\bm{\lambda}_\sigma, \bm{\lambda}_\mu$ and $\bm{\lambda}_k$ are fixed \cite{murphy2022probabilistic}. The specification of functions $\sigma, \mu$ and $k$ is problem-dependent and must be driven by knowledge of the underlying data. In our examples, we use a sum of $Q$ latent product kernels as per the linear model of coregionalisation structure \cite{alvarez2012kernels}:
\begin{equation}
    k((s,\bm{x}), (s',\bm{x}')) = \sum_{q=1}^Qk_{\Omega,q}(\bm{x}, \bm{x}') \cdot k_{\mathcal{S},q}(s,s') \;,
    \label{eq:sum_of_separable_kernels}
\end{equation}
where $k_{\Omega,q}(\bm{x}, \bm{x}')$ can be any standard kernel on continuous inputs (e.g., Matérn, radial basis function, etc.), and $k_{\mathcal{S},q}(s,s')$ are weights determining between-species similarity (\textit{Materials and Methods}). 

\begin{figure*}[ht]
\centering
    \includegraphics[width=17.8cm]{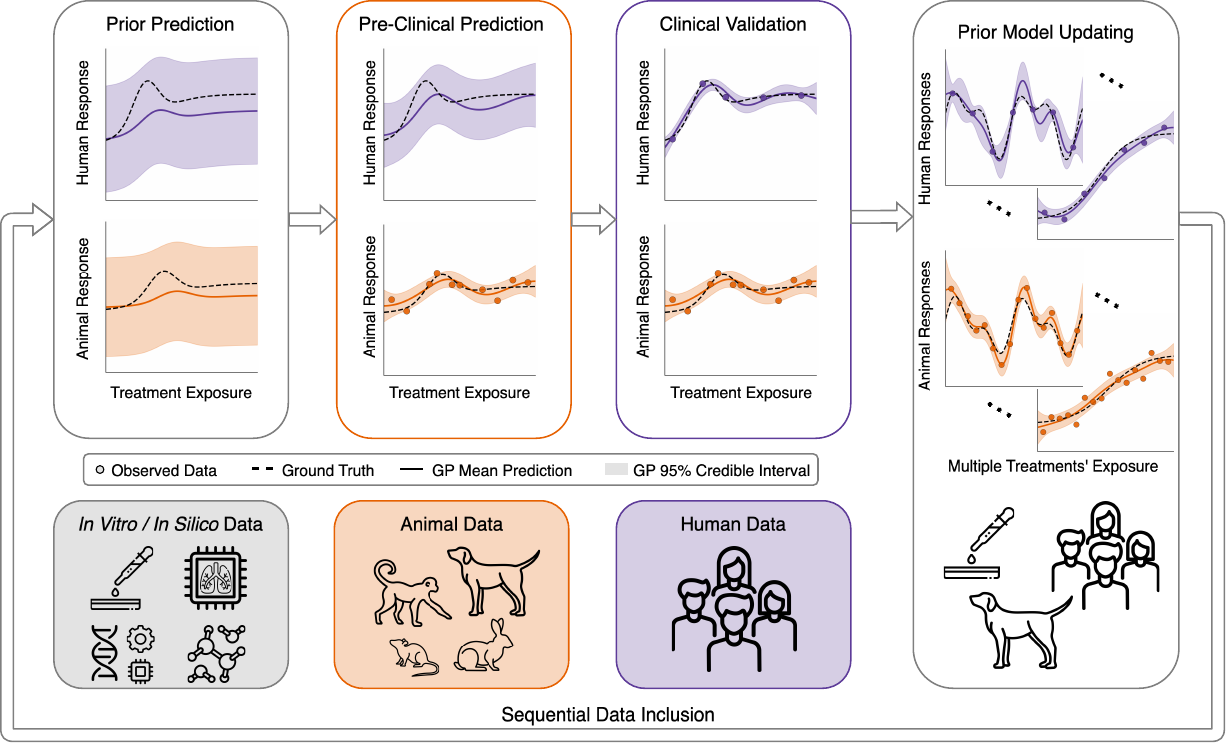}
\caption{Summary figure for the proposed multi-fidelity Gaussian Process (MFGP) method for translational modelling. First panel: using domain knowledge or previous data, a MFGP can provide a mapping from \textit{in vitro} experimental data to response variables in animal and human experiments. Once \textit{in vitro} data is collected for a new treatment of interest, prior predictions of animal and human responses can be made, with quantified uncertainty. Second panel: responses from animal experiments can be used to update the GP prediction for human response. Third panel: a reduced number of human experimental data is required to validate the GP prediction of human response to the treatment of interest. Fourth panel: The new animal and human response data collected can be used in prior predictions for subsequent treatments. The MFGP can easily incorporate the sequential inclusion of data at every stage via the Gaussian conditioning formula.}
\label{fig:summary_figure}
\end{figure*}

This modelling framework is summarised in figure \ref{fig:summary_figure}. Note that this structure is able to integrate all available experimental data into one probabilistic model: any continuous inputs (e.g., \textit{in vitro} drug information, subject covariates, etc.) can be included in the vector $\bm{x}$, and data from all species $s=1,\hdots,S$ can be modelled simultaneously. Moreover, data can be included in a sequential manner easily by the Gaussian conditioning formula. 

\section*{Results on Simulated Data}
To first explore this approach on simulated data, we will focus on a cardiac adverse event known as Torsades de Pointes (TdP). This is a rare but life-threatening tachycardia that can be inadvertently induced by drug exposure \cite{li2017drug}. A precursor to TdP can be detected using features of an ECG: a prolongation of the QTc interval \cite{yap2003drug, france2015role}. All applications for new chemical entities or repurposing of existing marketed pharmaceuticals must evaluate the risk of QTc prolongation, where 10 ms is taken as the threshold of concern \cite{ICH2005S7B, ICH2005E14, EMA2022E14S7B}. 

QTc prolongation risk is typically evaluated in three stages. \textit{In vitro} data are collected to quantify the propensity of the drug to block the cardiac ionic current (IC) most known to prolong QTc: \(I_{\mathrm{Kr}}\) \citep{ICH2005S7B}. More recent recommendations from Comprehensive In Vitro Proarrhythmia Assay (CiPA) initiative also suggest collecting similar data for six other ICs \citep{colatsky2016comprehensive}. \textit{In vivo} studies measuring drug-induced change in QTc on animal subjects are then carried out \citep{vargas2023improving}. Finally, a clinical trial with human subjects must usually take place \citep{ICH2005E14}: this involves either a hypothesis test on the QTc interval being prolonged by 10 ms or more on average \citep{darpo2010thorough}, or a linear mixed-effects model developed to describe the relationship between drug concentration in the bloodstream and change in QTc \citep{garnett2018scientific}. Due to the significant cost of these trials \cite{bouvy2012cost} and the importance of cardiac safety assessments, there has been much focus in literature on methods to improve this process \cite{vargas2023improving}.

Many authors have successfully used \textit{in vitro} drug data or electrophysiology (EP) simulators to predict TdP risk \textit{in silico} \cite{mirams2011simulation, sahli2020classifying, dutta2017optimization, passini2017human, romero2018silico}. Other works have shown how correlations between relevant quantities in animal and human experiments can be identified or modelled \cite{chain2013identifying, morotti2021quantitative}. While these methods are effective in solving distinct aspects of the problems in TdP risk prediction, they do not allow all evidence to be combined in one single probabilistic model.

For the purpose of investigating QTc prolongation, we use computational EP simulators to generate synthetic data, where the action potential duration at 90\% repolarisation in a ventricular myocyte ($\apd$) is taken as a surrogate for QTc (\textit{Materials and Methods}). To mimic between-subject variability (BSV), we perturb the physiological parameters of these simulators (\textit{Materials and Methods}). Since we have no subject covariates, we use the noise term $\sigma(s,\bm{x})$ to account for both BSV and measurement noise; $f(s, \cdot)$ is then the population average relative change in $\apd$ for a given species $s$. We take 10\% prolongation of $\apd$ to be the threshold of concern corresponding to prolongation of QTc by 10 ms, in line with previous studies \cite{romero2018silico, davies2012silico}. Supplementary figure S1 shows the assumed structure of the data-generating process for drug-induced change in QT interval in any species. 

We will take the continuous input $\bm{x}$ to be a vector of inhibition factors for $P=7$ cardiac ICs deemed relevant in accordance with CiPA \cite{colatsky2016comprehensive}. These can be determined from \textit{in vitro} drug data to represent the effect of any drug, $d$, at any concentration, $c$, on cardiac function by a single vector \(\bm{x}(d,c) \in [0,1]^7 = \Omega\) (\textit{Materials and Methods}) \cite{romero2018silico}. Hence the target drug and its relevant concentration range determine a subset of interest $X\subset [0,1]^7$. Given the observed data, $\mathcal{X}$, we wish to estimate $\max_{\bm{x} \in X} f(S, \bm{x})$. Our posterior $f^\mathcal{X}(S,\cdot)$ gives a predictive distribution over this maximum, which we can evaluate using the continuous rank probability score (CRPS) \cite{gneiting2007strictly} since the EP simulators can be used to derive a ground truth value for this quantity (\textit{Materials and Methods}). Later we will also estimate the exceedance probability of the threshold of concern, $\Prob\left[\max_{\bm{x} \in X} f(S, \bm{x}) \geq 0.1\right]$, though we will not evaluate this via a metric since we have no ground truth for this quantity.

\subsection*{Between-Species Learning Improves Prediction of Change in $\apd$ in Humans for Single Drugs}
As a base scenario, we consider evaluating the maximum change in simulated $\apd$ for a single drug, $d$, up to its listed therapeutic concentration, $c_d^*$, with humans as our HF species and one other species as our LF data. To isolate the effect of the data itself, we assume the GP hyperparameters $\bm{\lambda}_\sigma, \bm{\lambda}_\mu$ and $\bm{\lambda}_k$ are fixed to known, optimal values (\textit{Materials and Methods}). We take each of the twelve drugs from the CiPA dataset \cite{colatsky2016comprehensive} individually, and assume that we have observations of $\apd$ for three human subjects and nine animal subjects at concentrations $\frac{1}{3}c_d^*, \frac{2}{3}c_d^*$ and $ c_d^*$. We first condition just on human data, then on human and animal data, and evaluate the difference in CRPS on the maximum $\apd$ change up to $c_d^*$ between the two cases. Resampling these human and animal subjects 100 times to account for BSV, we examine 95\% empirical confidence intervals (CIs) for CRPS across experimental repeats.

\begin{figure*}[ht]
    \centering
    \begin{subfigure}{0.45\textwidth}
        \centering
        \includegraphics[width=\linewidth]{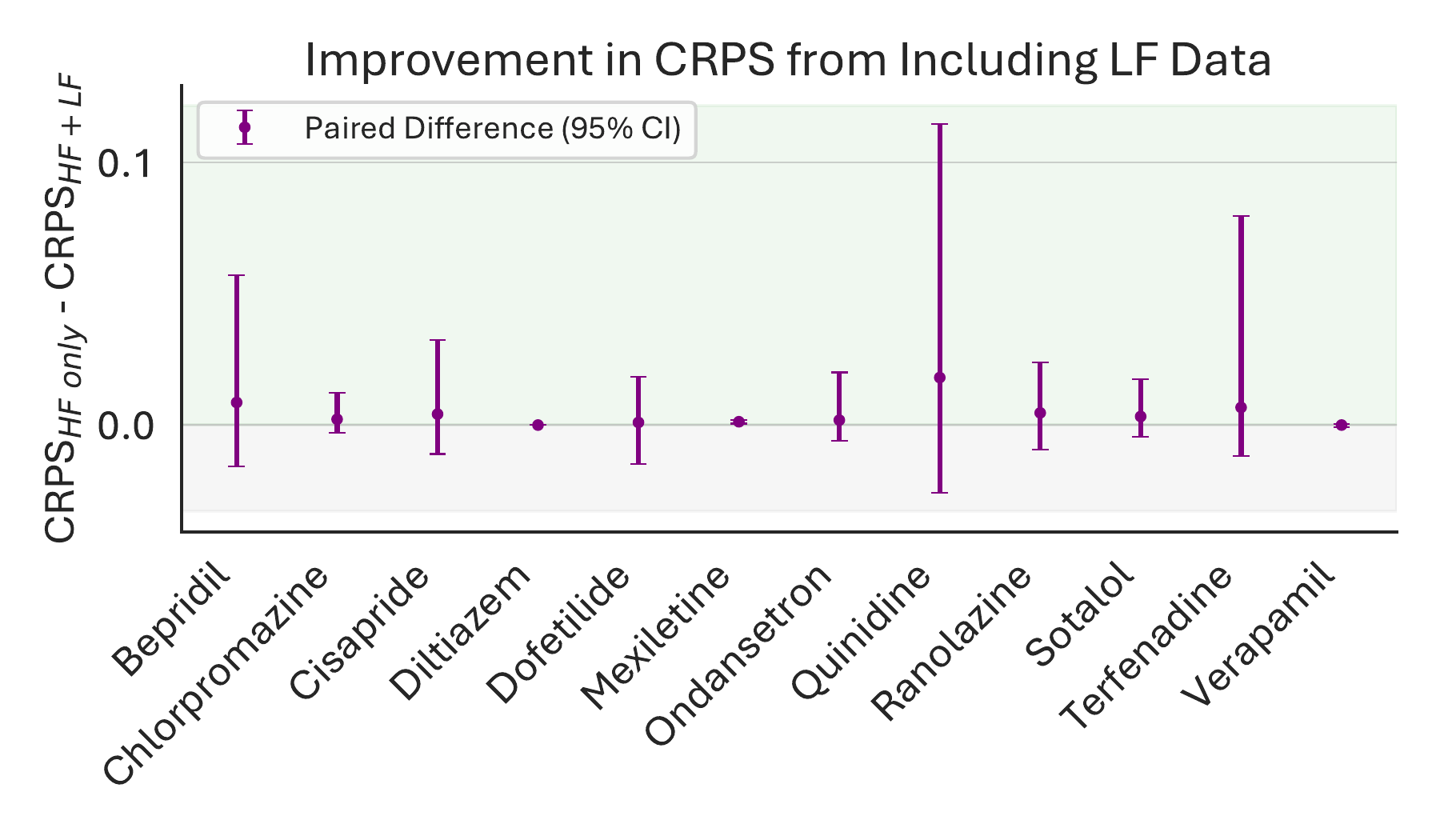}
        \caption{}
        \label{fig:single_drug_crps}
    \end{subfigure}
    \begin{subfigure}{0.45\textwidth}
        \centering
        \includegraphics[width=\linewidth]{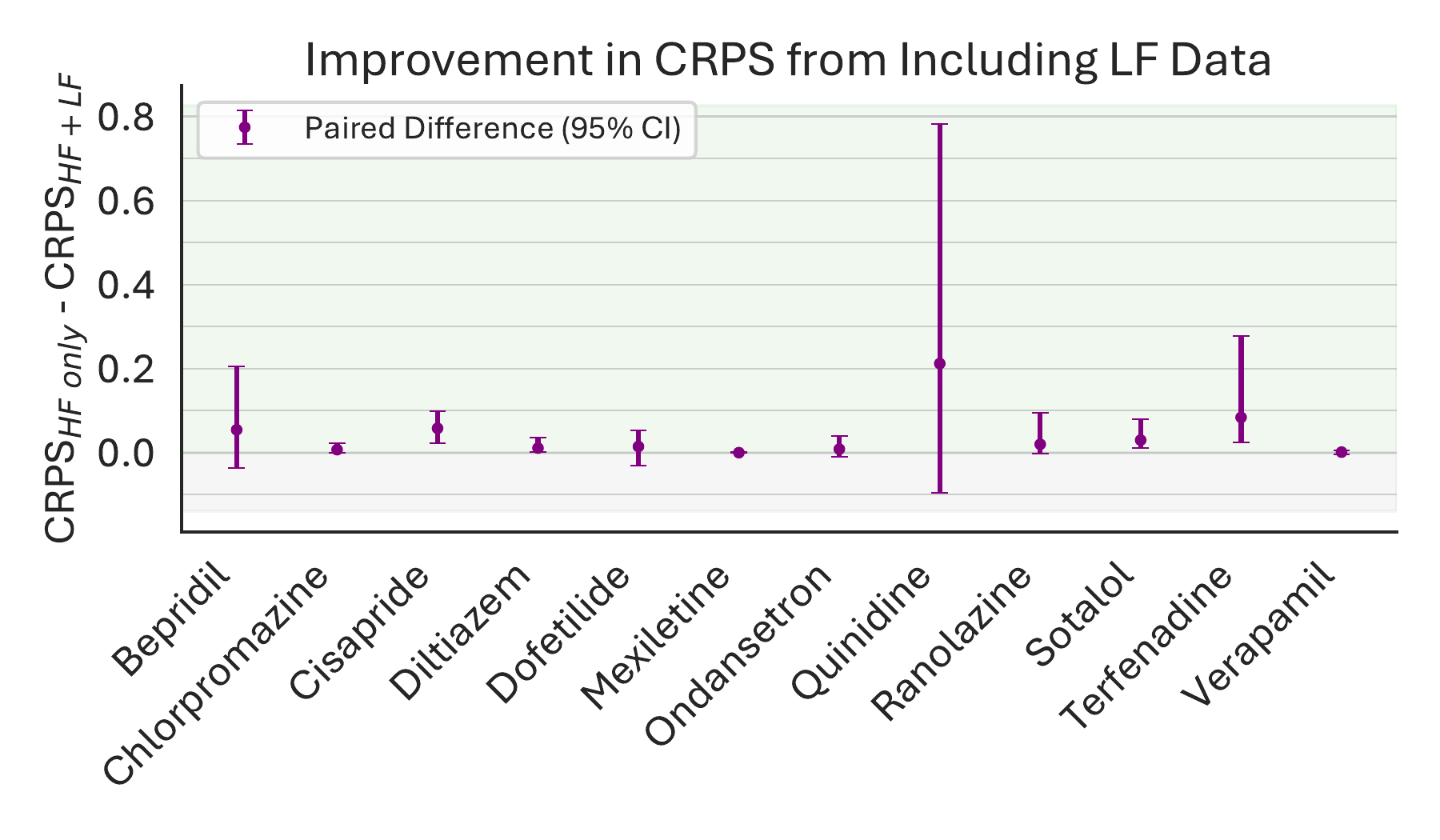}
        \caption{}
        \label{fig:extrapolate_crps}
    \end{subfigure}

    \vspace{0.5cm}
    
    \begin{subfigure}{\textwidth}
        \centering
        \includegraphics[width=17.8cm]{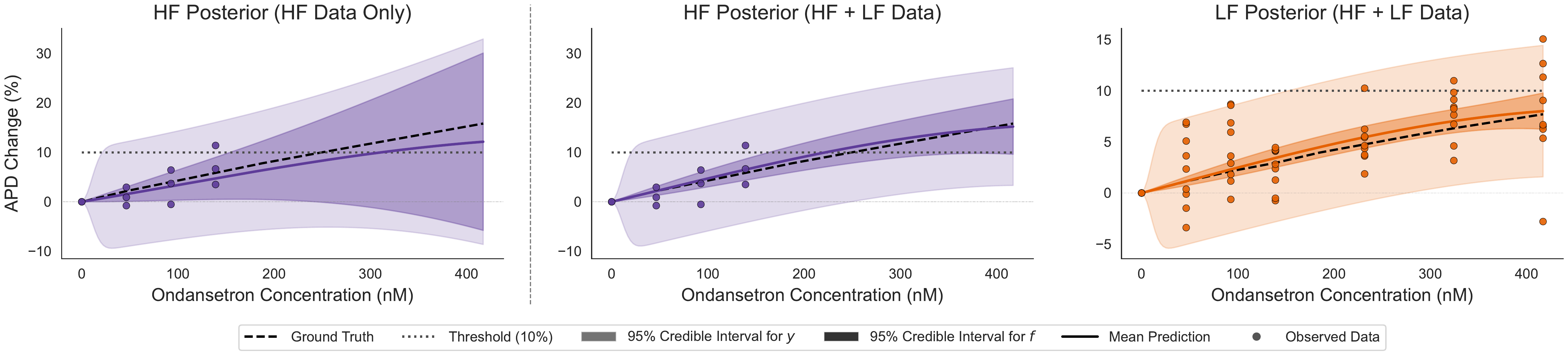}
        \caption{}
        \label{fig:single_drug_posteriors}
    \end{subfigure}
    % Shared caption
    \caption{\textbf{(a)} and \textbf{(b)}: Difference in continuous ranked probability score (CRPS) for prediction of drug-induced maximum prolongation of action potential duration (APD) in humans between a Gaussian process (GP) conditioned only on human data and the same GP further conditioned on animal data. 95\% confidence intervals are shown across 100 randomly sampled training datasets for each drug in the Comprehensive \textit{in vitro} Proarrhythmia Assay dataset. All GP hyperparameters are fixed to known values. In \textbf{(a)}, data from three human subjects and nine animal subjects up to the therapeutic concentrations of each drug are used, while in \textbf{(b)}, the same number of subjects as in (a) are used but human data is available only up to the therapeutic concentration, while animal data extends up to $3\times$ the therapeutic concentration. \textbf{(c)} An example of GP posterior predictive plots for ondansetron-induced change in APD in scenario \textbf{(b)}.}
\end{figure*}

Figure \ref{fig:single_drug_crps} shows the CIs for difference in CRPS for each drug between the HF-only data and the HF and LF data scenarios. While in a few cases CRPS is relatively unchanged (e.g., diltiazem, dofetilide, verapamil), in most cases the majority of the CI lies in the positive region, indicating an improvement in estimation using LF data. Examples of GP posterior predictive distributions with and without LF data are shown in supplementary figure S2.

We have used a 3:9 ratio of HF to LF subjects in order to mimic the clinical reality that it is easier to collect animal data in higher volumes. Examples of how the improvement in CRPS with the addition of LF data varies as the number of HF and LF subjects changes are shown in supplementary figures S3 and S4. For any fixed number of HF subjects, the addition of further LF data generally improves CRPS. This improvement plateaus as the number of HF subjects increases: LF data becomes less influential as HF data becomes richer.

\subsection*{Between-Species Learning Facilitates Predictions of Change in $\apd$ at High Drug Concentrations Not Observed in Humans}
As in the previous scenario for prediction of change in $\apd$, we consider one drug and one LF species at a time, but now assessing the ability of animal data to influence human predictions \textit{beyond} the concentration range tested in humans. Specifically, we assume that we have observations for three human subjects with $\apd$ samples at concentrations $[\frac{1}{3}c^*_d, \frac{2}{3}c^*_d, c^*_d]$ and nine animal subjects with $\apd$ samples at concentrations $[\frac{1}{3}c^*_d, \frac{2}{3}c^*_d, c^*_d, \frac{5}{3}c^*_d, \frac{7}{3}c^*_d, 3c^*_d]$, where we are now interested in the maximum prolongation of $\apd$ in humans up to $3c^*_d$. This imitates the clinical scenario in which a drug must be assessed up to ``significant multiples'' of the therapeutic concentration \citep{ICH2005E14} but these ranges cannot be reached in humans or are avoided for patient safety concerns \citep{vargas2023improving}.

Figure \ref{fig:extrapolate_crps} shows the difference in CRPS for each drug when dogs are used as the LF species. Again we see an improvement in CRPS in almost all cases, with larger improvements observed than in the previous case (figure \ref{fig:single_drug_crps}). Figure \ref{fig:single_drug_posteriors} shows examples of GP posterior predictive distributions for ondansetron with humans as the HF species and dogs as the LF one (examples for other drugs are shown in supplementary figure S5). The addition of dog data is able to improve the estimation of the latent human function, $f$, particularly in ranges where no human data was present. The median exceedance probability for humans increases from 74.0\% (with 95\% CI across repeats [26.4\%, 80.2\%]) in the HF-only model to 96.0\% ([65.4\%, 98.0\%]) with the addition of LF data. This indicates that the LF data is able to improve accuracy and reduce uncertainty in the estimation of the exceedance probability, even when the HF ground truth exceeds the threshold and the LF ground truth does not.

\subsection*{Multi-Fidelity Gaussian Processes Allow Transfer Learning Across Both Species and Drugs}
\begin{figure*}[!ht]
\centering
\begin{subfigure}{\textwidth}
    \centering
    \includegraphics[width=17.8cm]{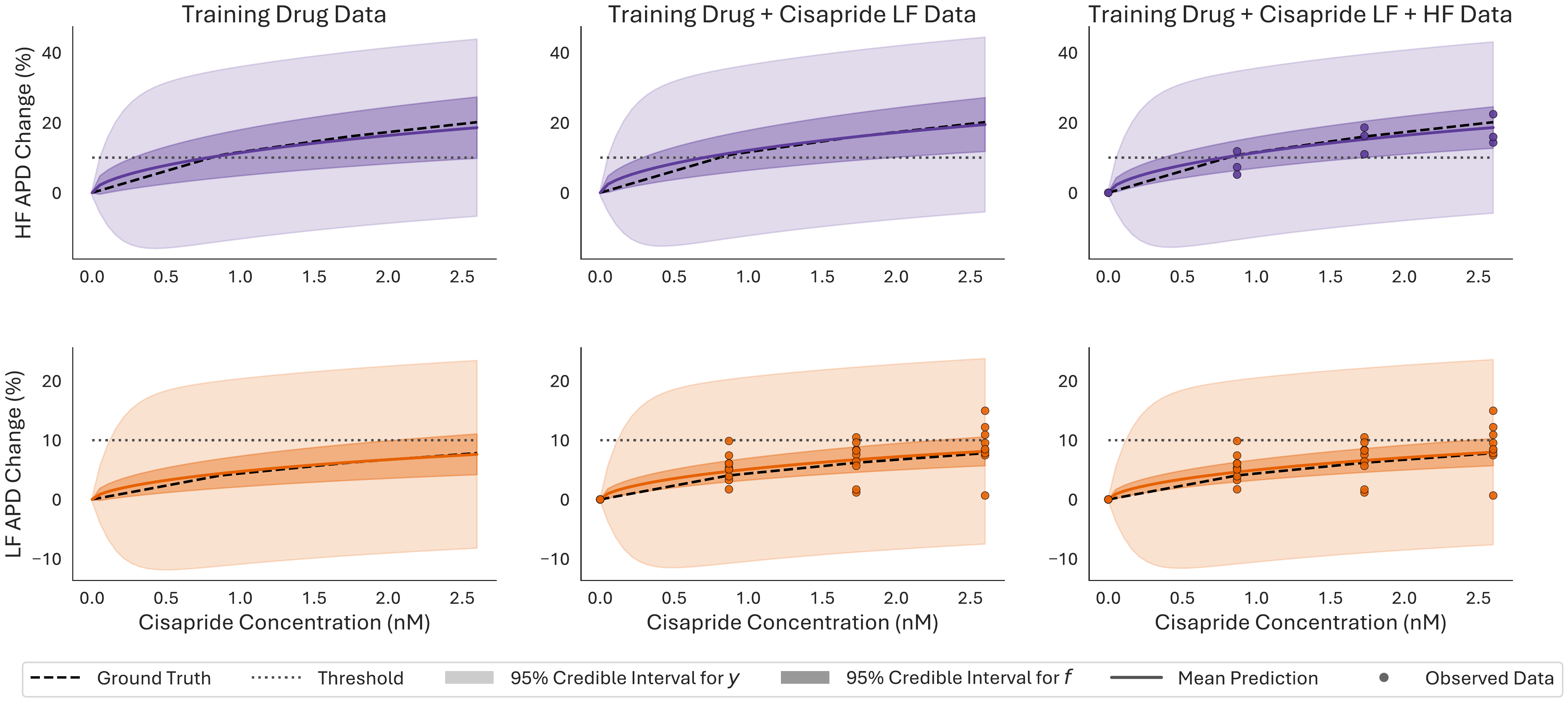}
\caption{}
\label{fig:iterative_posteriors}
\end{subfigure}
\begin{subfigure}{0.6\textwidth}
    \centering
    \includegraphics[width=11cm]{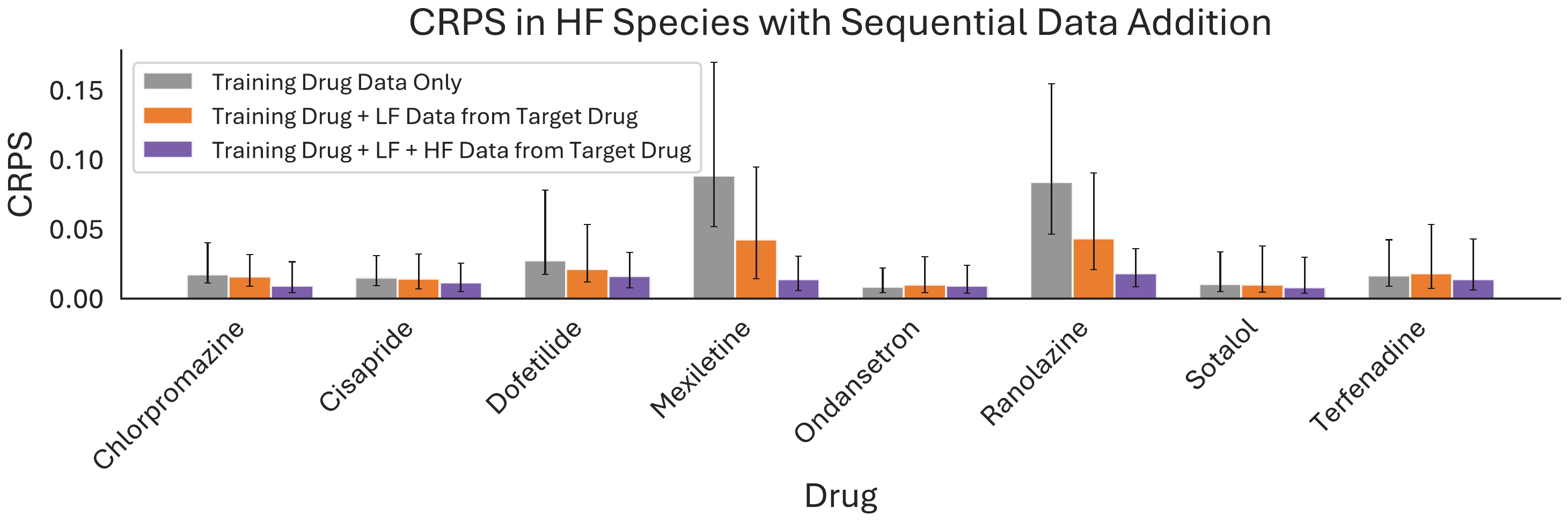}
\caption{}
\label{fig:iterative_crps}
\end{subfigure}
\begin{subfigure}{0.35\textwidth}
    \centering
    \includegraphics[width=4.5cm]{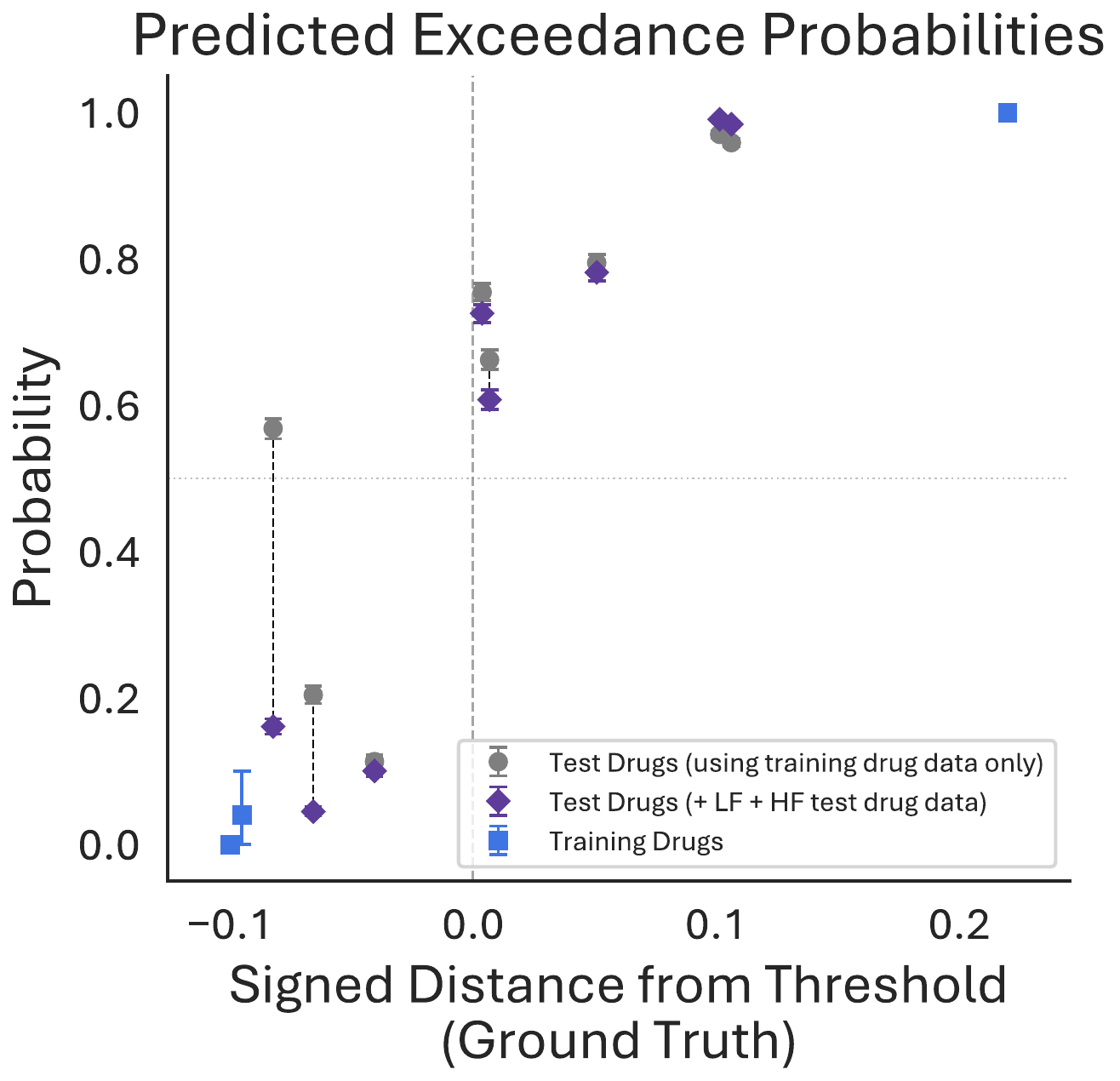}
\caption{}
\label{fig:iterative_exceedance}
\end{subfigure}
\caption{\textbf{(a)} Posterior predictive plots for cisapride-induced change in action potential duration (APD) in humans with three-stage data inclusion for one experimental repeat. Left: a Gaussian process (GP) is first trained on four previous drugs from the Comprehensive \textit{in vitro} Proarrhythmia Assay (CiPA) dataset (bepridil, diltiazem, quinidine and verapamil) with data from humans (high-fidelity (HF) species) and dogs (low-fidelity (LF)). Centre: the GP is then additionally conditioned on cisapride data from nine dog subjects at three concentration levels. Right: human cisapride data from three subjects is added at the same three concentration levels. \textbf{(b)} Continuous ranked probability score (CRPS) on maximum prolongation of APD for eight held-out test drugs in the CiPA dataset, first using only data from the four training drugs (grey), then including LF data for each drug (orange), then finally further including HF data for each drug (purple). 95\% confidence intervals (CIs) are shown across 100 randomly sampled training datasets. \textbf{(c)} Predicted probability of exceeding 10\% prolongation of APD against the true signed distance from this threshold for each of the eight held-out CiPA drugs. Predictions using only previous drug data are shown as grey circles, while purple diamonds indicate predictions after LF and HF data has been added from the test drug. 95\% CIs are shown across 100 randomly sampled training datasets. Quinidine at coordinates (0.8, 1.0) is not shown for readability.}
\label{fig:iterative_long}
\end{figure*}

We now assess the model's ability to simultaneously learn from data on change in $\apd$ across multiple drugs and multiple species. We again use only one LF species, taking three human subjects and nine animal subjects with $\apd$ samples at concentrations $[\frac{1}{3}c_{d}^*, \frac{2}{3}c_{d}^*, c_{d}^*]$ for four drugs, $d= 1, \hdots, 4$. This data is used to fit GP hyperparameters and obtain a GP posterior predictive (see supplementary figure S6). We then evaluate CRPS on predictions of the maximum $\apd$ change up to $c_{d}^*$ on the remaining eight CiPA compounds, i.e. $d=5,\hdots,12$, individually in three stages: (i) using no data from the held-out drug, (ii) after further conditioning the GP on animal data from the drug (again nine subjects with samples at concentration $[\frac{1}{3}c_{d}^*, \frac{2}{3}c_{d}^*, c_{d}^*]$), and finally (iii) after conditioning on the same data as in (ii), plus further human data from the drug (three subjects with samples at the same concentrations as in (ii)). This full process is repeated 100 times, and CRPS CIs are calculated across experimental repeats; these are shown in figure \ref{fig:iterative_crps} for dog as the LF species.

For most drugs we see a monotonic decrease in CRPS as data from both animals and humans are added, while posterior predictive distributions for the latent function, $f$, also converge towards the ground truth (e.g., cisapride in figure \ref{fig:iterative_posteriors}). In some cases (e.g., chlorpromazine, sotalol), the CRPS effectively plateaus across the three scenarios. This occurs when the predictive distribution from stage (i) already produces an accurate prediction with low variance: additional data in stages (ii) and (iii) will always reduce predictive uncertainty and typically reduce mean absolute error (MAE). The former effect increases CRPS (to penalise overconfidence) while the latter decreases it, meaning the CRPS remains unchanged if these effects are of approximately equal strength; these components are shown in supplementary figure S7, while posterior predictive distributions for all test drugs are shown in figures S8-S10.

Figure \ref{fig:iterative_exceedance} shows the median and 95\% CIs for estimated exceedance probability for each drug between stages (i) and (iii). These already follow a roughly sigmoidal relationship with signed distance from threshold using only stage (i) data. Most drugs remain unchanged with additional data, though certainty is increased (i.e., probabilities move closer to 0 or 1) with additional data in a few cases. For drugs very close to the threshold, the exceedance probability remains close to 0.5 as expected. We emphasise that GPs easily facilitate this sequential inclusion of data by the Gaussian conditioning formula, without needing to refit hyperparameters. 

The four training drugs (bepridil, diltiazem, quinidine and verapamil) were chosen due to their diverse effects on $\apd$ (supplementary figure S11) and their propensity to block different ICs to different degrees (supplementary figures S12 and S13). In fact, comparative results can be obtained using just quinidine and verapamil as training drugs (supplementary figure S14), since quinidine inhibits all seven ICs but predominantly blocks $I_{\mathrm{Kr}}$, the most influential IC in prolonging $\apd$, while verapamil inhibits $I_{\mathrm{Ca,L}}$, the IC which shortens $\apd$ the most, and $I_{\mathrm{Kr}}$ equally. Hence using these two drugs alone the GP is able to learn the effect of blocking each of these channels, and the interactions between them, across both HF and LF species. Training only on quindine and diltiazem, which effectively blocks only $I_{\mathrm{Ca,L}}$, we see poor predictions on verapamil (supplementary figure S14), since the interaction between $I_{\mathrm{Kr}}$- and $I_{\mathrm{Ca,L}}$-block differs between species (supplementary figure S11) and cannot be learned from training data which blocks them independently. 

\subsection*{Multi-Fidelity Gaussian Processes Can Quantify Between-Species Similarity}
So far we have shown results using only simulated dog data as the LF species, since these are most commonly used for QTc assessments \cite{vargas2023improving}. In the supplementary materials, we repeat the last experiment using simulators for various other species: guinea pig, mouse, pig and rabbit \cite{pasek2008model, li2010mathematical, gaur2021computational, mahajan2008rabbit}. Figures S15 and S16 show the change in CRPS on held-out test drugs as data is added, as in figure \ref{fig:iterative_crps}.

Guinea pig, pig and rabbit data show consistent improvements or little change in CRPS across all drugs (note dofetilide is included as a training drug when using guinea pig data, since the guinea pig simulators failed to converge under quinidine administration). When using mouse data, CRPS is mostly unchanged and slightly increases in a few cases (though, as when using dog data, this increase can be attributed to a greater decrease in predictive uncertainty than decrease in MAE, see figure S16). This is to be expected from the use of mouse data: trends in simulated mouse data differ the most from human data (figure S11), and regulations currently exclude rodents from QTc analyses due to significant difference in physiology \cite{ICH2005S7B}. 

This difference in between-species similarity can be seen in the optimised GP kernel parameters. Figure S17 shows the values of between-species kernels ($k_{\mathcal{S},q}(s,s')$ in equation \ref{eq:sum_of_separable_kernels}) after fitting the GP to data from all available HF and LF subjects for the four training drugs. Figure S18 shows the corresponding between-species correlation matrices for interpretability. These kernel parameters reflect the expected physiology: similarity to humans is highest for dogs, lowest for mice, while the other three species lie in between. Hence we see that with sufficient training data, the model is robust to negative transfer from LF data and is able to effectively quantify between-species similarity in an interpretable manner. Naturally, in a scenario where training data is insufficiently informative, we anticipate that HF predictions could be worsened by learning spurious relationships with LF data. We emphasise that biological domain knowledge is crucial in informing the selection of training data to avoid negative transfer, in the same way that it should inform kernel selection \cite{hayden2026kernels}.

\section*{Results on Clinical Data}
We now validate the MFGP approach on a small set of real clinical data which represents some of the challenges typically found in multi-species datasets. This data relates to change in heart rate (HR) after administration of an autonomic blocker \cite{dsouza2017targeting,soattin2026endurance}. Since this is only for a single drug, we aim to validate only the between-species learning of the GP approach. In this case, $P=1$ and $f(s, x)$ represents the population average HR after autonomic blockade in species $s$, for those with baseline HR $x > 0$. As in the simulated scenario, $\sigma(s,x)$ accounts for both BSV and measurement noise since we have no subject covariates. Here we will evaluate the posterior predictive distribution $y^\mathcal{X}(S, x) \sim \mathcal{GP}(\mu^\mathcal{X}(S, x), k^\mathcal{X}(S,x)+\sigma(S,x)^2)$ on held-out HF observations $(S,x,y)$ not included in $\mathcal{X}$, via CRPS. We consider two HF-LF combinations: humans as HF with dogs as LF, and dogs as HF with mice as LF (note in the latter case the response variable is the spontaneous beating rate of the isolated and denervated sinoatrial node rather than \textit{in vivo} HR post-autonomic blockade, see \textit{Materials and Methods}).  
\subsection*{Low-Fidelity Data Improves Prediction of Change in Heart Rate After Autonomic Blockade}
As a base scenario, we perform leave-$l$-out cross-validation on HF datapoints and examine the effect as $l$ varies. Specifically, for each $1 \leq l < N_S$ we randomly sample $l$ HF datapoints without replacement, fit GP hyperparameters to the remaining $N_S-l$ HF datapoints, then fit a second GP to the same HF datapoints alongside all available LF datapoints. For each held-out datapoint, we calculate the pointwise difference in CRPS between the two models before averaging across all held-out points. For each value of $l$ we repeat this experiment 100 times, randomly sampling $l$ held-out points each time. 

\begin{figure*}[ht]
\centering
% subfigure
    \begin{subfigure}{0.45\textwidth}
        \centering
        \includegraphics[width=\linewidth]{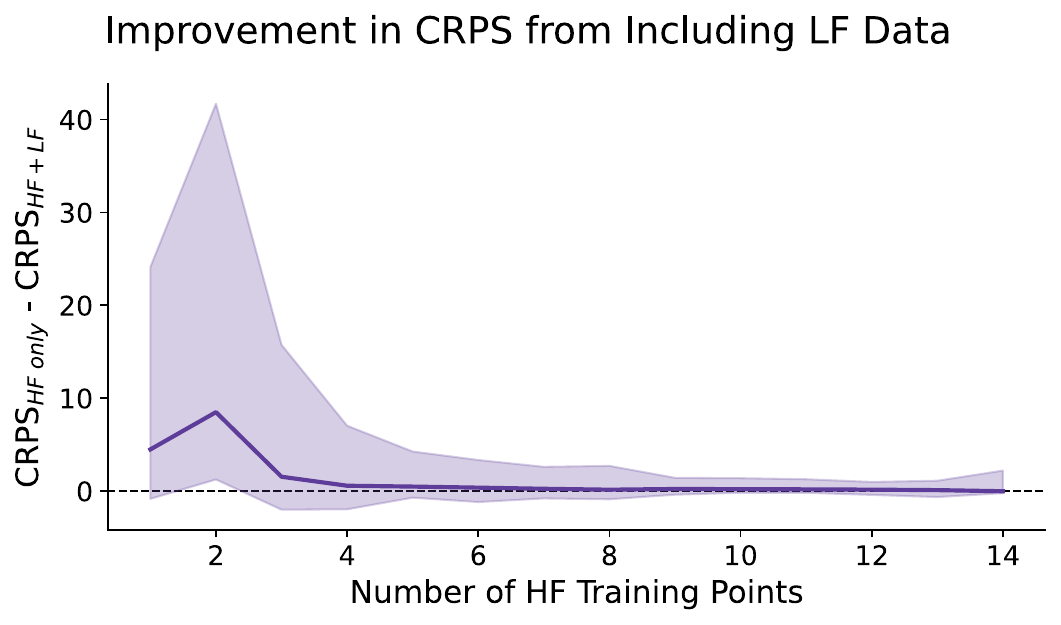}
        \caption{}
        \label{fig:human_hr_sweep}
    \end{subfigure}
    % % Second subfigure
    \begin{subfigure}{0.45\textwidth}
        \centering
        \includegraphics[width=\linewidth]{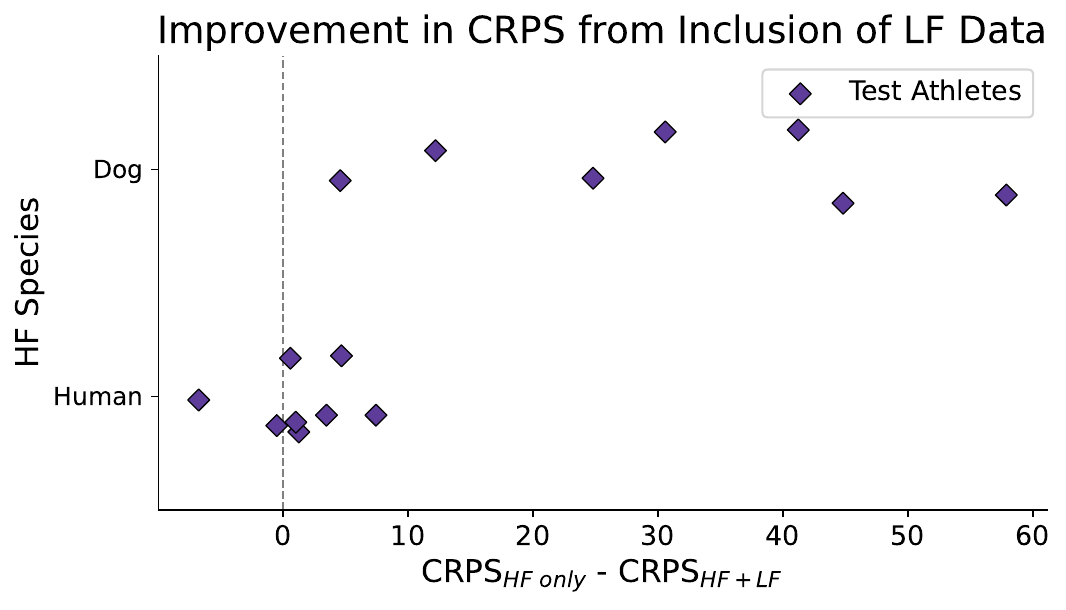}
        \caption{}
        \label{fig:athletes_crps}
    \end{subfigure}

    \vspace{0.5cm}
    
    \begin{subfigure}{\textwidth}
        \centering
        \includegraphics[width=17.8cm]{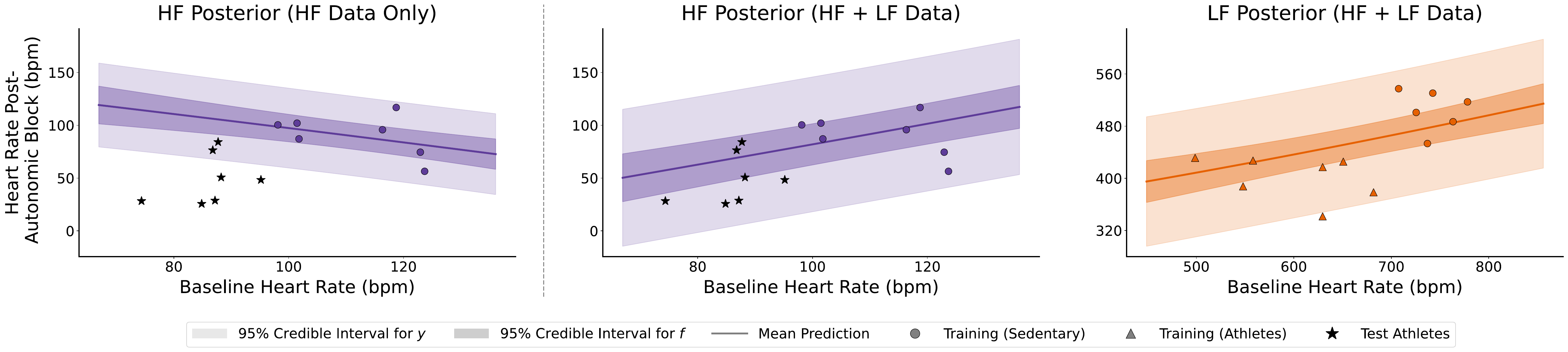}
        \caption{}
        \label{fig:hr_posteriors}
    \end{subfigure}

    % Shared caption
    \caption{Difference in continuous ranked probability score (CRPS) on held-out test data for prediction of drug-induced change in heart rate (HR) between a Gaussian process (GP) trained only on high-fidelity (HF) data versus a GP trained on both HF and low-fidelity (LF) data. In \textbf{(a)}, humans are taken as the HF species and dogs as LF. For each number of HF training points used, all remaining HF data points from the total set of 15 are taken as held-out test data. Median and 95\% confidence intervals for the average difference across all held-out data points are shown across 100 repeats of randomly sampled training data. In \textbf{(b)}, HF athletes are taken as the held-out dataset, while HF sedentary subjects and all LF subjects are used in training. Results are shown for two HF-LF species combinations: dog-mouse and human-dog. \textbf{(c)} Posterior predictive distributions for scenario \textbf{(b)} with dog-mouse species combination. The auxiliary LF data (right) improves estimation of HF trends (centre) when compared to using just the HF data alone (left).}
    \label{fig:hr_crps}
\end{figure*}

Figure \ref{fig:human_hr_sweep} shows the median and 95\% CI for this difference in CRPS across repeats for all values of $N_S-l$, for the human HF scenario. For all values of $N_S-l$ we see the median curve and the majority of the CI region lie above zero, indicating that the addition of LF data is consistently improving HF estimates. This improvement is largest for lower numbers of HF training points, and diminishes as $N_S-l$ increases. The same trend is present in the dog HF scenario, shown in supplementary figure S19.

Supplementary figures S20 and S21 show examples of the posterior predictive distributions in the human HF and dog HF cases respectively, for both models with and without LF data for several values of $N_S-l$. We see that the addition of LF data allows inference of a more representative trend which better generalises to predictions on held-out points. 

In supplementary figures S22 and S23, respectively human HF and dog HF cases, we also explore the effect of varying the number of LF data points used. As in the case with simulated $\apd$ data, for any fixed number of HF data points, additional LF data generally improves CRPS, however at least two LF data points are required for this, since otherwise no trend can be learned in the LF data. This improvement again plateaus as the number of HF data points increases, though this plateau occurs at a lower number of HF data points than in the simulated case. This is likely due to the simple, linear trend present in the clinical data requiring fewer data points to learn, as well as the influence of priors on hyperparameter optimisation (\textit{Materials and Methods}).

\subsection*{Low-Fidelity Data Improves Prediction of Heart Rate Change in High-Fidelity Subgroups with No Observed Data}
We now take a specific subgroup to be our held-out HF dataset. The HR subjects are divided into sedentary and athletic individuals of each species; we treat athletic HF individuals as our held-out data, and compare posterior predictions for these points when training a GP using only sedentary HF individuals versus training with the addition of LF individuals (sedentary and athletic). Since athletic subjects generally have a lower baseline HR than sedentary ones, this scenario is analogous to the prediction of change in $\apd$ at higher drug concentrations where no HF data is available.

Figure \ref{fig:athletes_crps} shows the change in CRPS for the held-out HF subjects in both the human HF case and dog HF case. We see an improvement in both metrics for almost every point in both cases (with two exceptions in the human HF case, though there is still an improvement on average). This improvement is also reflected in the GP posterior predictives (supplementary figure S24 for human HF and figure \ref{fig:hr_posteriors} for dog HF), most noticeably in the dog HF scenario, in which using the HF sedentary individuals alone leads to the erroneous prediction of a negative trend.

\section*{Discussion}
We have demonstrated how multi-fidelity GPs can facilitate learning between species to improve estimation of clinically-relevant QOIs in human beings, first in a simulated example where we can compare against ground truth values, before validating on a small set of clinical data. We chose to focus on drug exposure as a widely-studied and impactful use-case, but highlight that this is a method applicable to any experiments in which animals and human beings share the same measurable endpoint.

We used state-of-the-art simulators of ventricular myocytes (\textit{Materials and Methods}) in order to generate synthetic data for the effects of drug exposure on cardiac function. In doing so, we hoped to mimic as closely as possible the complexity of clinical measurements across multiple species, in a context where animals are consistently used as human surrogates. With fixed hyperparameters, we have shown that animal data can be used to improve estimation of drug-induced QTc prolongation for a single drug both when human and animal data lie in the same concentration range, as well as when human data lies only in a limited range but animal data is available over the full range of interest. We used fixed hyperparameters in these two cases to isolate the effect of the additional data: in reality the amount of data needed to learn hyperparameters is problem-dependent, though domain knowledge can be used to specify nominal values, prior distributions or feasible ranges. 

We have also shown how this method allows for transfer learning across different experiments over the same input space, specifically across different drugs in the case of simulated QTc prolongation. While providing a method for learning GP hyperparameters, training on previous drugs also has strong implications for the reduction of financial cost and experimentation on live subjects: once only \textit{in vitro} data has been collected for a new drug, the human response can already be predicted. For example, the exceedance probability of cisapride can already be estimated at 97\% (figure \ref{fig:iterative_posteriors}, top left) using only data from four previous drugs, meaning that its toxicity can be determined with high probability without any human or animal testing, and efforts can be concentrated on drug candidates with lower predicted exceedance probabilities. 

This drug-to-drug learning may prove more challenging outside of the simulated context: we have made the assumption that the drug's full effect on QTc prolongation can be represented by a single curve $\bm{x}(d,c) \in [0,1]^7$. While this may be a strong assumption, previous work has shown that drugs can be classified for their early afterdepolarisation risk (a related precursor to TdP) using ion channel inhibition fractions alone \cite{sahli2020classifying}. Furthermore, any other continuous inputs relevant to the drug or the test subjects (e.g., pharmacokinetic parameters, patient covariates) can be included as inputs to the GP to facilitate generalisability. Previous works have shown the capabilities of GPs in other areas of biology with more complex and high-dimensional continuous input data than that which we have considered here \cite{gutierrez2024multi, gardiner2020using}. The poor scalability of exact GP regression may also make learning from larger, historic datasets more challenging, having cubic computational complexity and quadratic memory complexity in terms of number of training data points \cite{williams2006gaussian}. However, many approximate methods exist to mediate this \cite{liu2020gaussian}.

As validation of our approach in the single-drug scenario, we used a small set of clinical data for drug-induced change in HR across multiple species. Here we are able to learn hyperparameters directly from observed data and improve estimation of this QOI, while placing principled priors on relevant hyperparameters to prevent overfitting to a small dataset (\textit{Materials and Methods}). In more complicated examples with, e.g., higher-dimensional and more varied inputs, prior specification becomes more challenging; further validation within these scenarios is required.

Future work could extend this approach to other surrogates for human experiments beyond animals. For example, other \textit{in vitro} systems such as organoids or organs-on-a-chip, as well as \textit{in silico} models such as digital twins \cite{mehta2025modernizing} can also be included as more fidelities in the GP model. Depending on the relationships between these fidelities, this may necessitate a more sophisticated between-fidelity kernel structure, which may present challenges: misspecification between the kernel and the underlying data structure can lead to negative transfer \cite{hayden2026kernels}. Despite this, we emphasise that this kernel specification also facilitates interpretability and provides a principled method to incorporate biological domain knowledge (\textit{Materials and Methods}).

We believe this framework presents a useful tool for translational research by balancing non-parametric flexibility with interpretable model specification, while retaining inbuilt uncertainty quantification. Moreover, by facilitating the integration of multiple sources of data, easily allowing sequential data inclusion and providing a probabilistic surrogate through which experimental design can be optimised \cite{folch2023combining}, this method has great potential for reducing the number of animal and human test subjects required in biomedical research.

\matmethods{
\subsection*{Gaussian Process Modelling}
\subsubsection*{Multi-Fidelity Architectures}
Given $P$ continuous inputs $\bm{x} \in \Omega \subset \Reals^P$ and available species \(s \in \mathcal{S} = \{1, \hdots, S\}\), we wish to characterise an underlying response function \(f(s,\bm{x}):\mathcal{S} \times \Reals^P \mapsto \Reals\). To model this as a Gaussian process (GP) requires specifying mean and kernel functions $\mu, k$ as per equation [\ref{eq:gp_fns}]. Specifying a kernel between species can be considered as a multi-output or multi-task objective, for which many different architectures exist \cite{alvarez2012kernels}. We use a sum of $Q$ latent separable kernels as per equation [\ref{eq:sum_of_separable_kernels}]. The between-species kernels are parameterised as 
\[k_{\mathcal{S},q}(s,s') = \left(\bm{b}_q\bm{b}_q^\top + \diag(\bm{v}_q)\right)_{s,s'} \;,\] 
for vectors $\bm{b}_q \in \Reals^{S}$ and $ \bm{v}_q  \in \Reals^{S}_{\geq 0}$, resulting in $Q$ coregionalisation matrices of dimension $S \times S$ \cite{gardner2018gpytorch}. This construction guarantees that the kernel is positive semi-definite, while also partially separating variance and covariance terms: $\bm{v}_q$ contributes only to the within-fidelity variance, while $\bm{b}_q$ determines both within-fidelity variance and between-fidelity covariance.

We specify different continuous kernels $k_{\Omega,q}(\bm{x}, \bm{x}')$ for each dataset. When modelling simulated $\apd$ data, we have the boundary constraint that $f(s,\bm{0})=0$. To incorporate this, we use a constant prior mean $\mu(s,\bm{x})=0$ and modified automatic relevance determination (ARD) kernel 
\[k_{\Omega,q}(\bm{x}, \bm{x}') = \bm{x}^\top \exp\left(-\sum_{p=1}^P (x_p - x_p' )^2 / (2\ell_{q,p}^2)\right)\bm{x}' \;.\]
This construction means that at $\bm{x}=\bm{0}$, $f$ cannot deviate from the prior mean and will always have zero variance. Note that the usual scale parameter is omitted from $k_{\Omega,q}$ since this is incorporated in the between-species kernel $k_{\mathcal{S},q}(s,s')$ when $s=s'$. The use of per-dimension lengthscales $\ell_{q,p}$ when the number of latents $Q>1$ allows between-species similarity to vary across the continuous input space $\Omega$, i.e. a non-separable kernel. When modelling data from a single drug, we use the concentration $c$ as a continuous input instead of the vector of channel inhibitions, $\bm{x}$, i.e. $k_{\Omega,q}(\bm{x}, \bm{x}') = k_{\Omega,q}(c, c')= c \exp(-\lvert c -c ' \rvert^2 / (2\ell_q^2))c'$. With a univariate continuous input, choosing $Q>1$ now only allows between-species similarity to vary across scales. We found $Q=2$ to be sufficient in all experiments using $\apd$ data. 

For the clinical heart rate data, we use a simpler GP structure, with $Q=1$, a linear prior mean $\mu(s,x)=mx$ (where $m \in \Reals$ is another learned hyperparameter, shared for all $s$) and standard RBF kernel $ k_{\Omega}(x, x') = \exp(-\lVert x - x' \rVert_2^2 / (2\ell^2))$.

\subsubsection*{Noise Models}
We use a Gaussian likelihood in all GP models trained. For $\apd$, we model the standard deviation $\sigma(s,\bm{x})$ as heteroskedastic noise in terms of $\bm{x}$, since more variability is observed for higher drug concentrations, i.e. $\bm{x}$ of larger magnitude. Specifically, we model noise as:
\begin{equation*}
    \sigma(s,\bm{x}) = \sigma_{\text{min}} + (\sigma_{s,\text{max}} -\sigma_{\text{min}})(1-\exp(-(\lVert \bm{x}  \rVert_2^2/\rho_s)^2) \;,
\end{equation*}
such that as $\lVert \bm{x}  \rVert$ increases, $\sigma(s,\bm{x})$ tends to homoskedastic noise with standard deviation $\sigma_{s,\text{max}}$, at a rate controlled by $\rho_s$. $\sigma_{\text{min}}$ is required only for numerical stability, and is fixed to a small value (e.g., 10$^{-4}$), while $\sigma_{s,\text{max}}$ is optimised numerically. 

We use homoskedastic noise when modelling heart rate data: $\sigma(s,x) = \sigma$, where $\sigma$ is one shared value across all species and is optimised numerically.

\subsubsection*{Hyperparameter Optimisation}
Several GP hyperparameters remain to be optimised numerically: lengthscales $\ell_{q,p}$, between-species coregionalisation parameters $\bm{b}_q, \bm{v}_q$, observation noise terms $\sigma_{s,\text{max}}$ (or $\sigma$ in the HR case) and $\rho_s$, and mean function coefficient $m$ in the HR model. We obtain point estimates under exact GP inference by maximising the log marginal likelihood augmented with the log-prior terms specified subsequently, i.e. maximum a posterior estimation \cite{williams2006gaussian}. Parameters for which we state no prior are assigned flat priors, so their estimates coincide with maximum likelihood (ML) estimates.

In the single-drug cases for $\apd$ data, all these hyperparameters are fixed to ``oracle'' values obtained by fitting the GP and optimising the hyperparameters to all available HF and LF data for the given experiment. This is done to isolate the effect of additional data (from animals and/or from other drugs) on the quantities of interest and remove the influence of varying hyperparameters. In all other scenarios, hyperparameters are optimised to the specified training data. We place no priors on any hyperparameters when determining the oracle values, since the data is sufficiently rich to avoid the need for them. 

When training on data from multiple drugs, hyperparameter identifiability becomes more challenging due to dimension-specific lengthscales $\ell_{q,p}$, heteroskedastic noise terms $\sigma_{s,\text{max}}$ and $\rho_s$ and diagonal terms in the coregionalisation matrix all simultaneously explaining variance. To address this, we place Lognormal priors of mean 0.1 and standard deviation 0.05 (location parameter $\approx -2.41$ and scale parameter $\approx 0.47$) on the noise growth rates $\rho_s$, since these are effectively nuisance parameters and have the least impact on the estimation of the latent function of interest, $f$.

Since the HR dataset is much smaller, ML estimation of the log marginal likelihood can be susceptible to multiple local optima \cite{williams2006gaussian}, hence we place stronger priors on more parameters to facilitate optimisation. We use a Lognormal prior with mean 10 and standard deviation 2 (location parameter $\approx 2.28$ and scale parameter $\approx0.20$) on the lengthscale $\ell$ and Lognormal prior with mean 1 and standard deviation 0.3 (location $\approx -0.04$ and scale $\approx0.29$) prior on each element of the variance vector $\bm{v}$ to prevent overfitting or strong deviation from a linear trend. We also place a Normal(0,1) prior on each element of the other coregionalisation vector $\bm{b}$ to avoid overfitting to LF trends (since there is more LF than HF data in training).

\subsection*{Simulated $\apd$ Data}
\subsubsection*{Ventricular Myocyte Electrophysiology Simulators} The length of the corrected QT (QTc) interval is mostly determined by the action potential duration (APD) of ventricular myocytes. In practice, it is easier to measure the time taken for the cell to repolarise by 90\% \cite{shaw1997electrophysiologic}, known as $\apd$. $\apd$ can be used as a surrogate for QTc in ventricular myocytes isolated in a laboratory. In the absence of real data, computational models of ventricular myocytes can be used to simulate $\apd$. We will refer to these as electrophysiology (EP) simulators.  

We use these EP simulators to generate synthetic dataset for each drug in each species. The EP simulators typically consist of systems of nonlinear ordinary differential equations (ODEs) which describe how states such as ionic currents (ICs) and ion concentrations in a cardiac muscle cell change over time as the heart beats \cite{mirams2011simulation}. Each model also contains an ODE for the membrane voltage of the cell dependent on all ICs, allowing us to calculate the $\apd$ for any beat. The pacing of each simulator can be standardised to 1Hz, i.e. one beat per second. Each ODE system is then run until a limit cycle is reached, defined as all ODE states being periodic with period one second. See supplementary section S1 for more details on how convergence to a limit cycle is assessed. 

From the CellML online repository \cite{lloyd2008cellml}, we took code for EP simulators of dogs \cite{benson2008canine}, guinea pigs \cite{pasek2008model}, mice \cite{li2010mathematical} and rabbits \cite{mahajan2008rabbit}. Code for the Gaur et al. pig simulator is available from the original publication's supplementary materials \cite{gaur2021computational}, while code for the Tomek et al. (2020) human simulator is available from the authors' GitHub repository \cite{tomek2019development, tomek2020tor}. Simulators were preferentially chosen based on code availability, ability to converge to a stable limit cycle, and numerical stability under modification of simulator parameters to mimic BSV and drug exposure (see later sections and supplementary section S2). 

\subsubsection*{Between-Subject Variability}
The EP simulators described above are published with fixed point estimates for the values of each of their internal parameters. In order to simulate BSV, we create a population of simulators by perturbing their parameters.

Let $\bm{\theta}_s = \left(\theta_{s,1}, \hdots, \theta_{s,P^\theta_s}\right) \in \Reals^{P^\theta_s}$ denote the ODE parameters for species $s$.  Parameters equal to zero are not perturbed since these indicate parameterisations previously removed from the model. For each nonzero parameter, we propose an alternative value $\tilde{\theta}_{s,p} = \text{sgn} (\theta_{s,p})\cdot \phi_{s,p}$, where $\text{sgn} \in \{-1,1\}$ denotes the sign of $\theta_{s,p}$ and  $\phi_{s,p}$ is drawn from a Lognormal distribution with mean $\lvert\theta_{s,p}\rvert$ and standard deviation $\nu_{s,p}$. $\nu_{s,p}$ is chosen such that $\mathbb{P}\left[\frac{1}{2}\lvert \theta_{s,p} \rvert \leq \phi_{s,p}\leq 2\lvert \theta_{s,p} \rvert\right] = 0.95$, to ensure a wide range of values are explored. 

Once all $\theta_{s,p}$ have been perturbed for $p=1,\hdots,P^\theta_s$, the simulator is run until a limit cycle is reached, if possible. The $\apd$ value and several other features of the EP simulator states are then compared to published feasible ranges in order to verify that the simulated subject's parameters are biologically plausible \cite{passini2019drug}. If the features are not plausible or the simulator does not converge to a stable limit cycle, new alternative values $\tilde{\theta}_{s,p}$ are proposed for each $p=1,\hdots,P^\theta_s$ and the process is repeated. See supplementary section S2 for more details on this verification process. We repeat this process until we obtain a population of 30 subjects for each species, denoting their parameters $\bm{\theta}_s^{(1)}, \hdots, \bm{\theta}_s^{(30)}$.

\subsubsection*{Ion Channel Inhibition Curves}
A drug affects the QT interval and the $\apd$ by reducing the conductance of one or more cardiac ICs. The exposure-response relationship between concentration, $c$, of drug $d$ and inhibition fraction of any cardiac channel, $i$, is modelled as a sigmoidal curve known as the Hill equation \cite{romero2018silico}:
\begin{equation*}
 x_{i}(d,c) = \mathds{1}(c >0)\left( 1 + \left(\frac{\mathrm{IC}_{50}(d,i)}{c}\right)^{b_{d,i}}\right)^{-1} \;,
\end{equation*}
where $ \mathds{1}$ is the indicator function, \(x_{i} \in [0,1]\) represents the proportion of the channel which is blocked, $\mathrm{IC}_{50}(d,i)$ is the concentration of drug $d$ which attains 50\% inhibition of channel $i$, and $b_{d,i}>0$ is the Hill coefficient. $\mathrm{IC}_{50}(d,i)$ and $b_{d,i}>0$ are estimated from \textit{in vitro} cell experiments; these are specific to each drug-channel pair $(d,i)$ but independent of species $s$. We use estimates for the values of $\mathrm{IC}_{50}(d,i)$ and $b_{d,i}$ as published by the CiPA Initiative for the seven cardiac ICs relevant to the QT interval and for twelve drugs of known cardiac risk level (supplementary section S3) \cite{crumb2016evaluation, strauss2019comprehensive}. This gives a seven-dimensional inhibition fraction vector for any drug at any concentration: $\bm{x}(d,c) = (x_1(d,c), \hdots, x_7(d,c)) \in [0,1]^7$. We assume the drug's full impact on $\apd$ / QTc is measured by these inhibition curves, and hence we are able to learn from data across different drugs simultaneously.

If the EP simulator for species $s$ contains a channel $i$, the maximum conductance of this channel exists as an explicit parameter in the simulator, i.e. $\theta_{s,j}$ for some $j\in\{1, \hdots, P^\theta_s\}$. To model the impact of a drug at concentration $c$, we scale $\theta_{s,j}$ by $(1-x_{i}(d,c))$ \cite{romero2018silico}.

\subsubsection*{Measuring Drug-Induced Change in $\apd$}
We define $h \left(s,\bm{x};\bm{\theta}_s \right) > 0$ as the $\apd$ for species $s$ with parameters $\bm{\theta}_s$ and inhibition vector $\bm{x}$. $\apd$ is only measured once the simulator has converged to a limit cycle; datapoints are removed if the simulator fails to converge to a stable limit cycle for the given parameters $\bm{\theta}_s$ and inhibition vector $\bm{x}$, since $h \left(s,\bm{x};\bm{\theta}_s \right)$ cannot be meaningfully measured. We calculate the relative change in $\apd$ from baseline as:
\begin{equation*}
    f\left(s,\bm{x};\bm{\theta}_s\right) \defeq \frac{h \left(s,\bm{x};\bm{\theta}_s \right) - h \left(s,\bm{0};\bm{\theta}_s \right)}{h \left(s,\bm{0};\bm{\theta}_s \right)} \;.
    \label{eq:f_definition}
\end{equation*}
Supplementary figure S11 shows a LOWESS curve of the relationship between drug concentration and relative change in $\apd$ for all CiPA drugs, using all simulated subjects for each species. 

Finally, Gaussian noise $\epsilon \sim \mathcal{N}(0,0.03^2)$ is added to each value of $f$ in order to simulate measurement uncertainty. The observed relative change, $y$, is then given as: $y(s,\bm{x};\bm{\theta}_s) \defeq f\left(s,\bm{x};\bm{\theta}_s\right) + \epsilon$. When using a GP to model the data, only $s$ and $\bm{x}$ are used as inputs, hence the latent (noiseless) GP approximates the population mean and the GP noise term accounts for variability due to both $\bm{\theta}_s$ and observation noise $\epsilon$. Continuous inputs $\bm{x}$ are z-standardised dimension-wise based on the mean and standard deviation within training data, since different channels reach different maximum inhibition levels in the $[0,1]$ interval within our training dataset. This standardisation also facilitates lengthscales in the ARD kernel in measuring relevance rather than accounting for scale. Observed datapoints $y$ are also z-standardised species-wise based on the mean and standard deviation within training data for each species, so that trends between species can be learned regardless of scaling (see supplementary figure S11).

\subsection*{Clinical Heart Rate Data}
In addition to the simulated $\apd$ data presented above, we use clinical data from two studies which measured intrinsic heart rate (HR) in athletic and sedentary cohorts in humans, dogs and mice \cite{dsouza2017targeting, soattin2026endurance}. Intrinsic HR is measured in humans via complete autonomic blockade (achieved by intravenous injection of 0.04 mg/kg atropine and 0.2 mg/kg propranolol followed by top-up doses). In mice it is measured as the spontaneous beating rate of the isolated and denervated sinoatrial node (SAN), while in dogs it is measured both via autonomic blockade (with the same dosing protocol as in human beings) and in the SAN. 

The dataset consists of 15 human subjects (7 sedentary, 8 athletic), 14 dogs (7 sedentary, 7 athletic) and 13 mice (6 sedentary, 7 athletic). Baseline HR is measured for all subjects, while HR post-autonomic blockade is recorded for all humans and all dogs (bar one athletic subject). HR in the SAN is measured for all dogs and mice. The GP inputs ($x>0$) and outputs ($y>0$) are the baseline and intrinsic HR, respectively. We consider two HF-LF pairings: humans as HF with dogs as LF (where intrinsic HR is measured via autonomic blockade) and dogs as HF with mice as LF (where HR in the SAN is the outcome of interest). This data is summarised in supplementary table S6; detailed descriptions of the data collection methods are available in the original publications \cite{dsouza2017targeting, soattin2026endurance}.

Since HR exists on different scales for each species, we z-standardise inputs $x$ and outputs $y$ separately for each species. For LF species, all their available data is used for z-standardisation in each experiment. In the leave-$l$-out experiments, HF inputs and outputs are standardised using the mean and standard deviation of the $N_S-l$ HF training data points. In the second experiment, when HF athletic subjects are used as the held-out data set, the HF output $y$ is again standardised using just the HF training data (sedentary individuals), while the HF input $x$ is standardised using the full HF dataset (sedentary and athletic). This facilitates between-species learning by aligning the standardised input spaces of LF and HF species. It relies on the assumption of knowing the baseline HRs for the held-out data. Even if these baseline HRs were not known explicitly, the same standardisation could be achieved by assuming that the range of baseline HRs for the athletic cohort is known, a realistic assumption since athletes are known to have lower baseline HRs.

\subsection*{Quantities of Interest}
\subsubsection*{Continuous Ranked Probability Score}
% The posterior of a GP gives us a probability distribution over the function of interest possible values a quantity of interest might take: 
We evaluate model predictions of two different quantities of interest (QOIs): for $\apd$ prolongation, this is the maximum of the function $f(S,\cdot)$ over a specific input range, while for change in HR this is simply the values of held-out observations. Since we have a single observed value for each of these QOIs, and the fitted GP gives us a probability distribution over the possible values these might take, the continuous ranked probability score (CRPS) is an appropriate metric to evaluate the quality of this predictive distribution. It has the following definition: for an observation $x$, and predictive distribution determined by cumulative distribution function (CDF) $F$:
\begin{align*}
    \text{CRPS}(F,x) = \Expectation_F\left[\lvert X-x \rvert \right] - \frac{1}{2} \Expectation_F \left[ \lvert X-X^* \rvert \right] \;,
\end{align*}
where $X$ and $X^*$ are independent copies of a random variable with CDF $F$. Note that this is effectively a generalisation of mean absolute error to non-deterministic predictions \cite{gneiting2007strictly}. Given a set of samples $\tilde{x}_1, \hdots, \tilde{x}_J \sim X$, the CRPS can be approximated as \cite{zamo2018estimation}:
\begin{equation}
    \text{CRPS}(\tilde{x},x) \approx \frac{1}{J}\sum_{j=1}^J \lvert \tilde{x}_j - x\rvert - \frac{1}{2J^2} \sum_{i,j=1}^J \lvert \tilde{x}_i - \tilde{x}_j\rvert \;.
    \label{eq:crps_definition}
\end{equation}
For any drug, $d$, we are interested in determining the maximum relative change in $\apd$ for the human population average in some concentration range $c\in[0,c_{d,\text{max}}]$ (where $c_{d,\text{max}}$ is taken to be either the therapeutic concentration, $c^*_d$, or $3c^*_d$ in our experiments), where we can observe the $\apd$ at some discrete set $\mathcal{C}=\{c_{d,1}, c_{d,2}, \hdots, c_{d,\text{max}}\}$. To obtain a ground truth, we use the average relative change in $\apd$ in the simulated population of 30 subjects at each observed concentration value, i.e. 
\begin{equation*}
    \bar{f}(S,\bm{x}(d,c))=\frac{1}{30}\sum_{j=1}^{30} f\left(S,\bm{x}(d,c);\bm{\theta}_S^{(j)}\right) \;,
\end{equation*} 
for $c \in \mathcal{C}$. Taking the maximum of these values over $c\in \mathcal{C}$, we have a ground truth maximum $\bar{f}_{\text{max}}(S,\bm{x}(d,c_{d,\text{max}}))$. We then approximate the distribution of the maximum of our GP posterior over this concentration range $c\in[0,c_{d,\text{max}}]$ via Monte Carlo (MC) sampling: for a given GP posterior obtained with a dataset $\mathcal{X}$, \(f | \mathcal{X} = f^\mathcal{X} \sim \mathcal{GP}(\mu^\mathcal{X}, k^\mathcal{X})\), we generate a large number of predictions for the curve $f^\mathcal{X}(S,\bm{x}(d,c))$ in the interval $c\in[0,c_{d,\text{max}}]$, and take the maximum of each curve as an estimate for the maximum of interest. Specifically, we take 50 linearly spaced concentrations values between $0$ and $c_{d,\text{max}}$, $\{0=\tilde{c}_{1}, \tilde{c}_{2}\hdots, \tilde{c}_{50}=c_{d,\text{max}}\}$, and draw 100 predictive curves at each point from $f^\mathcal{X}$: 
\begin{align*}
\{&\{f^\mathcal{X}_1(S,\bm{x}(d,\tilde{c}_1)), \hdots, f^\mathcal{X}_1(S,\bm{x}(d,\tilde{c}_{50}))\}, \hdots, \\ &\{f^\mathcal{X}_{100}(S,\bm{x}(d,\tilde{c}_1)), \hdots, f^\mathcal{X}_{100}(S,\bm{x}(d,\tilde{c}_{50}))\}\}
\end{align*}
Taking the maximum of each curve gives a predictive distribution for the maximum: $f^\mathcal{X}_{i,\text{max}} \defeq  \max_{j=1,\hdots,50} f^\mathcal{X}_i(S,\bm{x}(d,\tilde{c}_j))$ for $i=1, \hdots, 100$. These samples can then be used to approximate the CRPS$(f^\mathcal{X}_{\text{max}},\bar{f}_{\text{max}})$ of our predicted distribution for the maximum using equation [\ref{eq:crps_definition}].

When using clinical HR data, we instead evaluate CRPS on individual held out observations since we have no ground truth. For an observation $(x_{\text{obs}}, y_{\text{obs}})$ and GP posterior $y^\mathcal{X} \sim \mathcal{GP}(\mu^\mathcal{X}, k^\mathcal{X} + \sigma^2)$, we again take 100 MC samples from $y^\mathcal{X}$: $y^\mathcal{X}_1(S,x_{\text{obs}}),\hdots y^\mathcal{X}_{100}(S,x_{\text{obs}})$, and approximate the CRPS$(y^\mathcal{X}(S,x_{\text{obs}}), y_{\text{obs}})$ with equation [\ref{eq:crps_definition}].

\subsubsection*{Exceedance Probability of $\apd$}
When using $\apd$ as a surrogate for QT interval, previous studies have taken the threshold for TdP risk to be an $\apd$ increase of 10\% \cite{romero2018silico, davies2012silico}. We can use the predictive distribution for the maximum to estimate an exceedance probability for this threshold $t=0.1$: 
\begin{equation*}
\Expectation\left[\mathds{1}\left(\max_{c\in\left[0,c_{d,\text{max}}\right]}f^\mathcal{X}(S,\bm{x}(d,c)) > t\right)\right] \approx 
\frac{1}{100} \sum_{i=1}^{100} \mathds{1}\left( f^\mathcal{X}_{i,\text{max}} > t \right) \;.
\end{equation*}

\subsection*{Software}
All data generation and model fitting was carried out in Python 3.13.1 and assisted by Claude Sonnet 4.6 \cite{anthropic_claude_code_2026}. The CellML files for each EP simulator were converted into Python scripts via the OpenCOR software version 0.8.1 \cite{garny2015opencor}. ODE solutions were then approximated using the LSODA solver in SciPy 1.15.2 \cite{2020SciPy-NMeth}. All GPs were implemented in the GPyTorch package version 1.14 \cite{gardner2018gpytorch} and trained with the Adam optimiser \cite{kingma2017adammethodstochasticoptimization} for 100 iterations with learning rate $10^{-1}$. 
}

\showmatmethods{} % Display the Materials and Methods section

\dataavail{All code and data have been deposited in Github. Code for generating synthetic APD data can be found at \url{https://github.com/Isaac-Somerville/apd_drug_effects}, while all GP modelling code, simulated and clinical data are located at \url{https://github.com/Isaac-Somerville/translational_mfgps}.}

% \acknow{}

\showacknow{} % Display the acknowledgments section

% \section*{References}
% \bibsplit[3]
%Use \bibsplit to split the references from the body of the text. Value "[3]" represents the number of reference in the left column (Note: Please avoid single column figures & tables on this page.)

% Bibliography
\bibliography{references}

\ifarXiv
  \agappendix\section{Generation of Simulated Data}

Every action potential duration (APD) reported in this work is measured from a computational simulator of a ventricular myocyte of a given species \cite{benson2008canine, pasek2008model, tomek2019development, tomek2020tor, li2010mathematical, gaur2021computational, mahajan2008rabbit} that has been paced to its limit cycle. The same convergence criterion is applied in all settings, whether the cell is drug free or subject to ion channel block mimicking drug administration, and whether it uses the published parameter set or a perturbed one representing a new virtual subject.

\subsection{Pacing and Feature Extraction}

Each species-specific system of ordinary differential equations (ODEs) is integrated using the LSODA solver in SciPy \cite{2020SciPy-NMeth} under periodic stimulation at a cycle length of \SI{1000}{\milli\second} (1 Hz). Testing for convergence to a limit state takes place at the end of a block of $N_{\mathrm{block}} = 100$ beats. At the end of each block, a feature vector is extracted from the \emph{final beat of that block only}, sampled at $N_{\mathrm{steps}}$ points per beat (Table~\ref{tab:supp-tolerances}). Convergence is therefore assessed by comparing beat $n$ with beat $n - 100$, which makes the test sensitive to slow drift in the ionic concentrations, drift that a beat-to-beat comparison would not resolve.

The feature vector has two parts. The first is a set of nine AP features: the action potential durations at 40\%, 50\% and 90\% repolarisation ($\mathrm{APD}_{40}$, $\mathrm{APD}_{50}$ and $\mathrm{APD}_{90}$); the triangulation $\mathrm{Tri}_{90-40} = \mathrm{APD}_{90} - \mathrm{APD}_{40}$; the maximum upstroke velocity $(\mathrm{d}V/\mathrm{d}t)_{\max}$; the peak membrane potential $V_{\mathrm{peak}}$; the resting membrane potential $\mathrm{RMP}$; and the calcium transient durations at 50\% and 90\% repolarisation ($\mathrm{CTD}_{50}$ and $\mathrm{CTD}_{90}$). An $\mathrm{APD}_{p}$ is located by taking the time of maximum $\mathrm{d}V/\mathrm{d}t$ as the upstroke, setting the repolarisation threshold to $V_{\max} - \tfrac{p}{100}\left(V_{\max} - V_{\min}\right)$, and scanning forward from the peak; $\mathrm{CTD}_{p}$ is defined analogously on the intracellular calcium transient. If the threshold is never crossed within the beat (for instance, when repolarisation fails) the feature is undefined. These features are used since they are easily measurable using the AP waveform and intracellular calcium concentration, and all have published maxima and minima for physiological feasibility in human beings \cite{passini2019drug}.

The second part consists of the minimum and maximum, over the beat, of each intracellular concentration tracked by that species' model. These are $[\mathrm{Ca}^{2+}]_{i}$, $[\mathrm{Na}^{+}]_{i}$, $[\mathrm{K}^{+}]_{i}$ and $[\mathrm{Cl}^{-}]_{i}$, with the exact subset varying between models (Table~\ref{tab:supp-tolerances}); the Pig model resolves a second calcium compartment, $[\mathrm{Ca}^{2+}]_{i,2}$, which is included as well. We include these concentration terms since the membrane potential waveform can appear stationary for many beats while sodium and potassium loading continue to evolve.

\subsection{Convergence Criteria}

Write $g_{s,k}(n)$ for the $k$th feature evaluated on beat $n$ in the simulator for species $s$. The cell is considered to have reached its limit cycle at beat $n$ when
\begin{equation}
\bigl| g_{s,k}(n) - g_{s,k}(n - N_{\mathrm{block}}) \bigr| \;\le\; \tau_{s,k}
\qquad \text{for every feature } k ,
\label{eq:supp-convergence}
\end{equation}
where $\tau_{s,k}$ is the species-specific absolute tolerance of Table~\ref{tab:supp-tolerances}. A single feature outside tolerance is enough to reject convergence and trigger a further block of pacing. Any feature that is undefined on beat $n$ or $n-N_{block}$ likewise prevents convergence being declared.

Because the criterion requires two feature vectors, it can first be evaluated at the end of the second block; a minimum of $200$ beats is therefore always simulated. Pacing continues until~\eqref{eq:supp-convergence} is satisfied or a ceiling of $N_{\max} = 250$ blocks (i.e., $25\,000$ beats, equivalently $25\,000$ seconds of simulated time) is reached. If convergence has not been achieved after 250 blocks, then the resulting simulator is not used (i.e., if perturbing parameters to simulate new subjects, this subject is rejected and another parameter set is sampled; if running to limit state after drug administration, this APD value is not used).

\subsection{Derivation of the Tolerances}
\label{sec:tolerances}
The tolerances are calibrated per species and per feature, from the residual numerical drift that the unperturbed model exhibits once it is already at its limit cycle. The reference cell with  published parameter set is run for a number of beats until all elements of the feature vector are deemed to have converged by empirical examination of their values plotted against number of beats. Its states are then saved as its limit-cycle states. To determine the tolerances for each feature, this model is then paced for $10\,000$ beats from its stored limit-cycle state, with the feature vector sampled every $100$ beats, and $\tau_{s,k}$ is set to the largest successive difference observed across that reference run:
\begin{equation}
\tau_{s,k} \;=\; \max\Bigl( \max_{j} \bigl| g_{s,k}(100j) - g_{s,k}(100(j-1)) \bigr| ,\;
\tau_{s,k}^{\min} \Bigr) ,
\label{eq:supp-tau}
\end{equation}
subject to a floor $\tau_{s,k}^{\min}$ that prevents a tolerance from being set below the resolution at which the feature can be measured. For the time-valued features ($\mathrm{APD}_{p}$, $\mathrm{Tri}_{90-40}$, $\mathrm{CTD}_{p}$) that floor is the sampling interval, $1000 / N_{\mathrm{steps}}$ milliseconds (since no time-valued feature can be measured more finely than one time step); for all other features it is \num{1e-16}.

\subsection{Scenario-Specific Handling}
When simulating a drug block, the cell is warm-started from the stored limit-cycle state of the corresponding drug-free subject, so only the incremental effect of the channel inhibition must be equilibrated. When generating a new virtual subject, the model constants are first perturbed, as described in the main article, and the integration is warm-started from the stored state of the nearest already-simulated subject, measured by relative $L^{2}$ distance over the constant vector $\bm{\theta}_s$; the AP feasibility test of Section~\ref{sec:supp-feasibility} is then applied at each block, and a subject whose limit cycle is feasible is accepted while one that converges to an infeasible cycle is discarded and redrawn under a new random seed.

\begin{algorithm}[ht]
\caption{Pacing to the limit cycle. Shared by the drug-block and
new-subject settings; $N_{\mathrm{block}} = 100$, $N_{\max} = 250$, cycle length \SI{1000}{\milli\second}.}
\label{alg:supp-limit-cycle}
\begin{algorithmic}[1]
\Require initial state $\bm{z}_{s,0}$, constants $\bm{\theta}_s$,
         tolerances $\{\tau_{s,k}\}$
\State $\bm{z}_s \gets \bm{z}_{s,0}$; \quad $m \gets 0$; \quad
       \textit{converged} $\gets$ \textbf{false}
\While{\textbf{not} \textit{converged} \textbf{and} $m < N_{\max}$}
    \State integrate $N_{\mathrm{block}}$ beats from $\bm{z}_s$, sampling
           $N_{\mathrm{steps}}$ points per beat
    \If{the solver fails or the state is non-finite}
        \State \textbf{abort} this run
    \EndIf
    \State $\bm{z}_s \gets$ final state
    \State $\bm{g}_s^{(m)} \gets$ features of the \emph{final beat}
           \Comment{9 AP features $+$ concentration extrema}
    \If{$m > 0$}
        \State \textit{converged} $\gets \bigwedge_{k}
               \bigl[\, g^{(m)}_{s,k}, g^{(m-1)}_{s,k} \text{ both defined}
               \;\wedge\; |g^{(m)}_{s,k} - g^{(m-1)}_{s,k}| \le \tau_{s,k} \,\bigr]$
    \EndIf
    \State $m \gets m + 1$
\EndWhile
\State \Return $\bm{z}_s$, $\bm{g}_s^{(m-1)}$, \textit{converged}
\end{algorithmic}
\end{algorithm}

\begin{table}[ht]
\centering
\caption{Absolute tolerances $\tau_{s,k}$ used in the limit-cycle convergence criterion~\eqref{eq:supp-convergence}, by species and feature. Tolerances are calculated using equation~\eqref{eq:supp-tau}, full procedure described in section \ref{sec:tolerances}. $N_{\mathrm{steps}}$ is the number of samples per beat: this is adjusted per species depending on numerical stability and length of baseline APD at pacing of 1 Hz. A dash indicates a concentration not tracked by that model. Concentration tolerances are given in the units used internally by each model (see note).}
\label{tab:supp-tolerances}
\footnotesize
\sisetup{table-format = 1.2e2}
\begin{tabular}{l l S[table-format=1.2e2] S[table-format=1.2e2] S[table-format=1.2e2] S[table-format=1.2e2] S[table-format=1.2e2] S[table-format=1.2e2]}
\toprule
\textbf{Feature} & \textbf{Unit} & {\textbf{Dog} \cite{benson2008canine}}  & {\textbf{Guinea Pig} \cite{pasek2008model}} & {\textbf{Human} \cite{tomek2019development, tomek2020tor}} & {\textbf{Mouse} \cite{li2010mathematical}} & {\textbf{Pig} \cite{gaur2021computational}} & {\textbf{Rabbit} \cite{mahajan2008rabbit}} \\
\midrule
$N_{\mathrm{steps}}$ per beat & & {1000} & {2500} & {3000} & {7000} & {1000} & {2500} \\
\midrule
$\mathrm{APD}_{40}$ & \si{\milli\second} & 1.00e+00 & 4.00e-01 & 3.33e-01 & 1.43e-01 & 1.00e+00 & 4.00e-01 \\
$\mathrm{APD}_{50}$ & \si{\milli\second} & 1.00e+00 & 4.00e-01 & 3.33e-01 & 1.43e-01 & 1.00e+00 & 4.00e-01 \\
$\mathrm{APD}_{90}$ & \si{\milli\second} & 1.00e+00 & 4.00e-01 & 3.33e-01 & 1.43e-01 & 1.00e+00 & 4.00e-01 \\
$\mathrm{Tri}_{90-40}$ & \si{\milli\second} & 1.00e+00 & 4.00e-01 & 3.33e-01 & 1.43e-01 & 1.00e+00 & 4.00e-01 \\
$(\mathrm{d}V/\mathrm{d}t)_{\max}$ & \si{\milli\volt\per\milli\second} & 7.38e-03 & 2.54e-02 & 1.00e-02 & 2.11e-03 & 1.00e-04 & 6.78e-02 \\
$V_{\mathrm{peak}}$ & \si{\milli\volt} & 2.99e-04 & 1.42e-04 & 1.00e-03 & 7.64e-05 & 1.00e-02 & 3.67e-04 \\
$\mathrm{RMP}$ & \si{\milli\volt} & 1.37e-04 & 7.97e-05 & 1.00e-04 & 3.93e-05 & 1.00e-03 & 3.65e-05 \\
$\mathrm{CTD}_{50}$ & \si{\milli\second} & 1.00e+00 & 4.00e-01 & 3.33e-01 & 1.43e-01 & 1.00e+00 & 4.00e-01 \\
$\mathrm{CTD}_{90}$ & \si{\milli\second} & 1.00e+00 & 4.00e-01 & 3.33e-01 & 1.43e-01 & 1.00e+00 & 4.00e-01 \\
\midrule
$[\mathrm{Ca}^{2+}]_{i}^{\min}$ & & 9.57e-10 & 1.21e-10 & 1.72e-10 & 2.75e-08 & 2.36e-10 & 2.94e-08 \\
$[\mathrm{Ca}^{2+}]_{i}^{\max}$ & & 2.03e-08 & 6.66e-09 & 1.23e-08 & 4.87e-07 & 2.10e-09 & 2.26e-07 \\
$[\mathrm{Ca}^{2+}]_{i,2}^{\min}$ & & {--} & {--} & {--} & {--} & 2.45e-10 & {--} \\
$[\mathrm{Ca}^{2+}]_{i,2}^{\max}$ & & {--} & {--} & {--} & {--} & 1.71e-09 & {--} \\
$[\mathrm{Na}^{+}]_{i}^{\min}$ & & 1.52e-05 & 4.94e-06 & 6.04e-06 & 1.90e-03 & 7.44e-05 & 4.34e-06 \\
$[\mathrm{Na}^{+}]_{i}^{\max}$ & & 1.51e-05 & 4.95e-06 & 6.03e-06 & 1.90e-03 & 7.43e-05 & 4.29e-06 \\
$[\mathrm{K}^{+}]_{i}^{\min}$ & & 1.43e-04 & 2.24e-05 & 8.31e-06 & 1.87e-03 & 7.99e-05 & {--} \\
$[\mathrm{K}^{+}]_{i}^{\max}$ & & 1.43e-04 & 2.24e-05 & 8.33e-06 & 1.88e-03 & 7.99e-05 & {--} \\
$[\mathrm{Cl}^{-}]_{i}^{\min}$ & & 9.04e-05 & {--} & 2.77e-06 & {--} & {--} & {--} \\
$[\mathrm{Cl}^{-}]_{i}^{\max}$ & & 9.04e-05 & {--} & 2.77e-06 & {--} & {--} & {--} \\
\bottomrule
\end{tabular}

\vspace{0.5em}
\begin{minipage}{\textwidth}
\footnotesize
\textit{Note.} Concentration tolerances are expressed in each model's native units:
mM for Dog, Guinea Pig, Human and Pig; \textmu{}M for all Mouse
concentrations and for Rabbit $[\mathrm{Ca}^{2+}]_{i}$, with Rabbit
$[\mathrm{Na}^{+}]_{i}$ in mM.
\end{minipage}

\end{table}

\section{Action Potential Feasibility and Baseline Feature Values}
\label{sec:supp-feasibility}

Perturbing the model constants to generate a new virtual subject can drive the cell into a regime that is numerically stable (i.e., converges to a limit cycle by~\eqref{eq:supp-convergence}) but physiologically implausible, e.g., one that fails to repolarise or that plateaus far from the expected RMP. Each converged limit cycle is therefore additionally screened against a set of feasibility bands. An AP is declared \emph{infeasible} if any of the nine AP features is undefined or falls outside its band.

The bands, shown in Table~\ref{tab:supp-ap-limits}, are constructed by scaling a single human reference band (taken from work by Passini et al. \cite{passini2019drug}) across species. For every animal species, each bound is scaled by the ratio of that species' drug-free baseline value of the feature to the human baseline value of the same feature,
\begin{equation}
\bigl[\, L_{s,k},\, U_{s,k} \,\bigr]
\;=\;
\frac{g^{0}_{s,k}}{g^{0}_{S,k}}
\cdot
\bigl[\, L_{S,k},\, U_{S,k} \,\bigr] ,
\end{equation}
where $g^{0}_{s,k}$ denotes feature $k$ of species $s$ evaluated on the drug-free baseline cell at limit state, with $s=S$ and $s<S$ denoting human and animal species respectively. This keeps the band the same relative width in every species (for instance, $(\mathrm{d}V/\mathrm{d}t)_{\max}$ is always permitted to lie between roughly $0.34$ and $3.4$ times its baseline) while respecting the original scales of each feature for each species. The feature values at limit state for the baseline cell of each species (i.e., $g^0_{s,k}$ for all $s$ and each $k$ which comes from AP features) are shown in Table~\ref{tab:supp-baseline}.

\begin{table}[ht]
\centering
\caption{Drug-free action potential features of the baseline cell simulator (i.e. with unperturbed, published parameter set) for each species, evaluated on the final beat after pacing to the limit cycle at a cycle length of \SI{1000}{\milli\second}.}
\label{tab:supp-baseline}
\small
\sisetup{table-format = -3.2}
\begin{tabular}{l l S S S S S S}
\toprule
\textbf{Feature} & \textbf{Unit} & {\textbf{Dog} \cite{benson2008canine}}  & {\textbf{Guinea Pig} \cite{pasek2008model}} & {\textbf{Human} \cite{tomek2019development, tomek2020tor}} & {\textbf{Mouse} \cite{li2010mathematical}} & {\textbf{Pig} \cite{gaur2021computational}} & {\textbf{Rabbit} \cite{mahajan2008rabbit}} \\
\midrule
$\mathrm{APD}_{40}$ & \si{\milli\second} & 151.15 & 114.42 & 179.06 & 2.86 & 206.06 & 233.29 \\
$\mathrm{APD}_{50}$ & \si{\milli\second} & 175.18 & 129.43 & 195.73 & 3.86 & 221.21 & 256.50 \\
$\mathrm{APD}_{90}$ & \si{\milli\second} & 216.22 & 157.63 & 236.75 & 15.72 & 253.22 & 305.32 \\
$\mathrm{Tri}_{90-40}$ & \si{\milli\second} & 65.07 & 43.21 & 57.69 & 12.86 & 47.16 & 72.03 \\
$(\mathrm{d}V/\mathrm{d}t)_{\max}$ & \si{\milli\volt\per\milli\second} & 52.58 & 147.44 & 297.22 & 167.34 & 121.79 & 268.46 \\
$V_{\mathrm{peak}}$ & \si{\milli\volt} & 35.13 & 47.73 & 31.94 & 34.24 & 45.84 & 48.60 \\
$\mathrm{RMP}$ & \si{\milli\volt} & -85.79 & -85.14 & -90.75 & -86.25 & -87.33 & -87.05 \\
$\mathrm{CTD}_{50}$ & \si{\milli\second} & 220.22 & 23.40 & 146.05 & 139.88 & 355.82 & 402.56 \\
$\mathrm{CTD}_{90}$ & \si{\milli\second} & 523.52 & 158.43 & 346.12 & 509.22 & 672.76 & 733.49 \\
\bottomrule
\end{tabular}
\end{table}

% If this overflows the text block, wrap it in \begin{sidewaystable}[p]
% ... \end{sidewaystable} instead (requires the rotating package).
\begin{table}[ht]
\centering
\caption{Feasible ranges for each action potential feature, by species. A converged limit cycle is rejected as infeasible if any feature cannot be evaluated or falls outside its range (bounds inclusive). The Human range is the fixed reference band \cite{passini2019drug}; all others are obtained by scaling it by the ratio of the species' baseline feature value (Table~\ref{tab:supp-baseline}) to the human baseline value.}
\label{tab:supp-ap-limits}
\footnotesize
\setlength{\tabcolsep}{3pt}
\begin{tabular}{l l c c c c c c}
\toprule
\textbf{Feature} & \textbf{Unit} & {\textbf{Dog} \cite{benson2008canine}}  & {\textbf{Guinea Pig} \cite{pasek2008model}} & {\textbf{Human} \cite{tomek2019development, tomek2020tor}} & {\textbf{Mouse} \cite{li2010mathematical}} & {\textbf{Pig} \cite{gaur2021computational}} & {\textbf{Rabbit} \cite{mahajan2008rabbit}} \\
\midrule
$\mathrm{APD}_{40}$ & \si{\milli\second} & $[72.0,\, 270.9]$ & $[54.5,\, 205.1]$ & $[85.0,\, 320.0]$ & $[1.4,\, 5.1]$ & $[98.1,\, 369.3]$ & $[111.1,\, 418.1]$ \\
$\mathrm{APD}_{50}$ & \si{\milli\second} & $[98.7,\, 313.9]$ & $[72.9,\, 231.9]$ & $[110.0,\, 350.0]$ & $[2.2,\, 6.9]$ & $[124.6,\, 396.4]$ & $[144.5,\, 459.7]$ \\
$\mathrm{APD}_{90}$ & \si{\milli\second} & $[164.5,\, 402.2]$ & $[119.9,\, 293.2]$ & $[180.0,\, 440.0]$ & $[11.9,\, 29.2]$ & $[192.7,\, 471.0]$ & $[232.3,\, 567.9]$ \\
$\mathrm{Tri}_{90-40}$ & \si{\milli\second} & $[56.1,\, 168.2]$ & $[37.2,\, 111.7]$ & $[50.0,\, 150.0]$ & $[11.1,\, 33.4]$ & $[40.6,\, 121.9]$ & $[62.1,\, 186.2]$ \\
$(\mathrm{d}V/\mathrm{d}t)_{\max}$ & \si{\milli\volt\per\milli\second} & $[17.7,\, 176.9]$ & $[49.6,\, 496.0]$ & $[100.0,\, 1000.0]$ & $[56.3,\, 563.0]$ & $[41.0,\, 409.7]$ & $[90.3,\, 903.1]$ \\
$V_{\mathrm{peak}}$ & \si{\milli\volt} & $[10.8,\, 59.5]$ & $[14.7,\, 80.8]$ & $[10.0,\, 55.0]$ & $[10.7,\, 59.0]$ & $[14.1,\, 77.6]$ & $[15.0,\, 82.3]$ \\
$\mathrm{RMP}$ & \si{\milli\volt} & $[-89.8,\, -75.6]$ & $[-89.1,\, -75.1]$ & $[-95.0,\, -80.0]$ & $[-90.3,\, -76.0]$ & $[-91.4,\, -77.0]$ & $[-91.1,\, -76.7]$ \\
$\mathrm{CTD}_{50}$ & \si{\milli\second} & $[181.0,\, 633.4]$ & $[19.2,\, 67.3]$ & $[120.0,\, 420.0]$ & $[114.9,\, 402.3]$ & $[292.4,\, 1023.3]$ & $[330.8,\, 1157.8]$ \\
$\mathrm{CTD}_{90}$ & \si{\milli\second} & $[332.8,\, 1187.5]$ & $[100.7,\, 359.4]$ & $[220.0,\, 785.0]$ & $[323.7,\, 1154.9]$ & $[427.7,\, 1526.0]$ & $[466.3,\, 1663.7]$ \\
\bottomrule
\end{tabular}
\end{table}

\section{Drug Block Parameters}
\label{sec:supp-drug}

Pharmacological block is modelled with a concentration--response (Hill) relationship applied independently to each ion channel. For drug $d$ at free concentration $c$, the fraction of channel $i$ blocked is
\begin{equation}
x_{i}(d,c) = \mathds{1}(c >0)\left( 1 + \left(\frac{\mathrm{IC}_{50}(d,i)}{c}\right)^{b_{d,i}}\right)^{-1} \;,
\label{eq:supp-hill}
\end{equation}
where $ \mathds{1}$ is the indicator function. A channel for which no concentration-response data are available is assigned a multiplier of $1$: i.e., no block. The parameters $\mathrm{IC}_{50}(d,i)$ and $b_{d,i}$ are the optimal fits released by the Comprehensive \textit{in vitro} Proarrhythmia Assay (CiPA) initiative and are reported in Tables~\ref{tab:supp-ic50} and~\ref{tab:supp-hill} \cite{crumb2016evaluation, strauss2019comprehensive}. Concentrations are in nanomolar throughout, and $b_{d,i}$ is dimensionless. The two tables share the same pattern of missing entries: an $\mathrm{IC}_{50}$ and its Hill coefficient are always either both fitted or both absent.

Which of the seven channels actually contributes to the APD depends on the species model, since not every model contains every current. The Human and Dog models include all seven. The Pig model has no $I_{\mathrm{to}}$; the Mouse and Rabbit models have no $I_{\mathrm{NaL}}$; the Guinea Pig model has neither $I_{\mathrm{NaL}}$ nor $I_{\mathrm{to}}$. The effective number of blocked channels for a given drug--species pair is therefore the intersection of the drug's fitted channels with the species' channel set.

\begin{landscape}
\begin{table}[ht]
\centering
\caption{$\mathrm{IC}_{50}$ values (nM) for each drug and ion channel, from the CiPA optimal Hill fits \cite{crumb2016evaluation, strauss2019comprehensive}. $C_{\mathrm{ther}}$ is the free therapeutic plasma concentration. $C_{\mathrm{assay}}$ is the highest concentration screened in the CiPA patch-clamp assay for that drug, and is a property of the experiment rather than a clinical exposure. A dash indicates that no concentration--response fit is available, in which case the channel is left unblocked. $I_{\mathrm{Kr}}$ is the human ether-à-go-go-related gene (hERG) current.}
\label{tab:supp-ic50}
\footnotesize
\sisetup{table-format = 1.3e2}
\begin{tabular}{l S S S S S S S S[table-format=5.1] S[table-format=1.2e2] l}
\toprule
\textbf{Drug} & {$I_{\mathrm{CaL}}$} & {$I_{\mathrm{K1}}$} & {$I_{\mathrm{Ks}}$} & {$I_{\mathrm{Na}}$} & {$I_{\mathrm{NaL}}$} & {$I_{\mathrm{to}}$} & {$I_{\mathrm{Kr}}$} & {$C_{\mathrm{ther}}$} & {$C_{\mathrm{assay}}$} & \textbf{CiPA risk} \\
\midrule
Bepridil & 2.808e+03 & {--} & 2.863e+04 & 2.929e+03 & 1.814e+03 & 8.594e+03 & 5.134e+01 & 33 & 3.00e+03 & High \\
Chlorpromazine & 8.192e+03 & 9.270e+03 & {--} & 4.536e+03 & 4.560e+03 & 1.762e+07 & 9.752e+02 & 38 & 1.05e+04 & Intermediate \\
Cisapride & 9.267e+06 & 2.948e+04 & 8.117e+07 & {--} & {--} & 2.191e+05 & 1.127e+01 & 2.6 & 3.00e+02 & Intermediate \\
Diltiazem & 1.121e+02 & {--} & {--} & 1.109e+05 & 2.187e+04 & 2.822e+09 & 1.314e+04 & 122 & 1.00e+05 & Low \\
Dofetilide & 2.603e+02 & 3.944e+02 & {--} & 3.805e+02 & 7.531e+05 & 1.882e+01 & 6.125e+00 & 2 & 3.00e+01 & High \\
Mexiletine & 3.824e+04 & {--} & {--} & {--} & 8.957e+03 & {--} & 2.907e+04 & 4129 & 3.00e+05 & Low \\
Ondansetron & 2.255e+04 & {--} & 5.698e+05 & 5.767e+04 & 1.918e+04 & 1.023e+06 & 1.325e+03 & 139 & 2.00e+04 & Intermediate \\
Quinidine & 5.159e+04 & 3.959e+07 & 4.899e+03 & 1.233e+04 & 9.417e+03 & 3.487e+03 & 9.863e+02 & 3237 & 1.00e+04 & High \\
Ranolazine & {--} & {--} & 3.616e+07 & 6.877e+04 & 7.884e+03 & {--} & 8.208e+03 & 1948.2 & 1.00e+05 & Low \\
Sotalol & 7.062e+06 & 3.050e+06 & 4.222e+06 & 1.144e+09 & {--} & 4.314e+07 & 1.071e+05 & 14690 & 2.10e+06 & High \\
Terfenadine & 7.004e+02 & {--} & 3.998e+05 & 4.803e+03 & 2.006e+04 & 2.400e+05 & 2.040e+01 & 4 & 8.00e+02 & Intermediate \\
Verapamil & 2.018e+02 & 3.488e+08 & {--} & {--} & 7.028e+03 & 1.343e+04 & 2.956e+02 & 81 & 1.00e+03 & Low \\
\bottomrule
\end{tabular}
\end{table}
\end{landscape}

\begin{table}[ht]
\centering
\caption{Hill coefficients $b$ (dimensionless) for each drug and ion channel, from the CiPA optimal fits \cite{crumb2016evaluation, strauss2019comprehensive}. The pattern of missing entries matches Table~\ref{tab:supp-ic50} exactly: $\mathrm{IC}_{50}$ and $b$ are always fitted jointly.}
\label{tab:supp-hill}
\small
\sisetup{table-format = 1.4}
\begin{tabular}{l S S S S S S S}
\toprule
\textbf{Drug} & {$I_{\mathrm{CaL}}$} & {$I_{\mathrm{K1}}$} & {$I_{\mathrm{Ks}}$} & {$I_{\mathrm{Na}}$} & {$I_{\mathrm{NaL}}$} & {$I_{\mathrm{to}}$} & {$I_{\mathrm{Kr}}$} \\
\midrule
Bepridil & 0.6486 & {--} & 0.7061 & 1.1640 & 1.4160 & 3.5410 & 0.9293 \\
Chlorpromazine & 0.8441 & 0.6878 & {--} & 1.9950 & 0.9379 & 0.3654 & 0.8281 \\
Cisapride & 0.4261 & 0.5133 & 0.2921 & {--} & {--} & 0.2430 & 0.6210 \\
Diltiazem & 0.7142 & {--} & {--} & 0.7022 & 0.6779 & 0.1696 & 0.9119 \\
Dofetilide & 1.1630 & 0.7650 & {--} & 0.8920 & 0.2597 & 0.7712 & 1.0790 \\
Mexiletine & 1.0310 & {--} & {--} & {--} & 1.4090 & {--} & 0.8928 \\
Ondansetron & 0.7526 & {--} & 0.6535 & 1.0200 & 1.0350 & 0.9891 & 0.9210 \\
Quinidine & 0.5892 & 0.3468 & 1.3630 & 1.4940 & 1.3370 & 1.2820 & 0.8404 \\
Ranolazine & {--} & {--} & 0.5191 & 1.4250 & 0.9450 & {--} & 0.8576 \\
Sotalol & 0.8651 & 1.2040 & 1.1670 & 0.5089 & {--} & 0.6632 & 0.7850 \\
Terfenadine & 0.6601 & {--} & 0.5430 & 1.0150 & 0.6011 & 0.2559 & 0.6118 \\
Verapamil & 1.0970 & 0.2728 & {--} & {--} & 1.0310 & 0.8222 & 0.9378 \\
\bottomrule
\end{tabular}
\end{table}

\begin{figure}[ht]
\centering
    \includegraphics[width=12cm]{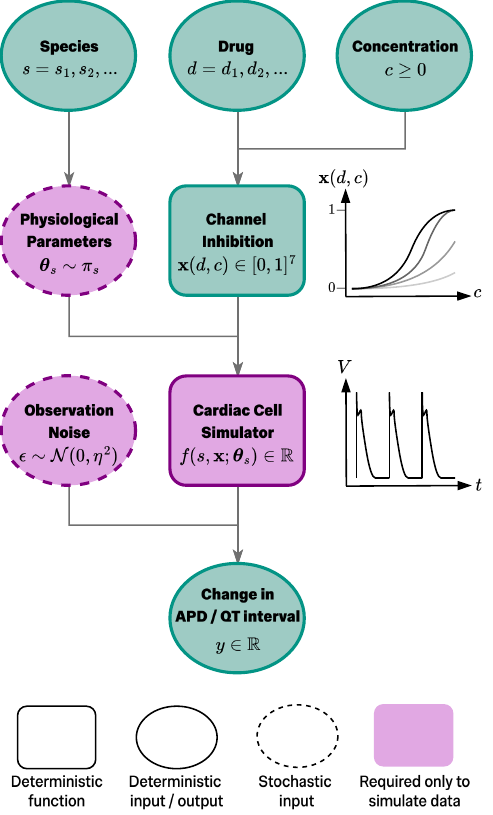}
\caption{Structure of the data-generating process for drug-induced change in QT interval in any species. We assume the effect on QT of any drug at any concentration can be characterised by the Hill concentration-inhibition curves of the seven ionic currents studied in the Comprehensive \textit{in vitro} Proarrhythmia Assay (CiPA) dataset. In the absence of clinical QT data, we use ventricular myocyte simulators to generate synthetic action potential durations (APDs) as a surrogate for QT interval. To mimic between-subject variability, we create virtual subjects by sampling new values for the parameters of these simulators from specified Lognormal distributions. Finally, we add i.i.d. Gaussian noise to each simulated change in APD in order to mimic measurement uncertainty.}
\label{fig:data_generation}
\end{figure}

\begin{figure}[ht]
\centering
    \includegraphics[width=\textwidth]{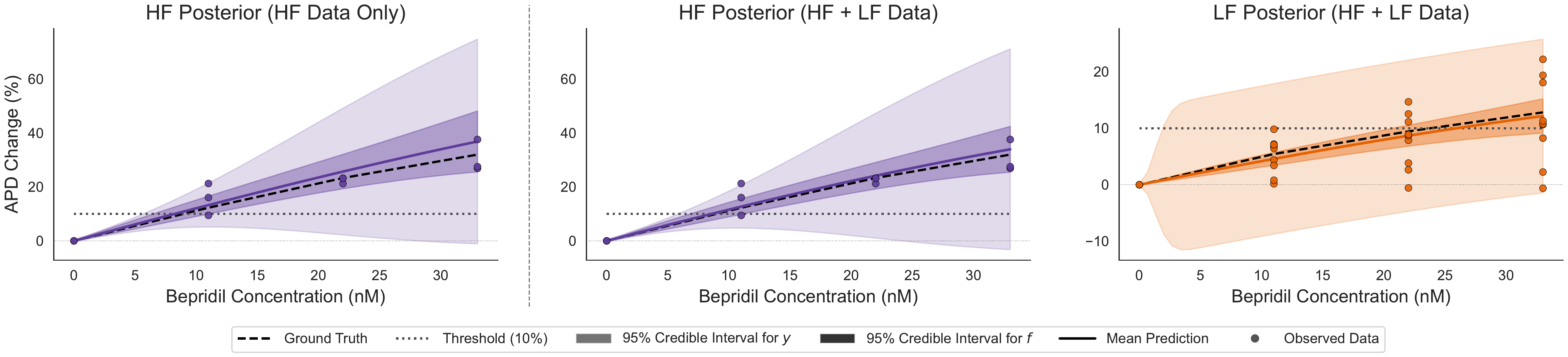}

    \vspace{1cm}

    \includegraphics[width=\textwidth]{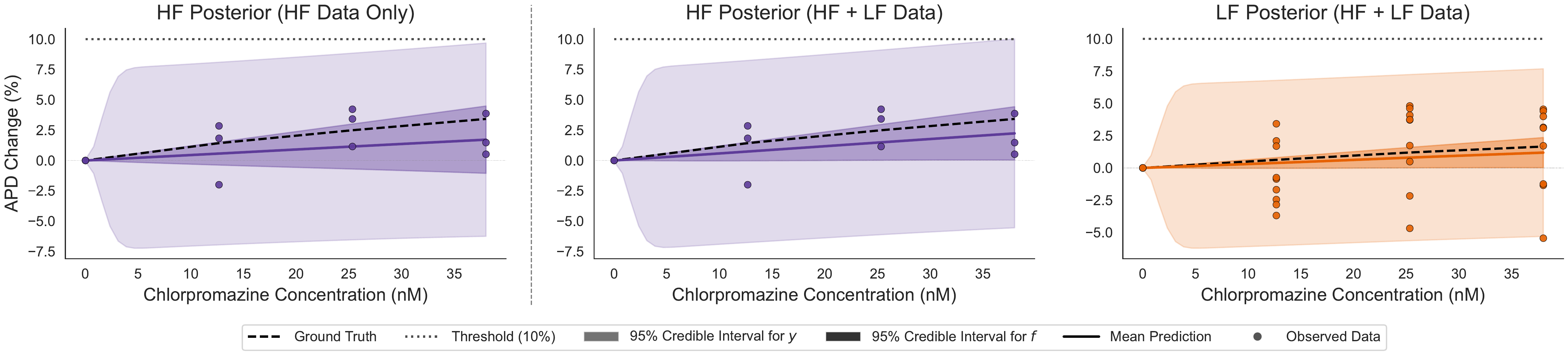}

    \vspace{1cm}
    
    \includegraphics[width=\textwidth]{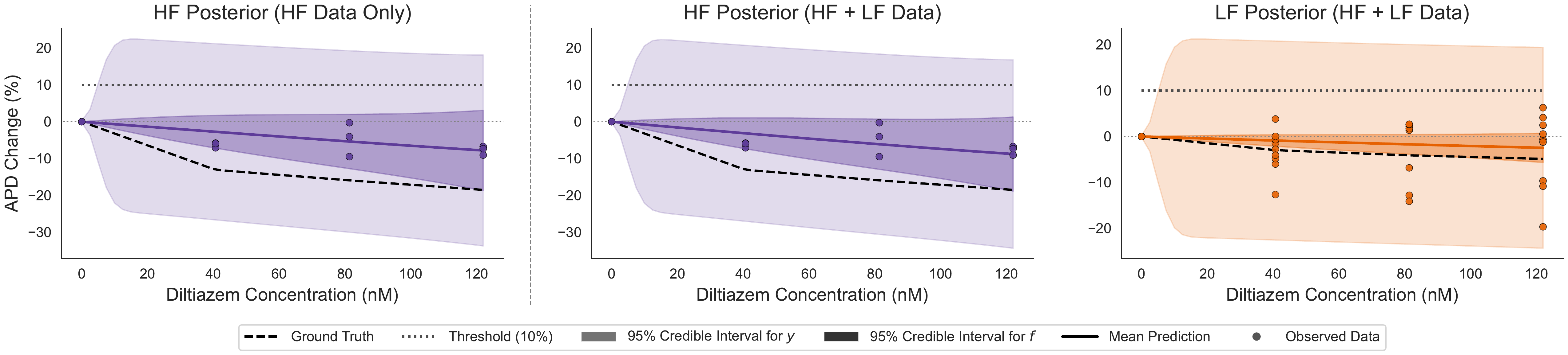}

    \vspace{1cm}

    \includegraphics[width=\textwidth]{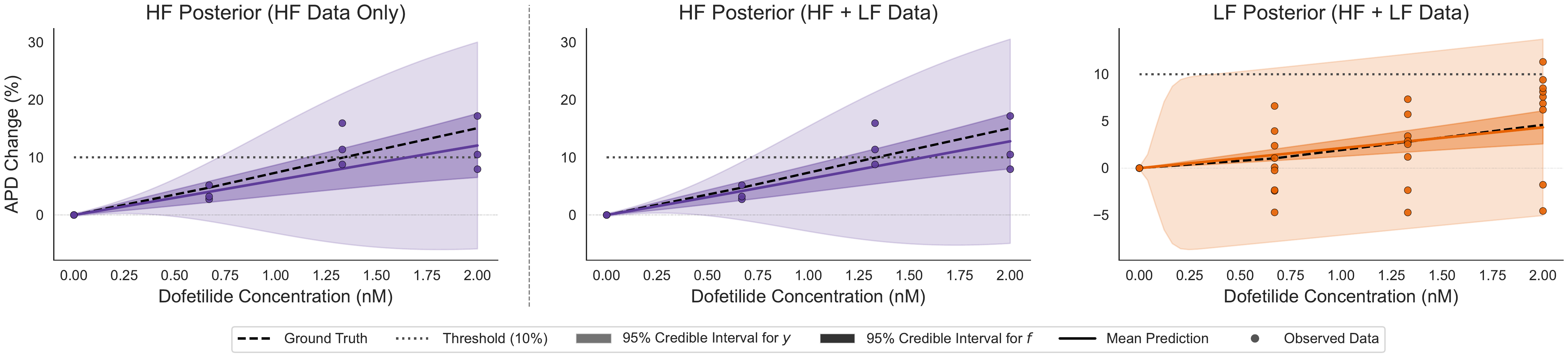}
\caption{Gaussian process (GP) posterior predictive plots for drug-induced change in action potential duration (APD) for four example drugs: bepridil, chlorpromazine, diltiazem and dofetilide. A high-fidelity (HF) only model is conditioned only on data from three human subjects (left), while the HF and low-fidelity (LF) model is conditioned on the same human data (centre) as well as data from nine dog subjects (right). All GP hyperparameters are fixed to known values.}
\label{fig:chlorpromazine}
\end{figure}

\begin{figure}[ht]
\centering
    \includegraphics[width=0.95\textwidth]{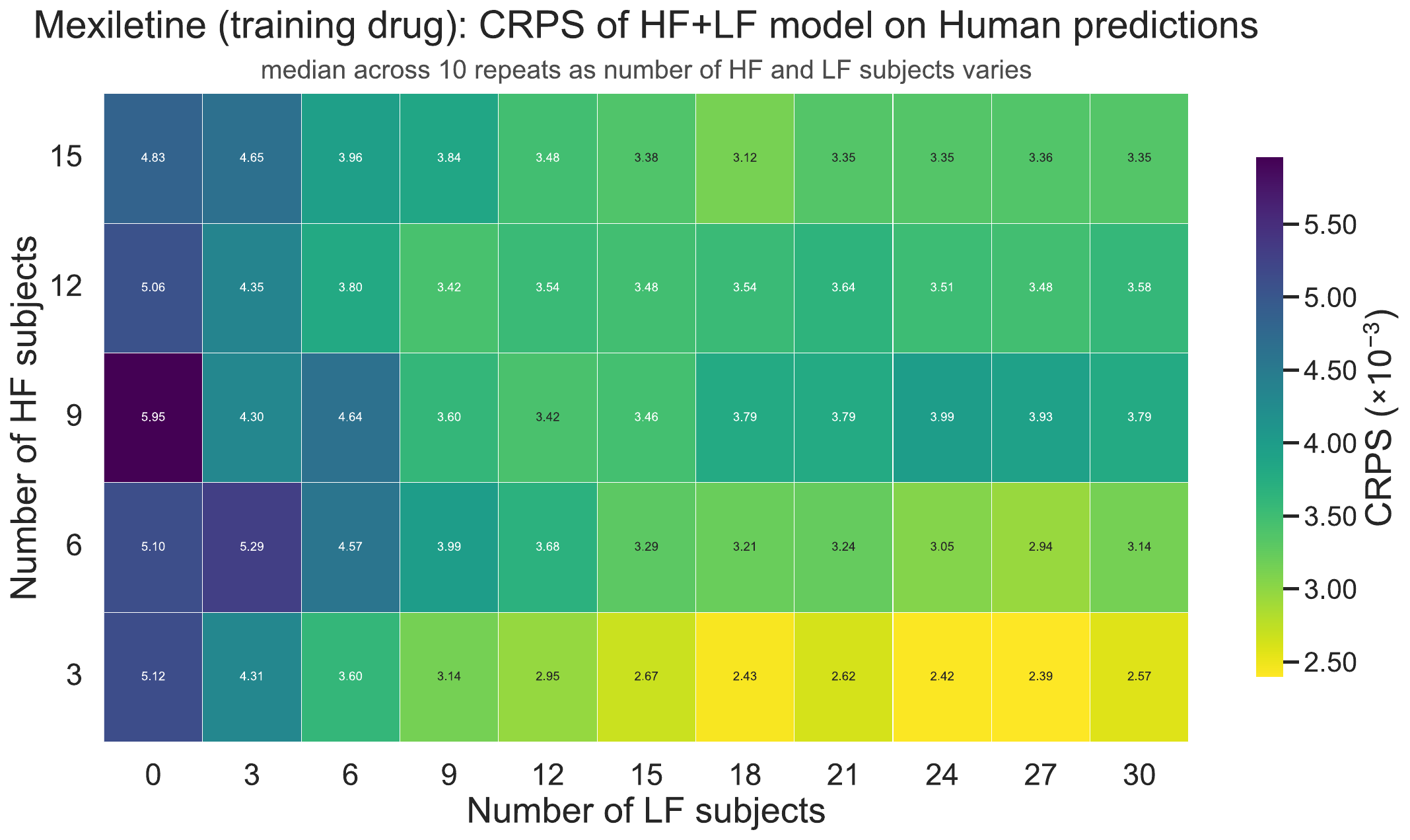}

    \vspace{1cm}

    \includegraphics[width=0.95\textwidth]{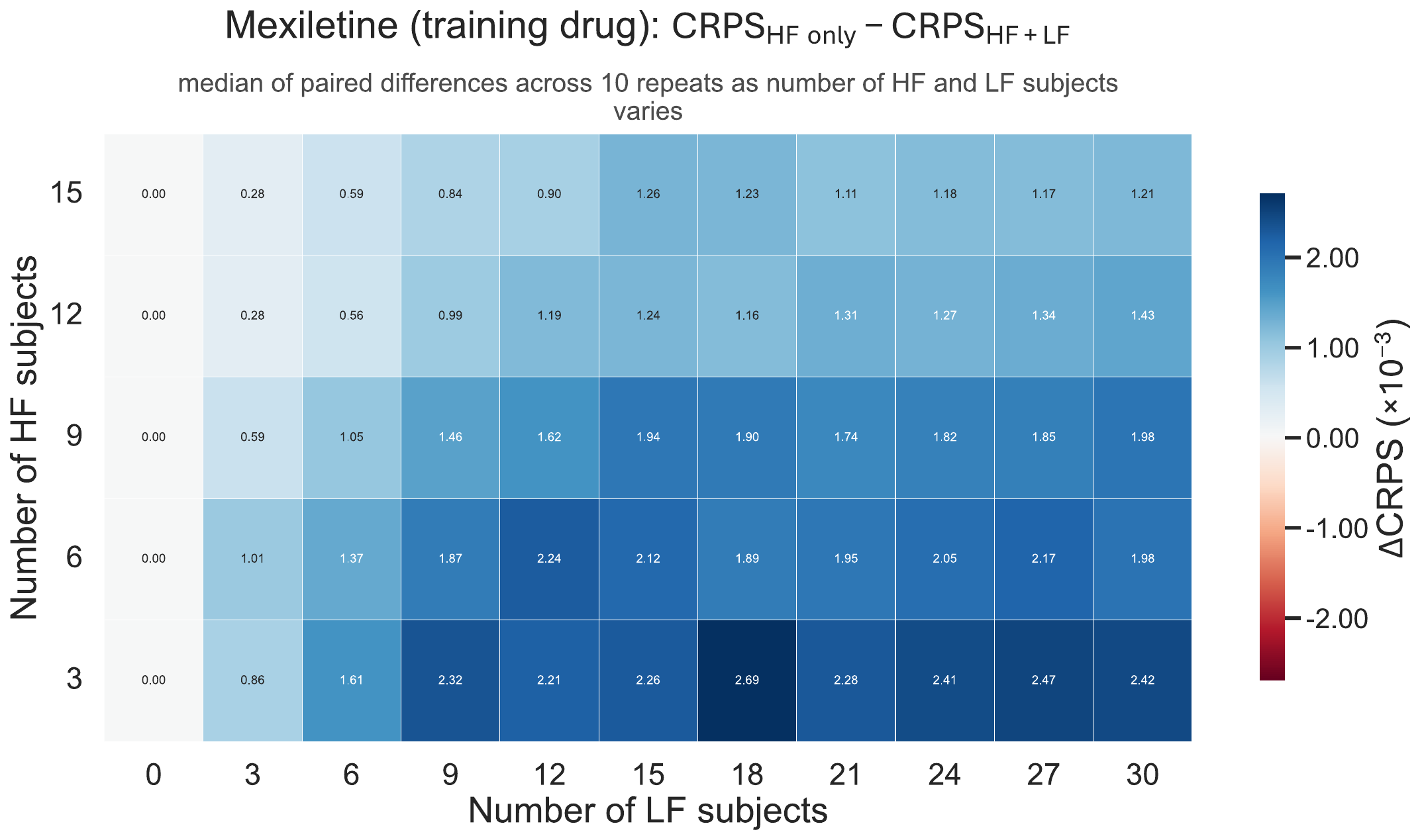}
\caption{\textbf{Above}: Continuous ranked probability score (CRPS) for prediction of mexiletine-induced maximum prolongation of action potential duration in humans (high-fidelity (HF) species) as the ratio of dog subjects (low-fidelity (LF)) to human subjects varies. \textbf{Below}: paired difference in CRPS in the HF-only vs HF+LF case for each fixed number of HF subjects. For each combination of number of HF and LF subjects, the median CRPS across 10 experimental repeats is shown, with HF and LF subjects randomly sampled each time. All GP hyperparameters are fixed to known values. For any fixed number of HF subjects, the addition of further LF data generally improves CRPS. This improvement plateaus as the number of HF subjects increases: LF data becomes less influential as HF data becomes richer.}
\label{fig:mexiletine_sweep}
\end{figure}

\begin{figure}[ht]
\centering
    \includegraphics[width=0.95\textwidth]{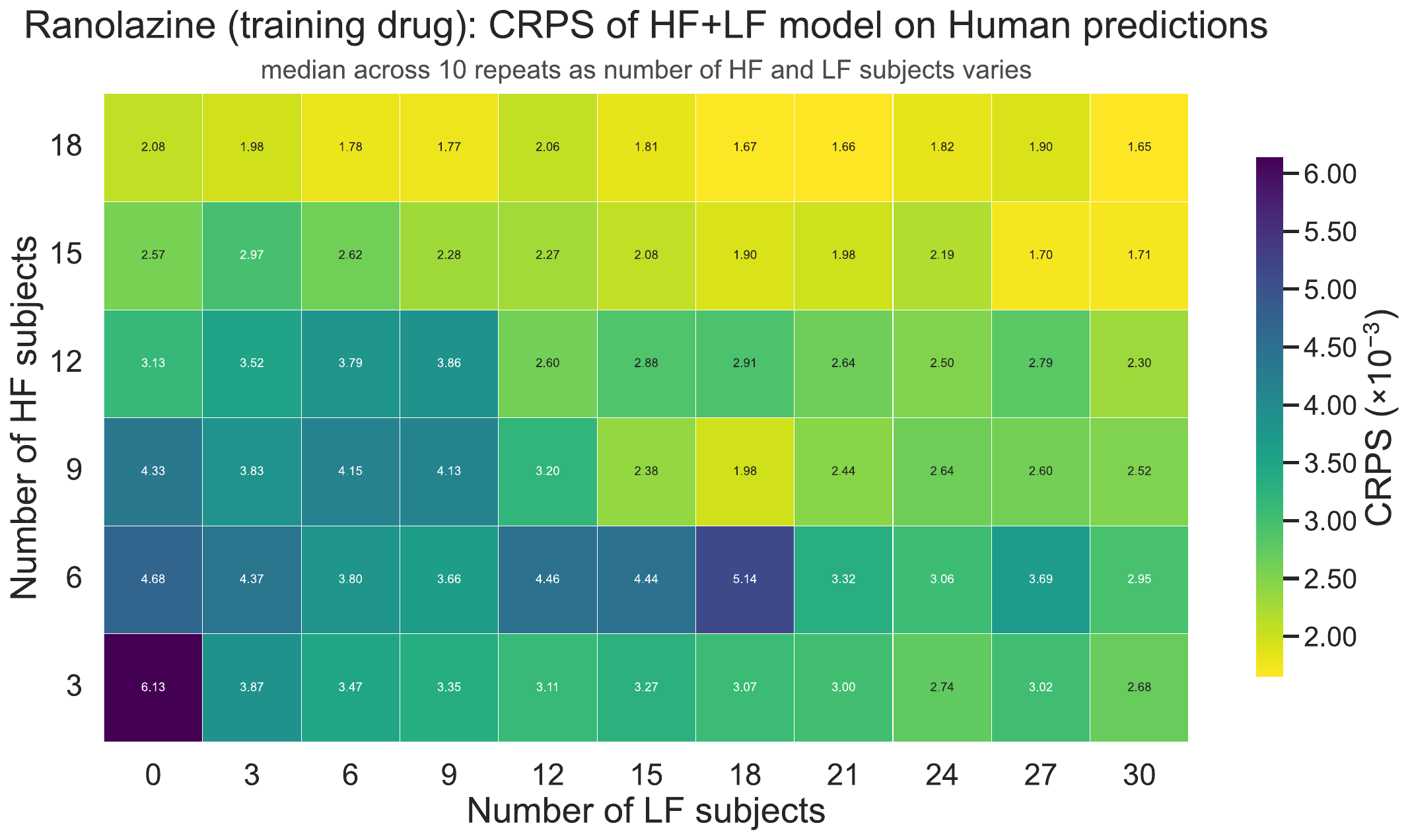}
    
    \vspace{1cm}

    \includegraphics[width=0.95\textwidth]{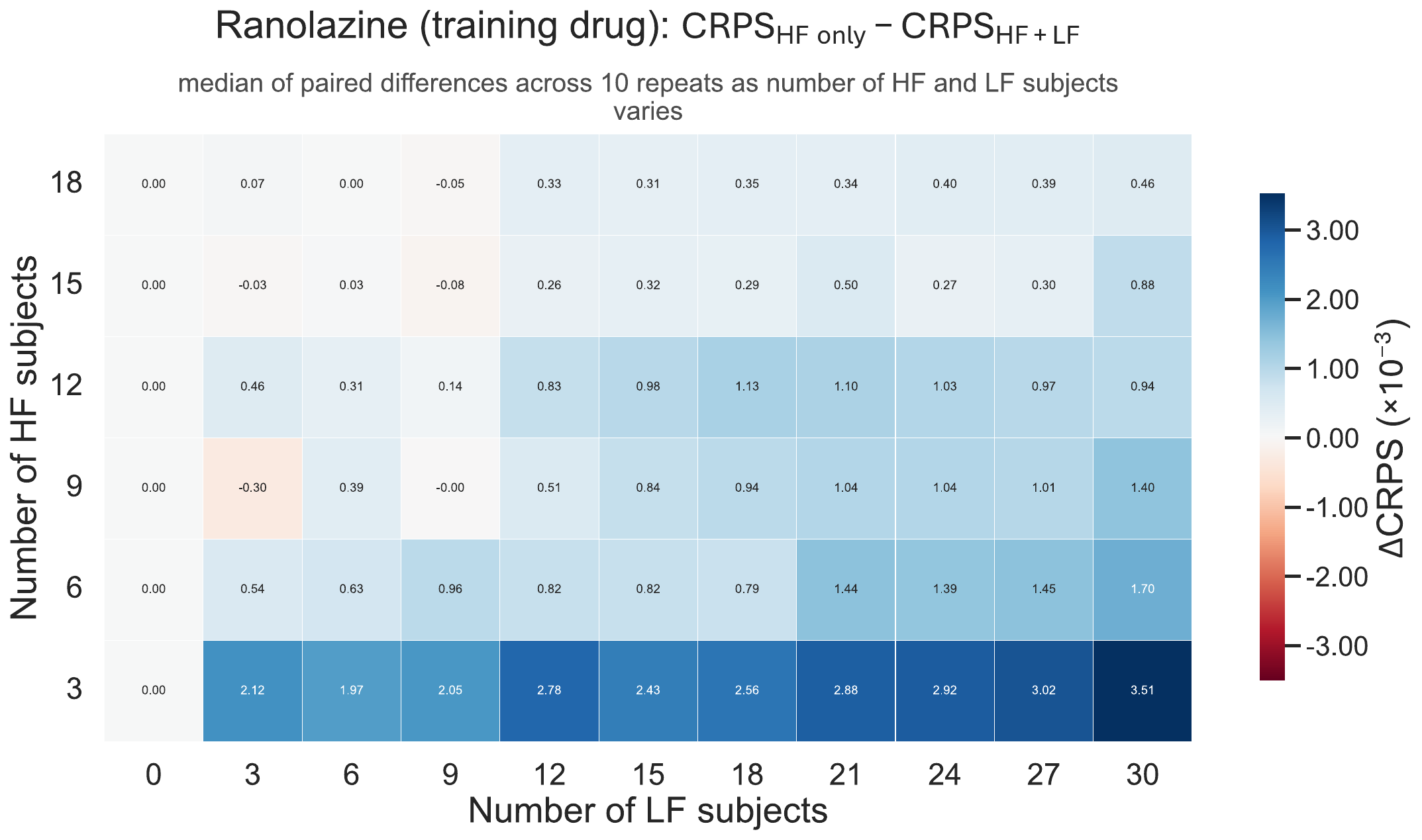}
\caption{\textbf{Above}: Continuous ranked probability score (CRPS) for prediction of ranolazine-induced maximum prolongation of action potential duration in humans (high-fidelity (HF) species) as the ratio of dog subjects (low-fidelity (LF)) to human subjects varies. \textbf{Below}: paired difference in CRPS in the HF-only vs HF+LF case for each fixed number of HF subjects. For each combination of number of HF and LF subjects, the median CRPS across 10 experimental repeats is shown, with HF and LF subjects randomly sampled each time. All GP hyperparameters are fixed to known values. For any fixed number of HF subjects, the addition of further LF data generally improves CRPS. This improvement plateaus as the number of HF subjects increases: LF data becomes less influential as HF data becomes richer.}
\label{fig:ranolazine_sweep}
\end{figure}

\begin{figure}[ht]
\centering
    \includegraphics[width=\textwidth]{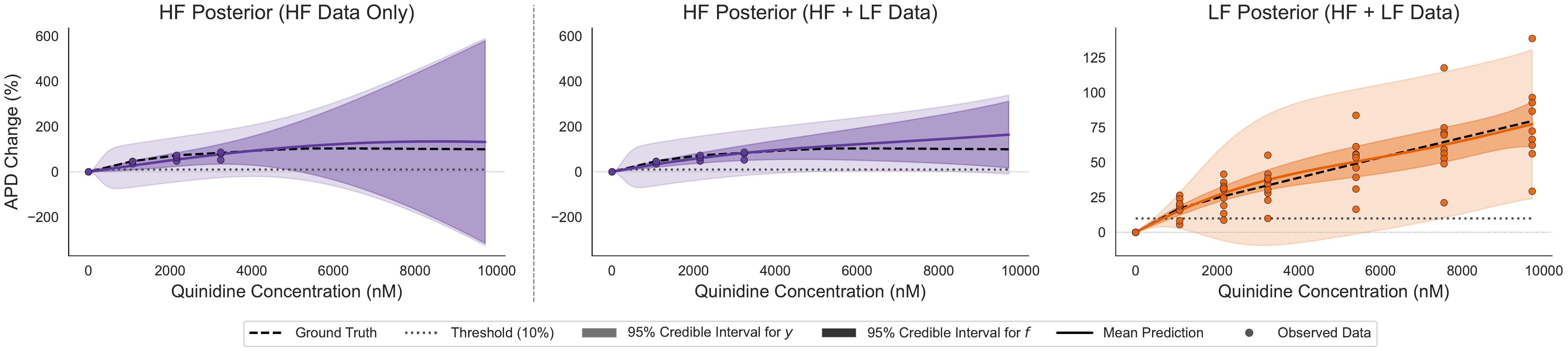}
    
    \vspace{1cm}

    \includegraphics[width=\textwidth]{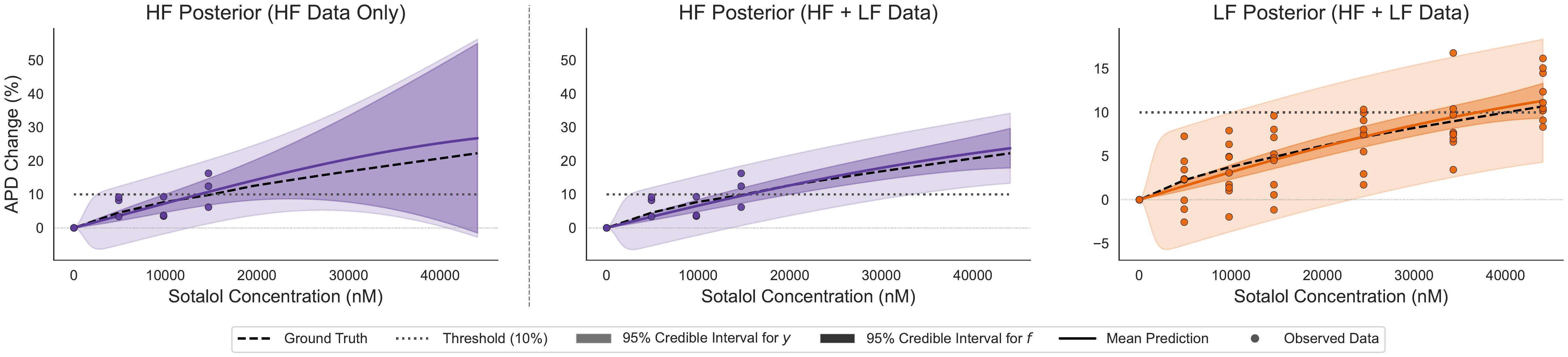}

    \vspace{1cm}
    
    \includegraphics[width=\textwidth]{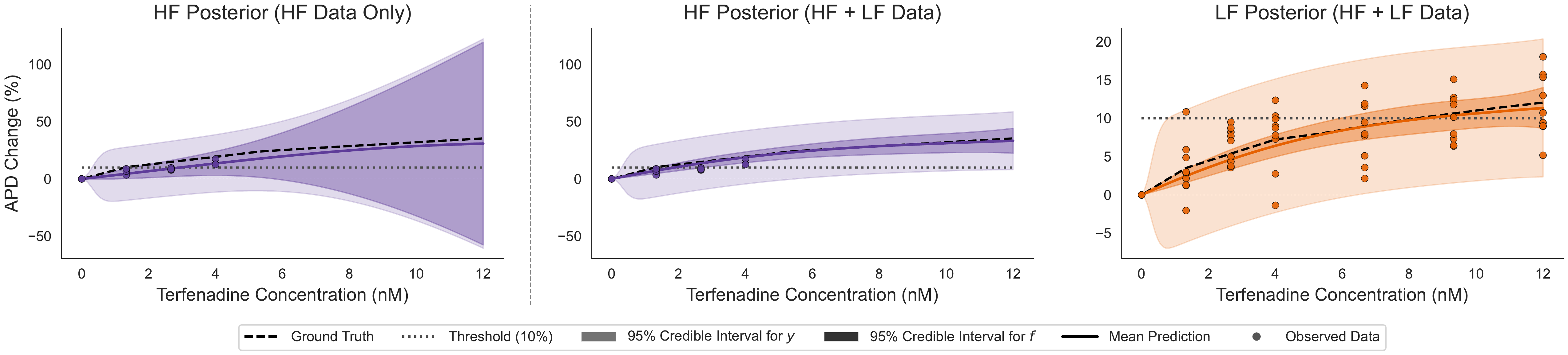}
    
    \vspace{1cm}

    \includegraphics[width=\textwidth]{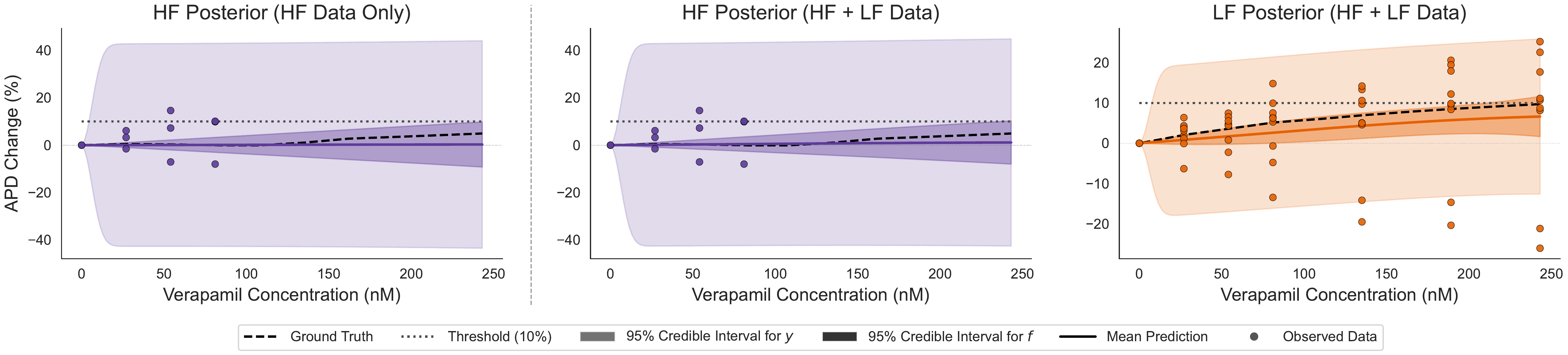}
\caption{Gaussian process (GP) posterior predictive plots for drug-induced change in action potential duration (APD) for four example drugs: quinidine, sotalol, terfenadine and verapamil. A high-fidelity (HF) only model is conditioned only on data from three human subjects in a limited input range (left), while the HF and low-fidelity (LF) model is conditioned on the same human data (centre) as well as data from nine dog subjects covering the full input range of interest (right). All GP hyperparameters are fixed to known values.}
\label{fig:exp_2_posteriors}
\end{figure}

\begin{figure}[ht]
\centering
    \includegraphics[width=\textwidth]{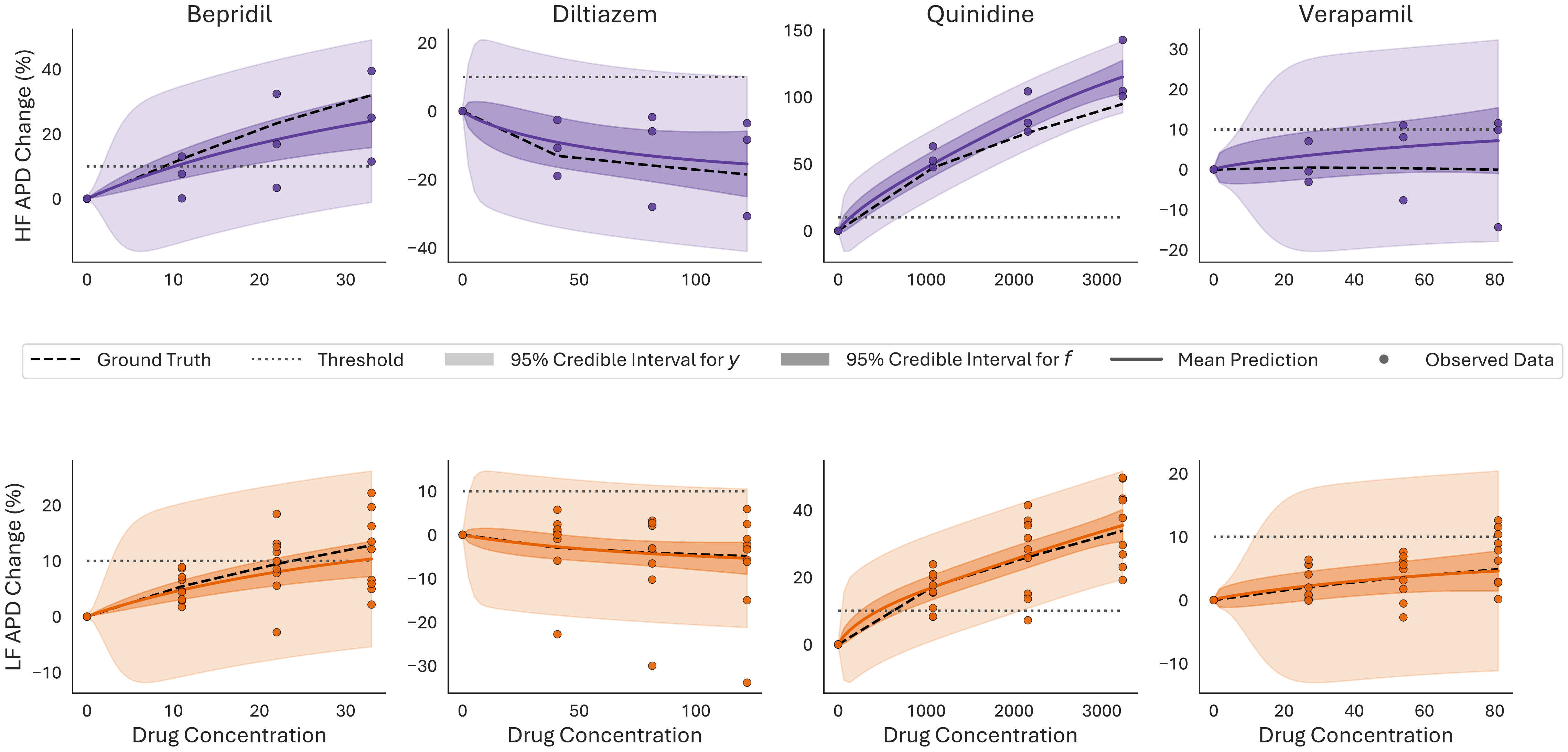}
\caption{Gaussian process (GP) posterior predictive plots for drug-induced change in action potential duration (APD) for four training drugs: quinidine, sotalol, terfenadine and verapamil, where humans are the high-fidelity (HF) and dogs are taken as the low-fidelity (LF) species. Data from these four drugs are used to fit GP hyperparameters and obtain a GP posterior predictive, to allow prediction for drugs before animal or human data is available.}
\end{figure}

\begin{figure}[ht]
\centering
    \includegraphics[width=\textwidth]{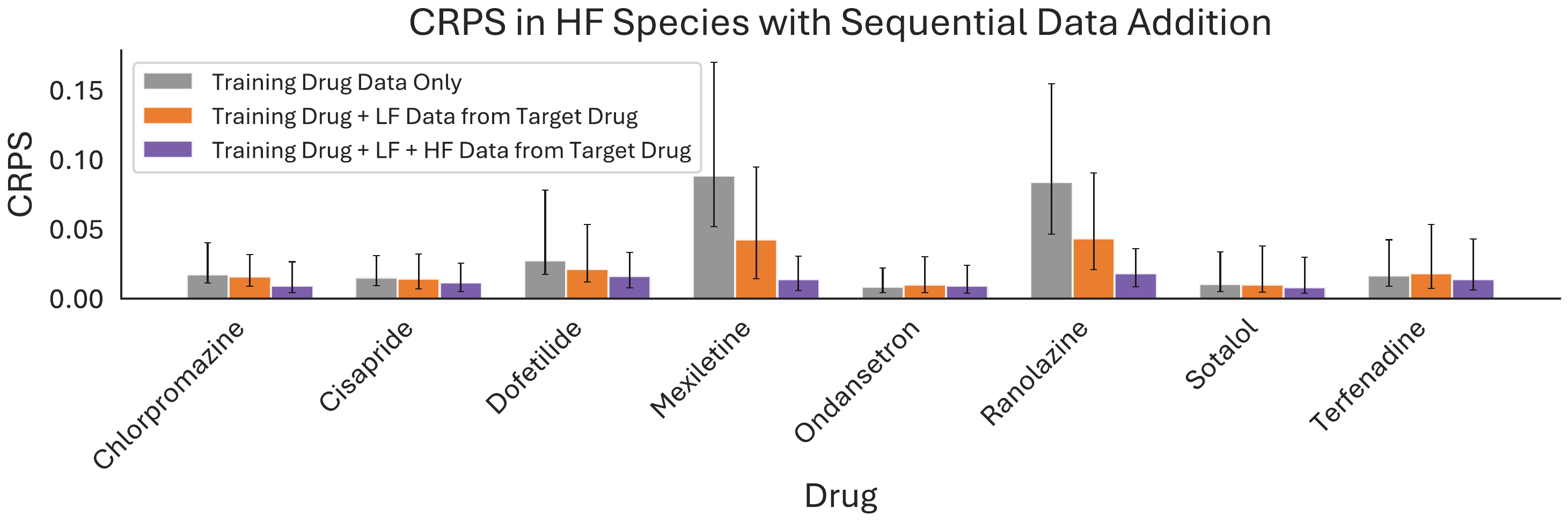}
    
    \vspace{1cm}

    \includegraphics[width=\textwidth]{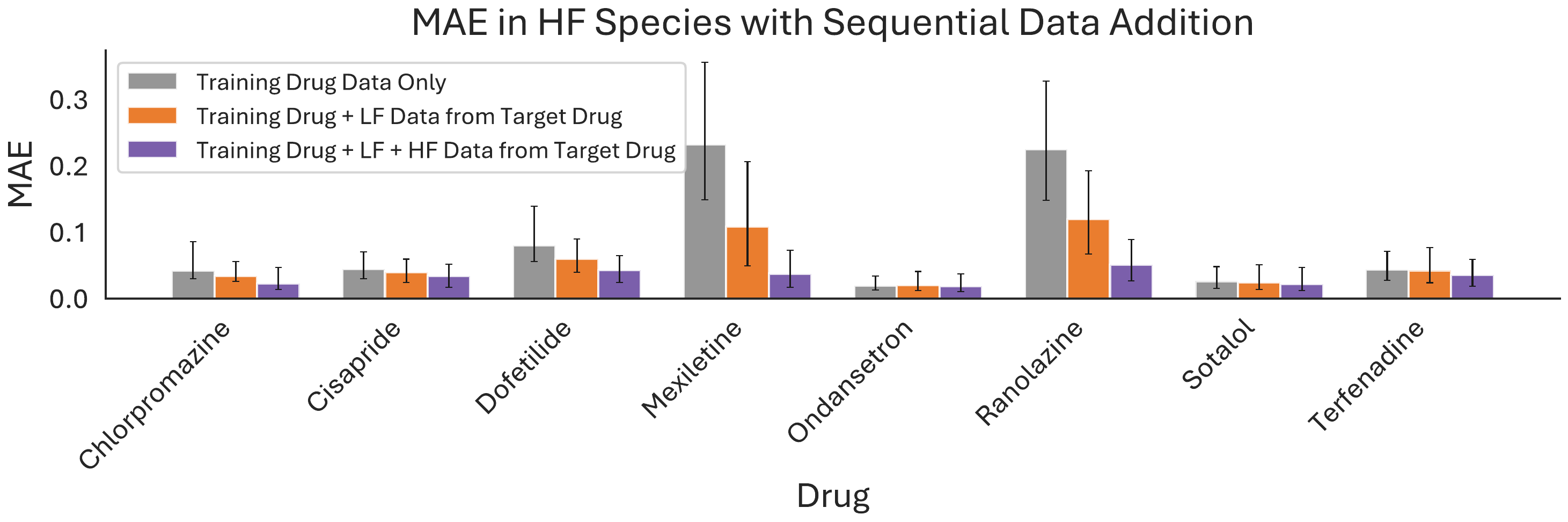}
    
    \vspace{1cm}

    \includegraphics[width=\textwidth]{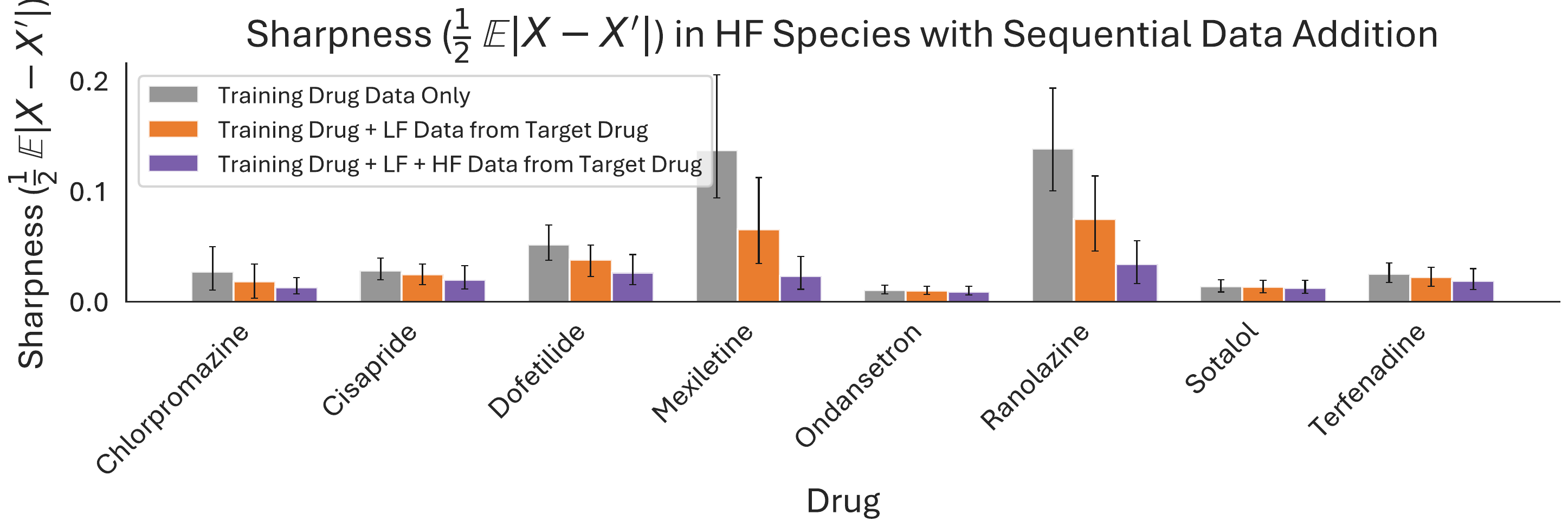}
\caption{Metrics assessing the prediction of maximum prolongation of action potential duration (APD) in humans (high-fidelity (HF) species) for eight held-out test drugs in the Comprehensive \textit{in vitro} Proarrhythmia Assay (CiPA) dataset with three-stage data inclusion. A Gaussian process (GP) is first trained on HF and low-fidelity (LF) data from four previous drugs in the CiPA dataset (grey), then including LF data for each test drug (orange), then finally further including HF data for each test drug (purple). 95\% confidence intervals (CIs) are shown across 100 randomly sampled training datasets. Dogs are taken as LF species; bepridil, diltiazem, quinidine and verapamil are used as training drugs. \textbf{Top:} continuous ranked probability score (CRPS); \textbf{middle:} mean absolute error (MAE); \textbf{bottom:} sharpness. CRPS is the difference between MAE and sharpness, hence rewards predictive accuracy but penalises predictions with low uncertainty. For some drugs (e.g., terfenadine), CRPS appears to change little, since the reduction in MAE and sharpness are almost equal.}
\end{figure}

\begin{figure}[ht]
\centering
    \includegraphics[width=\textwidth]{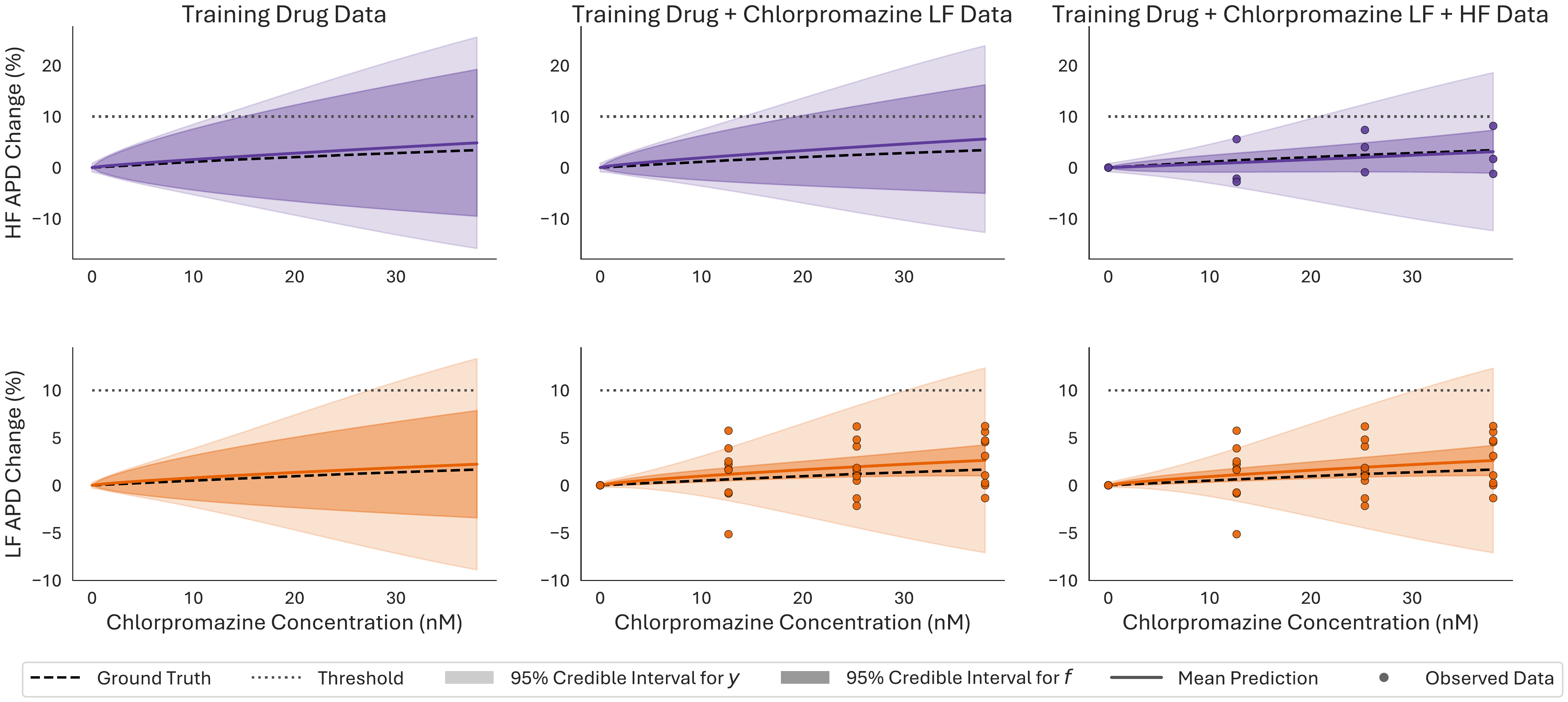}
    
    \vspace{1cm}

    \includegraphics[width=\textwidth]{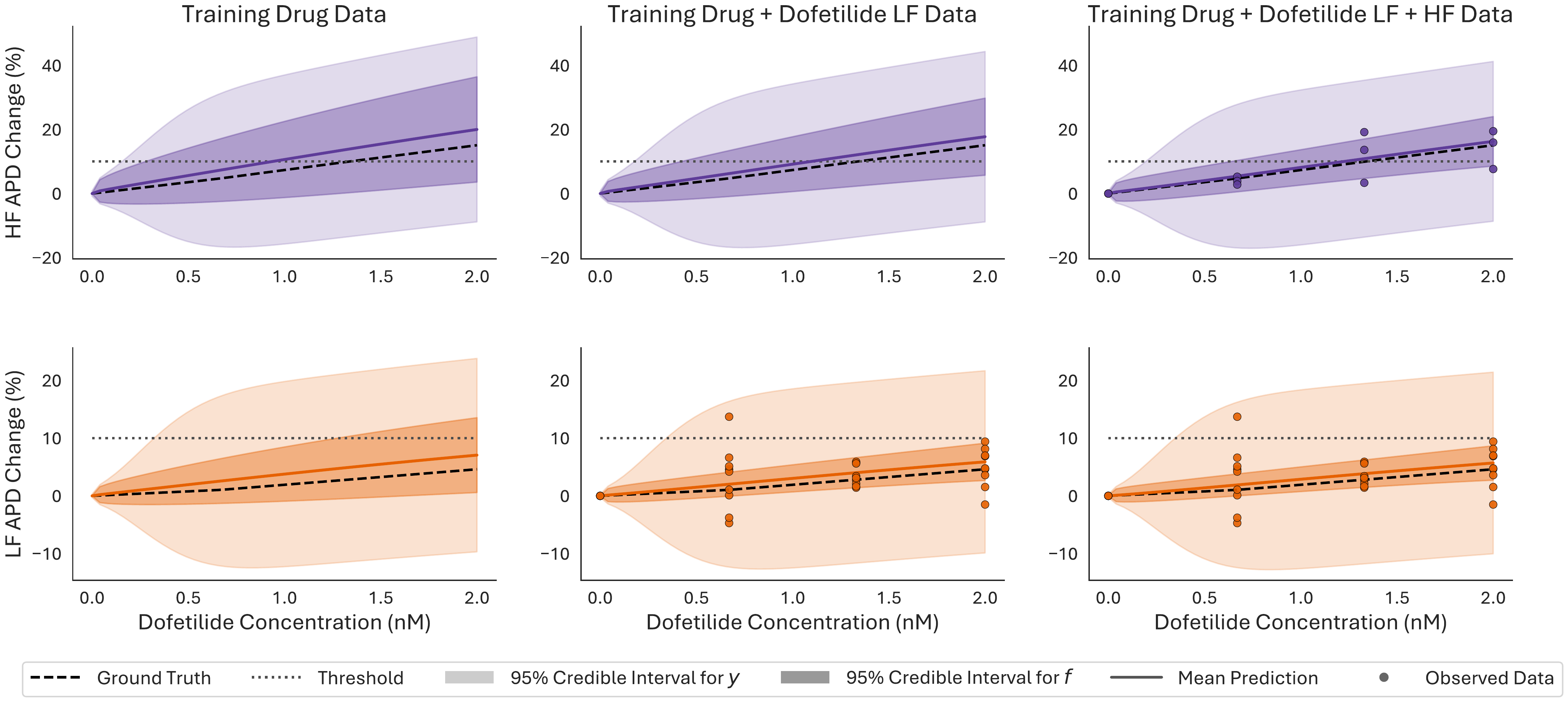}
\caption{Posterior predictive plots for drug-induced change in action potential duration (APD) in humans with three-stage data inclusion. Left: a Gaussian process (GP) is first trained on four previous drugs from the Comprehensive \textit{in vitro} Proarrhythmia Assay (CiPA) dataset (bepridil, diltiazem, quinidine and verapamil) with data from humans (high-fidelity (HF) species) and dogs (low-fidelity (LF)). Centre: the GP is then additionally conditioned on test drug data from nine dog subjects at three concentration levels. Right: human test drug data from three subjects is added at the same three concentration levels. Chlorpromazine (\textbf{top}) and dofetilide (\textbf{bottom}) are used as test drugs.}
\end{figure}

\begin{figure}[ht]
\centering
    \includegraphics[width=\textwidth]{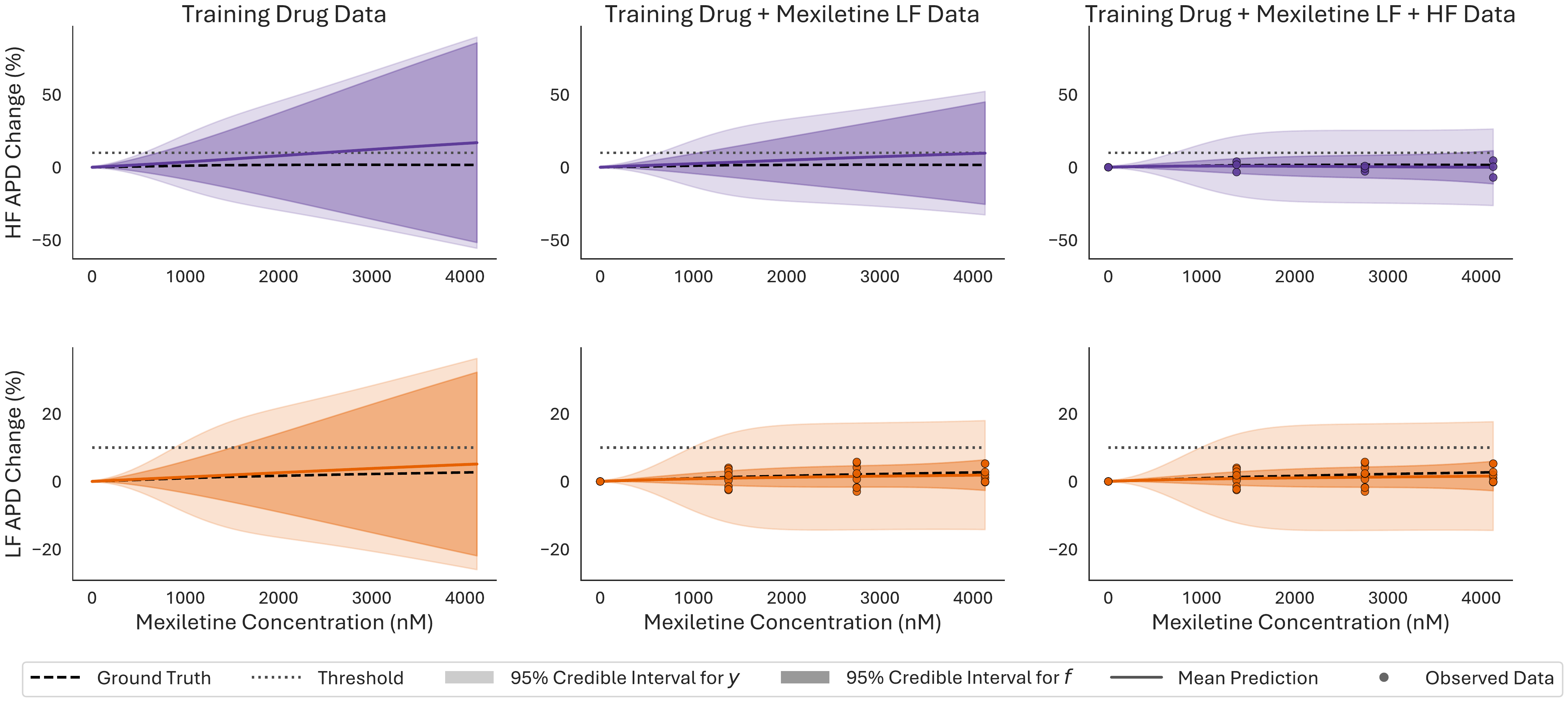}

    \vspace{1cm}

    \includegraphics[width=\textwidth]{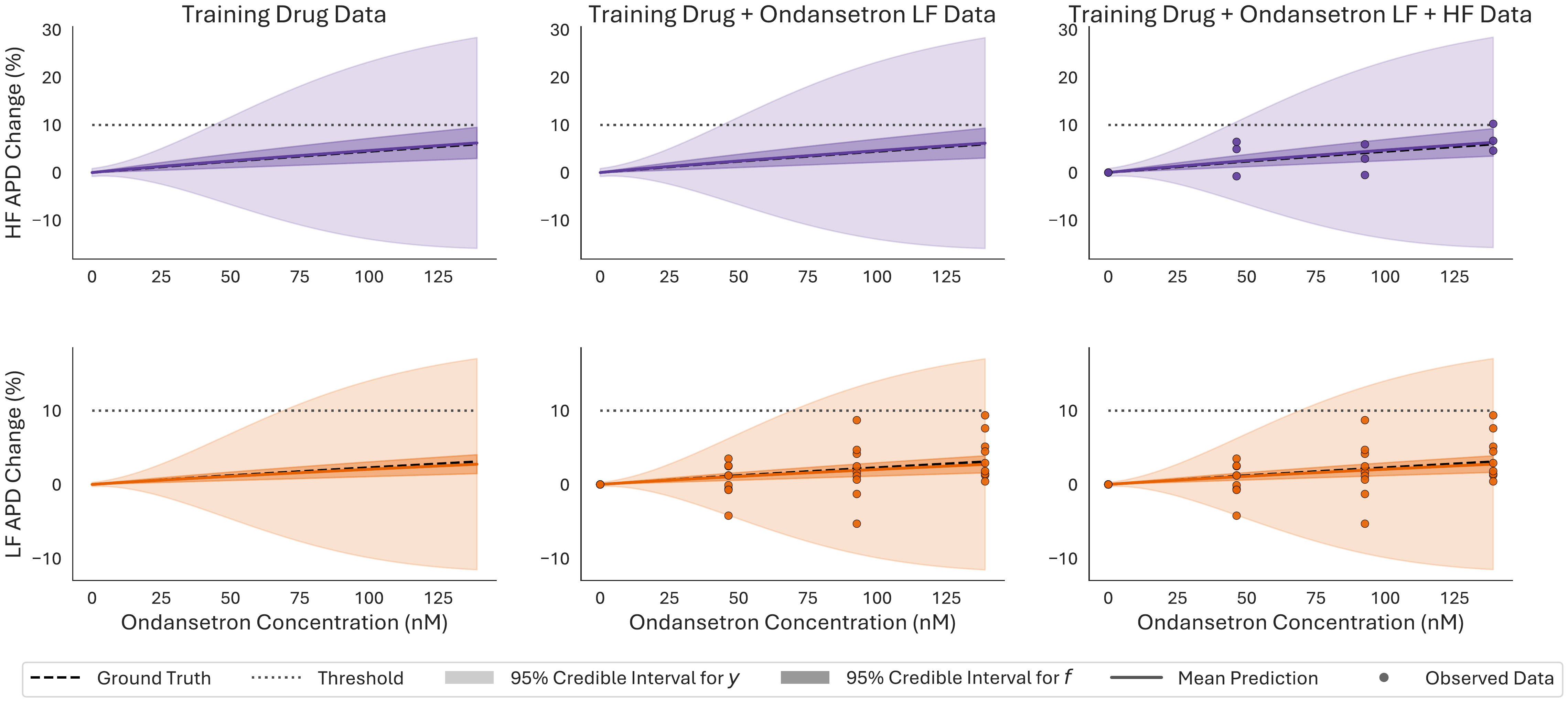}
\caption{Posterior predictive plots for drug-induced change in action potential duration (APD) in humans with three-stage data inclusion. Left: a Gaussian process (GP) is first trained on four previous drugs from the Comprehensive \textit{in vitro} Proarrhythmia Assay (CiPA) dataset (bepridil, diltiazem, quinidine and verapamil) with data from humans (high-fidelity (HF) species) and dogs (low-fidelity (LF)). Centre: the GP is then additionally conditioned on test drug data from nine dog subjects at three concentration levels. Right: human test drug data from three subjects is added at the same three concentration levels. Mexiletine (\textbf{top}) and ondansetron (\textbf{bottom}) are used as test drugs.}
\end{figure}

\begin{figure}[ht]
\centering
    \includegraphics[width=0.8\textwidth]{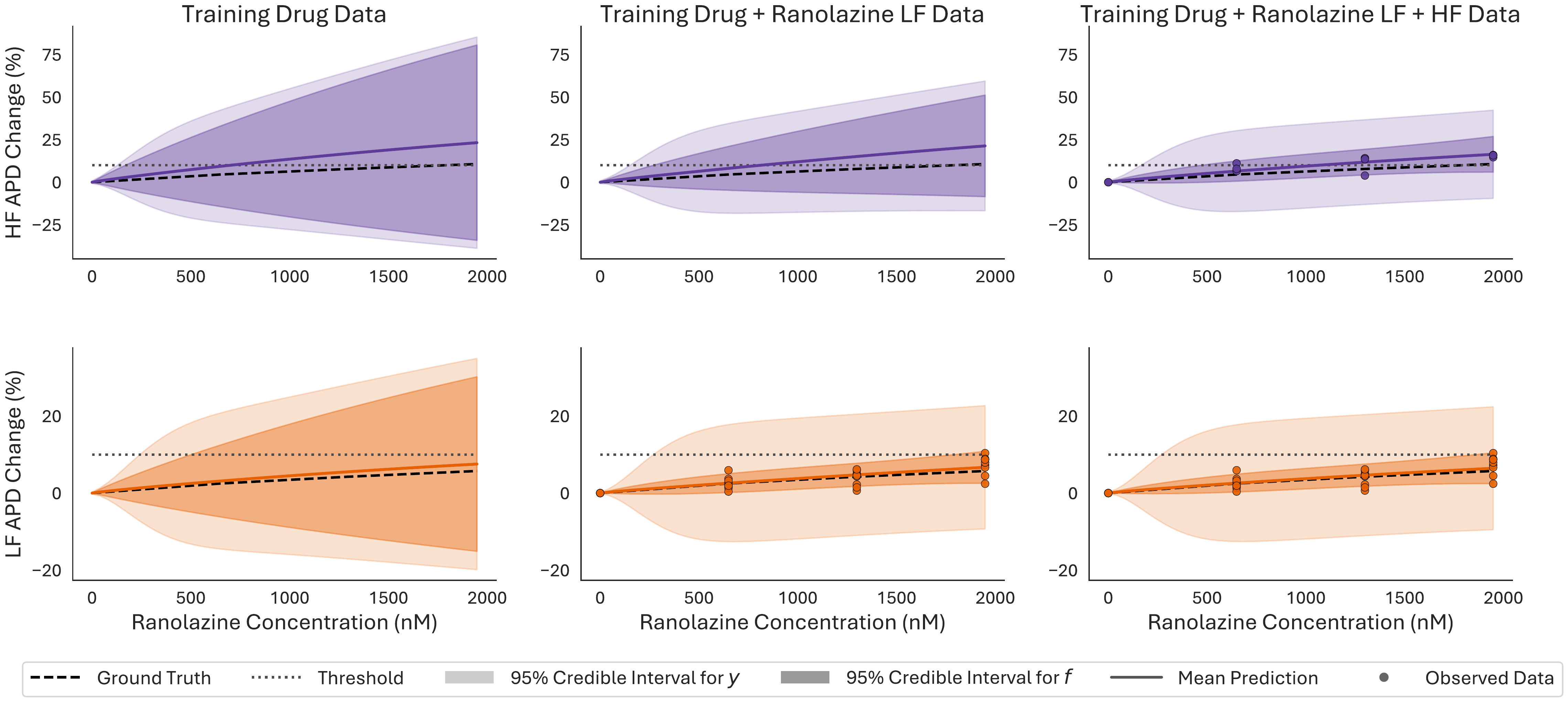}
    
    \vspace{1cm}

    \includegraphics[width=0.8\textwidth]{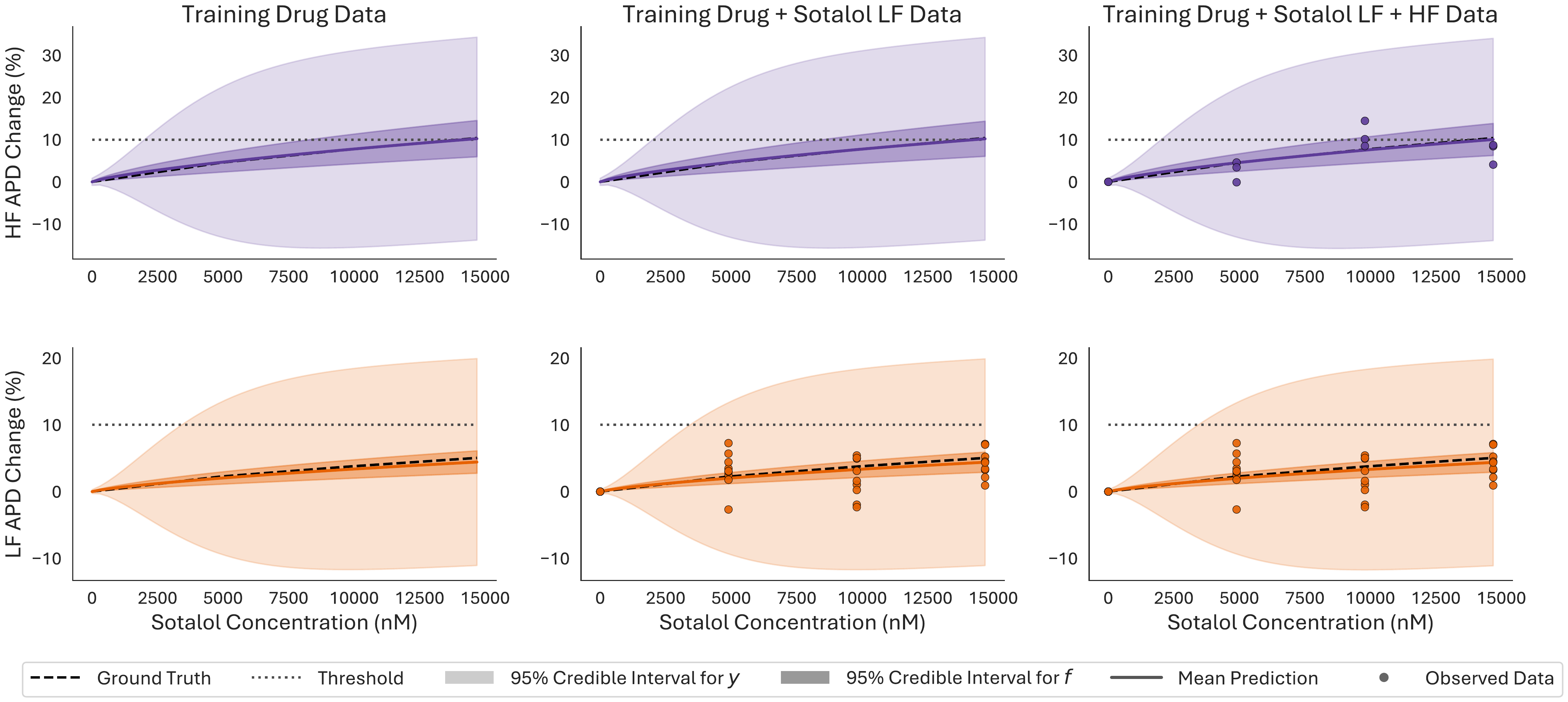}
    
    \vspace{1cm}

    \includegraphics[width=0.8\textwidth]{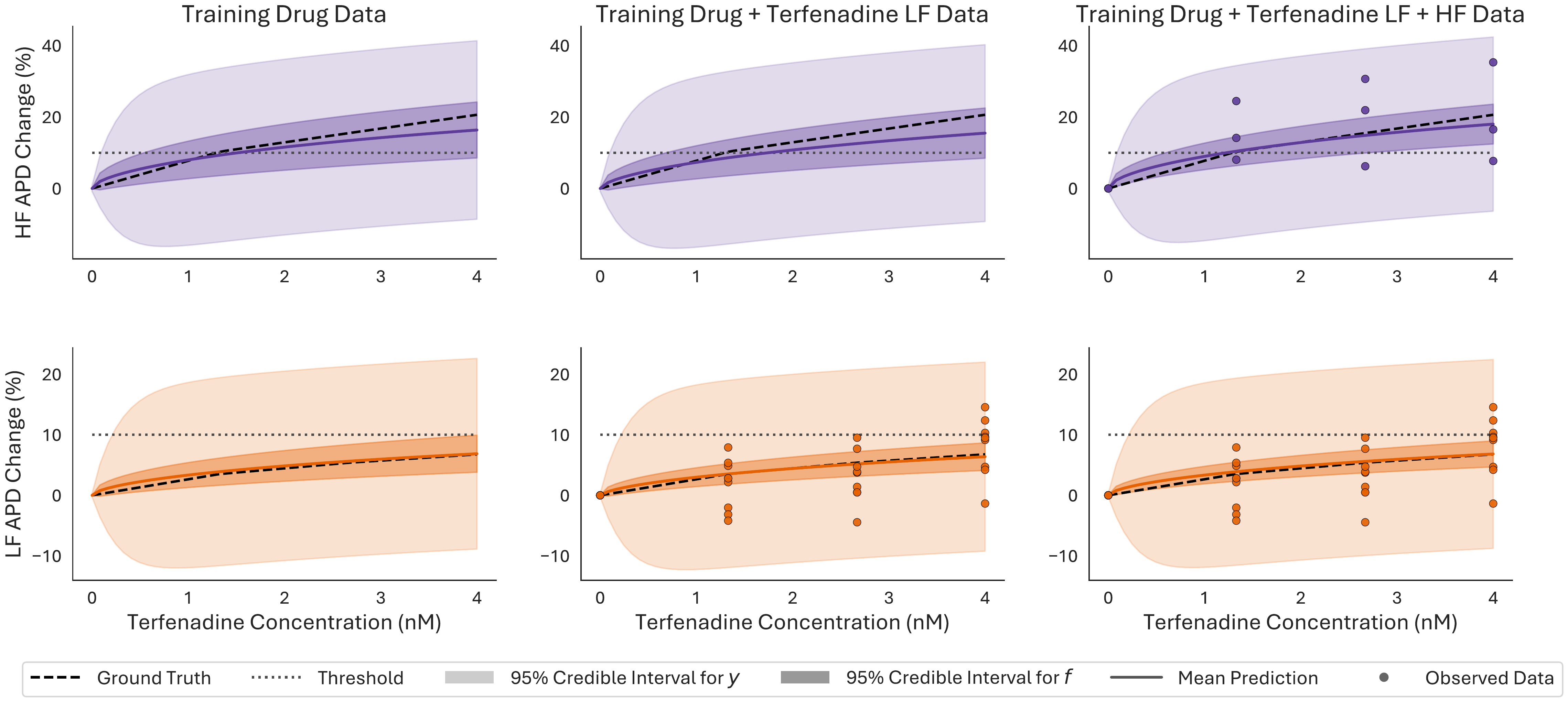}
\caption{Posterior predictive plots for drug-induced change in action potential duration (APD) in humans with three-stage data inclusion. Left: a Gaussian process (GP) is first trained on four previous drugs from the Comprehensive \textit{in vitro} Proarrhythmia Assay (CiPA) dataset (bepridil, diltiazem, quinidine and verapamil) with data from humans (high-fidelity (HF) species) and dogs (low-fidelity (LF)). Centre: the GP is then additionally conditioned on test drug data from nine dog subjects at three concentration levels. Right: human test drug data from three subjects is added at the same three concentration levels. Ranolazine (\textbf{top}), sotalol (\textbf{middle}) and terfenadine (\textbf{bottom}) are used as test drugs.}
\end{figure}

\begin{figure}
\centering
\includegraphics[width=\textwidth]{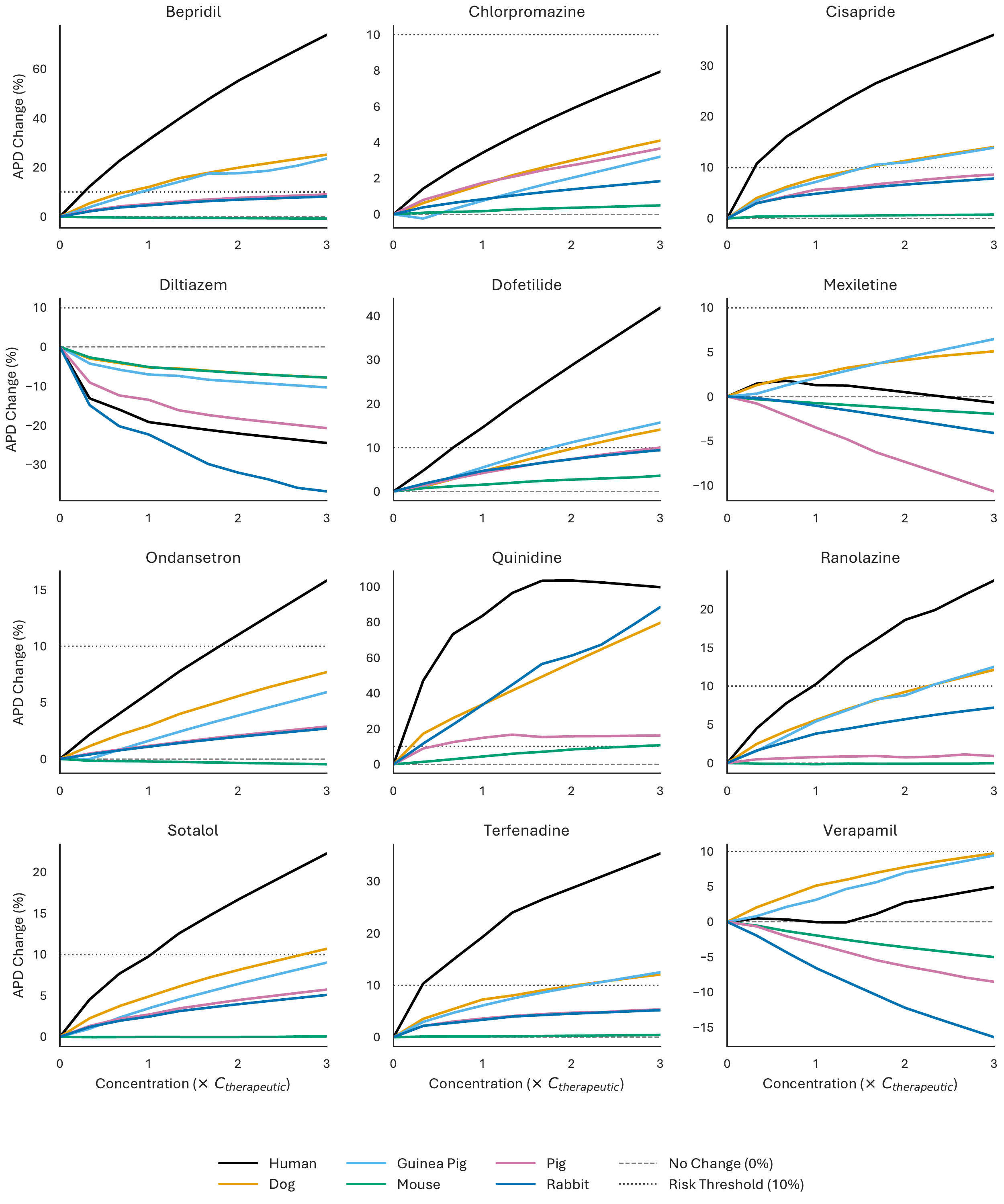}
\caption{Locally Weighted Scatterplot Smoothing (LOWESS) curves for simulated change in action potential duration (APD) data for each drug in the Comprehensive \textit{in vitro} Proarrhythmia Assay (CiPA) dataset, using all virtual subjects in each species studied. The LOWESS fraction parameter is set to 0.5. Drug effects vary in degrees of linearity and monotonicity; between-species similarity also varies drug-to-drug. Note that no quinidine data is available for guinea pigs since the simulators failed to converge to a stable limit cycle under quinidine administration. $C_{\text{therapeutic}}$: the drug's listed therapeutic free concentration.}
\label{fig:raw_data_egs}
\end{figure}

\begin{figure}
\centering
\includegraphics[width=0.45\textwidth]{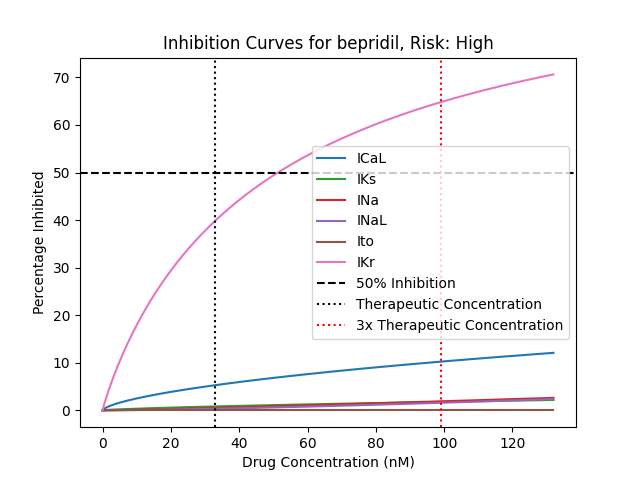}
\includegraphics[width=0.45\textwidth]{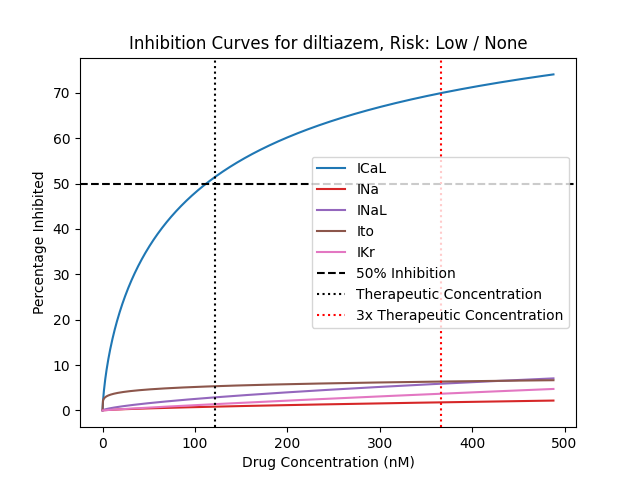}
\includegraphics[width=0.45\textwidth]{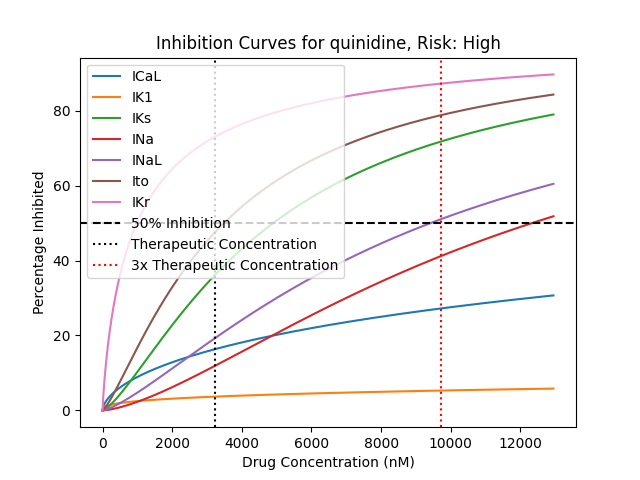}
\includegraphics[width=0.45\textwidth]{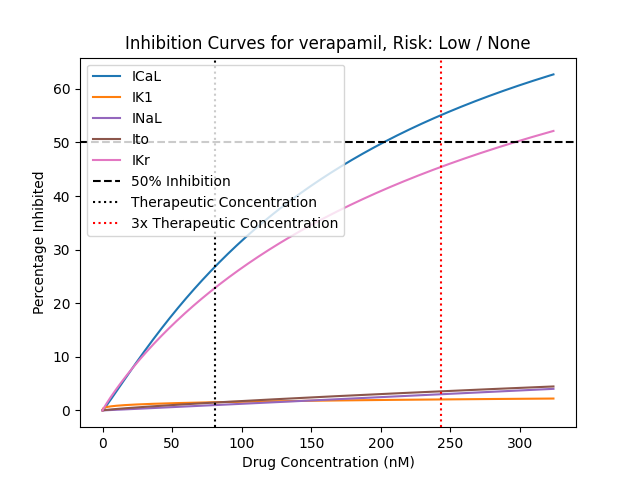}
\caption{Hill inhibition curves for the seven ionic currents studied in the Comprehensive \textit{in vitro} Proarrhythmia Assay (CiPA) dataset. Examples are shown for four drugs: bepridil, diltiazem, quinidine, verapamil.}
\label{fig:hill_curves}
\end{figure}

\begin{figure}
\centering
\includegraphics[width=0.8\textwidth]{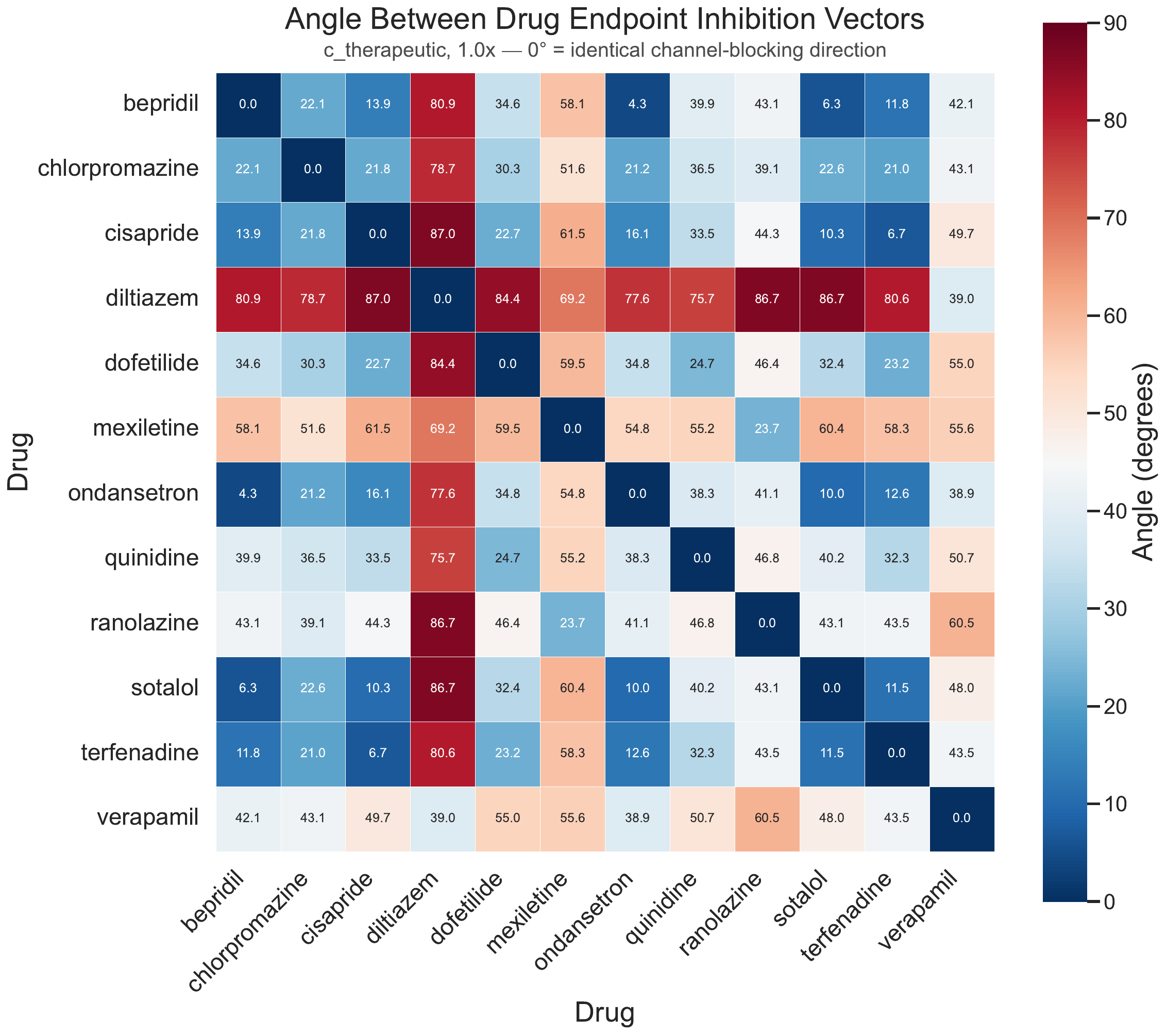}

    \vspace{1cm}

\includegraphics[width=0.7\textwidth]{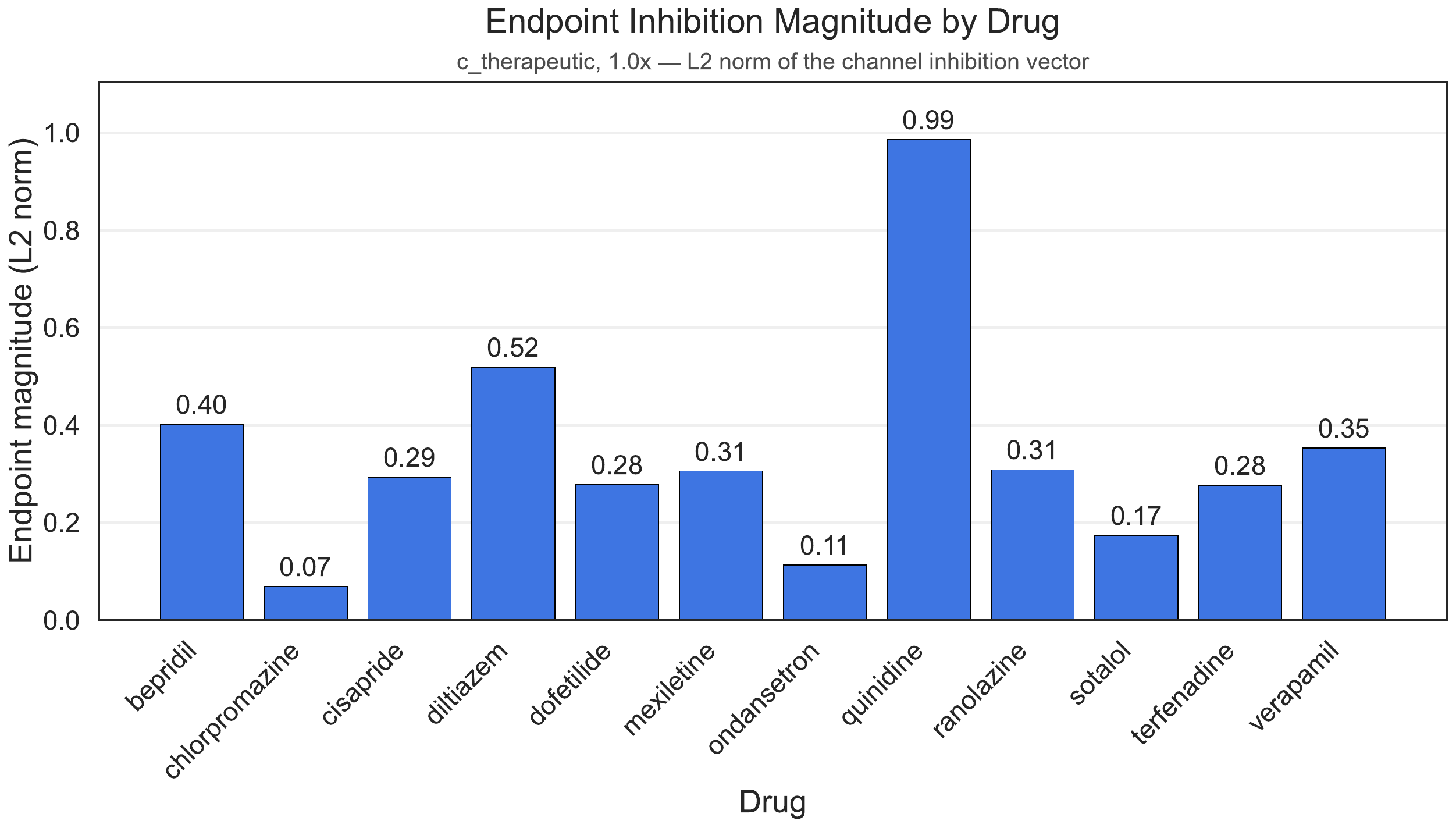}
\caption{\textbf{Above:} angles between the vectors of channel inhibition in $[0,1]^7$ for each drug in the Comprehensive \textit{in vitro} Proarrhythmia Assay (CiPA) dataset at its listed therapeutic concentration. \textbf{Below:} magnitude of each of the channel inhibition vectors for each drug at its listed therapeutic concentration, measured using the Euclidean norm. The angle between two drugs and the respective magnitudes of the vectors capture two distinct elements of drug-to-drug similarity. Smaller angles indicate drugs have similar propensities to inhibit the same ionic currents, and hence similar effects on the action potential duration. A drug with a larger magnitude is able to facilitate predictions for drugs with smaller magnitudes, if the angle between them is small. For example, bepridil has a magnitude of 0.4 compared to ondansetron with 0.11, while the angle between their trajectories is only 4.3 degrees. This means predicting for ondansetron having trained on bepridil data is effectively interpolating within the training set, while predicting for bepridil having trained on ondansetron is extrapolation.}
\end{figure}

\begin{figure}
\centering
\includegraphics[width=\textwidth]{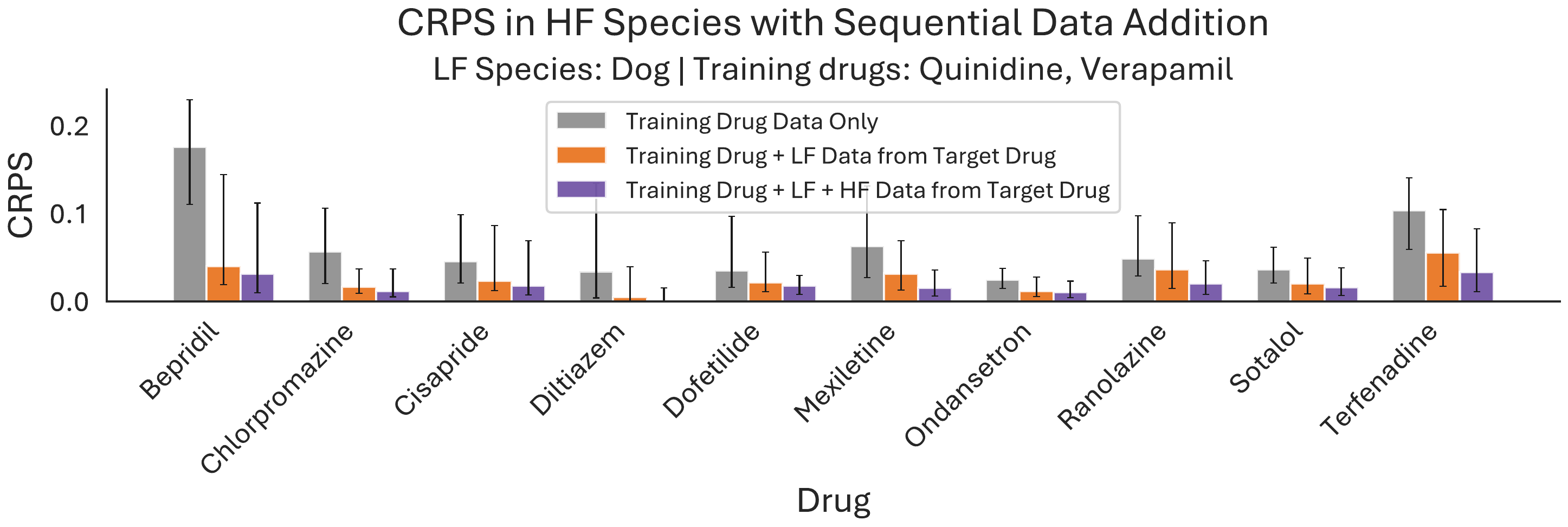}

\vspace{1cm}

\includegraphics[width=\textwidth]{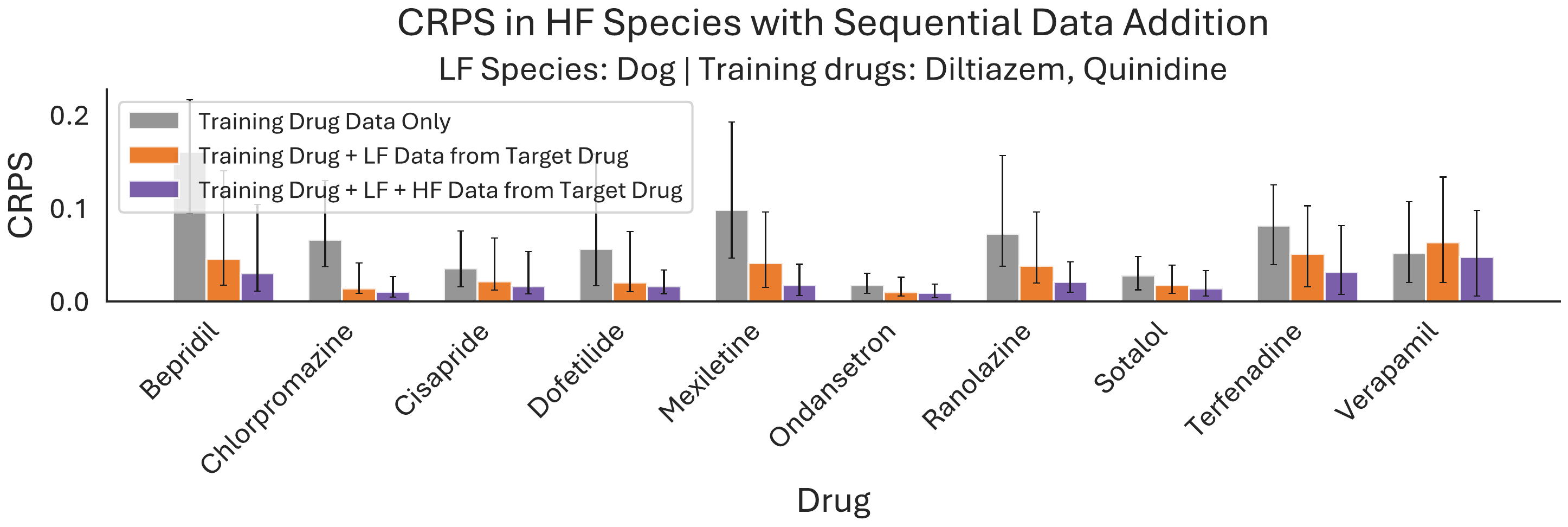}
\caption{Continuous ranked probability score (CRPS) for the prediction of maximum prolongation of action potential duration (APD) in humans (high-fidelity (HF) species) for eight held-out test drugs in the Comprehensive \textit{in vitro} Proarrhythmia Assay (CiPA) dataset with three-stage data inclusion. A Gaussian process (GP) is first trained on HF and low-fidelity (LF) data from four previous drugs in the CiPA dataset (grey), then including LF data for each test drug (orange), then finally further including HF data for each test drug (purple). 95\% confidence intervals (CIs) are shown across 100 randomly sampled training datasets. Dogs are taken as the LF species. \textbf{Top:} quinidine and verapamil are used as training drugs. \textbf{Bottom}: diltiazem and quinidine are used as training drugs, leading to increase in CRPS when adding dog data for verapamil. This occurs since verapamil blocks $I_{\mathrm{Ca,L}}$ and $I_{\mathrm{Kr}}$ almost equally (figure \ref{fig:hill_curves}), and the interaction between the block of these two channels differs between species, causing different verapamil response in humans and dogs (figure \ref{fig:raw_data_egs}). This interaction cannot be sufficiently learned from diltiazem and quinidine alone, which block one of the two channels significantly more than the other (figure \ref{fig:hill_curves}).}
\end{figure}

\begin{figure}
\centering
\includegraphics[width=\textwidth]{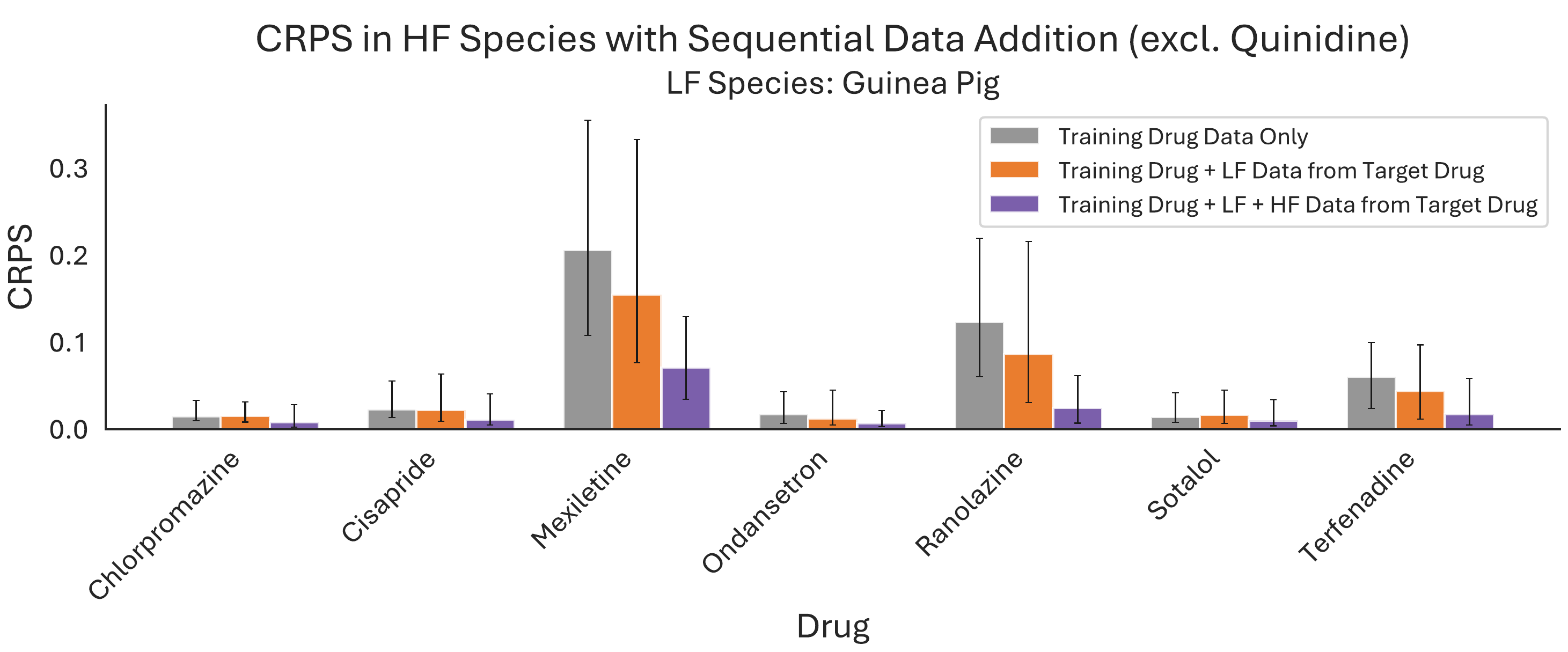}

    \vspace{1cm}

\includegraphics[width=\textwidth]{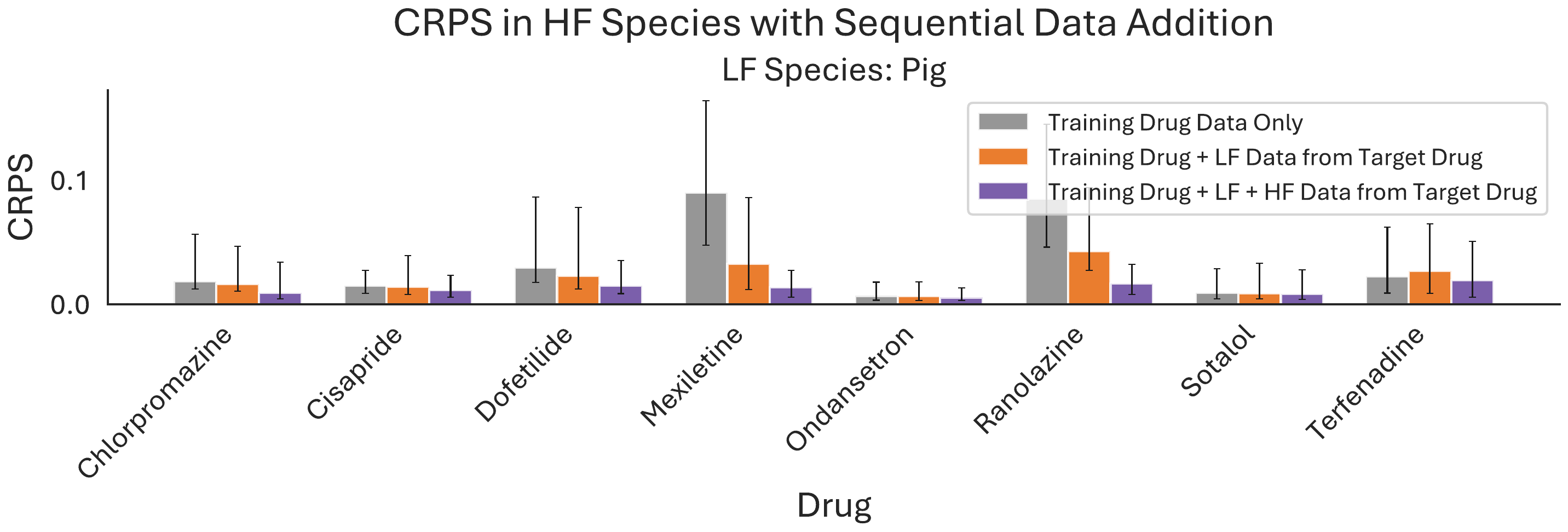}

    \vspace{1cm}

\includegraphics[width=\textwidth]{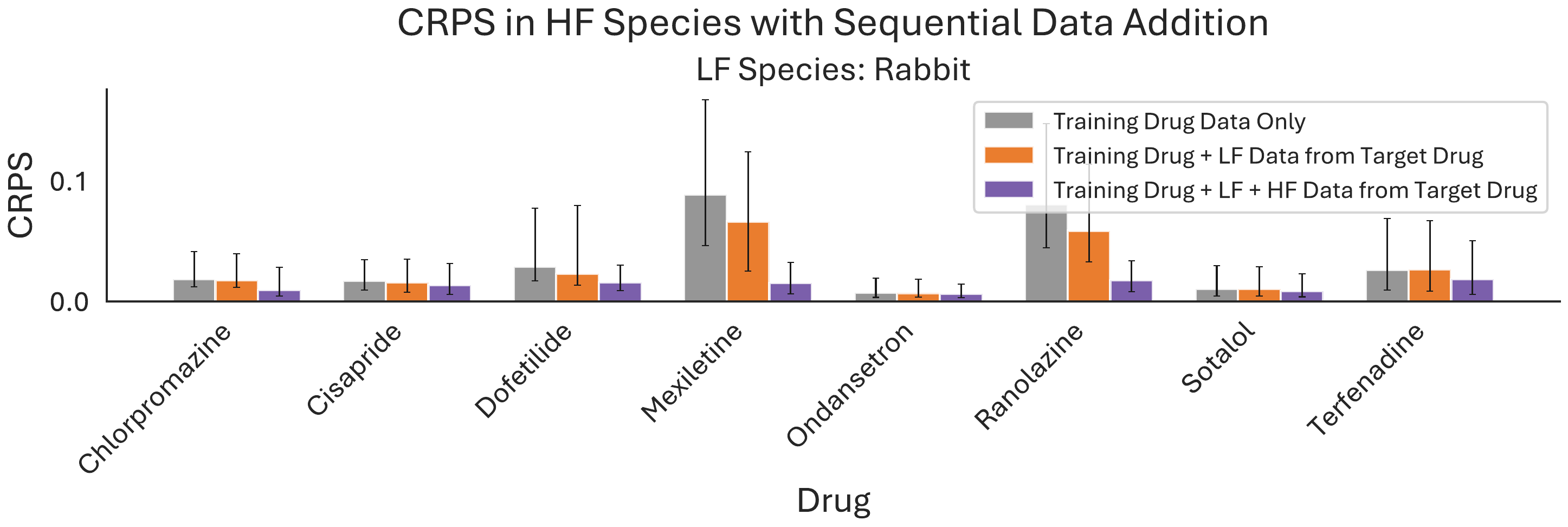}
\caption{Continuous ranked probability score (CRPS) for the prediction of maximum prolongation of action potential duration (APD) in humans (high-fidelity (HF) species) for eight held-out test drugs in the Comprehensive \textit{in vitro} Proarrhythmia Assay (CiPA) dataset with three-stage data inclusion. A Gaussian process (GP) is first trained on HF and low-fidelity (LF) data from four previous drugs in the CiPA dataset (grey), then including LF data for each test drug (orange), then finally further including HF data for each test drug (purple). 95\% confidence intervals (CIs) are shown across 100 randomly sampled training datasets. Bepridil, diltiazem, quinidine and verapamil are used as training drugs. Guinea pigs (\textbf{top}), pigs (\textbf{middle}) and rabbits (\textbf{bottom}) are taken as LF species.}
\end{figure}

\begin{figure}
\centering
\includegraphics[width=\textwidth]{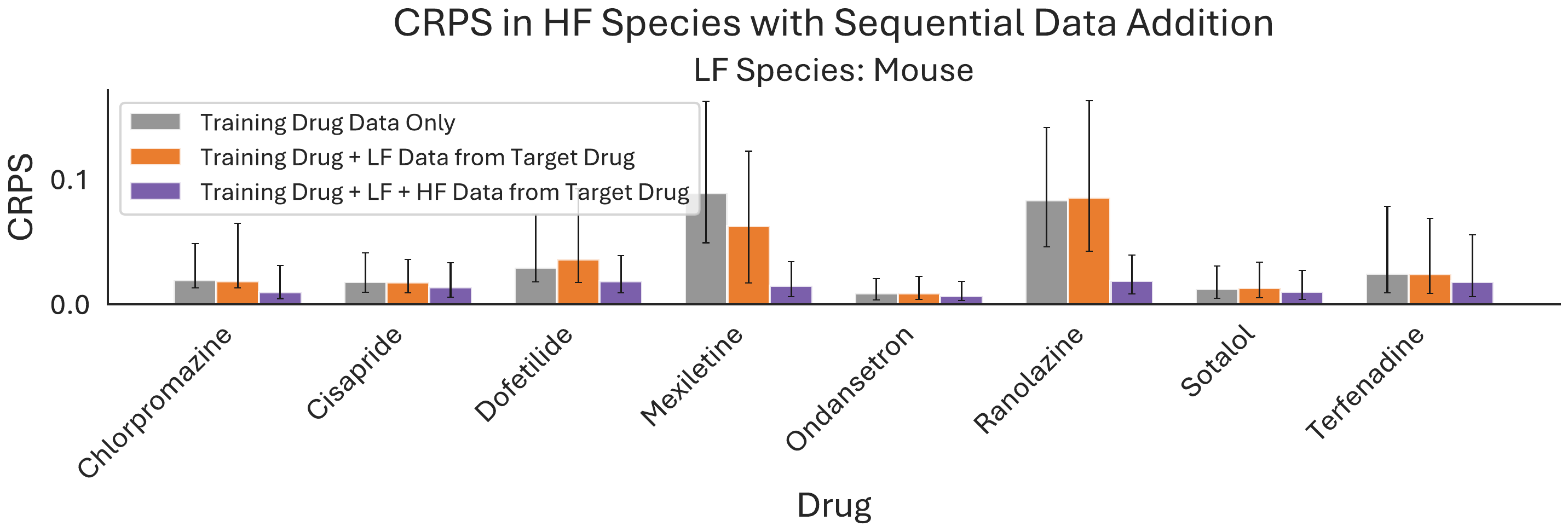}

    \vspace{1cm}

\includegraphics[width=\textwidth]{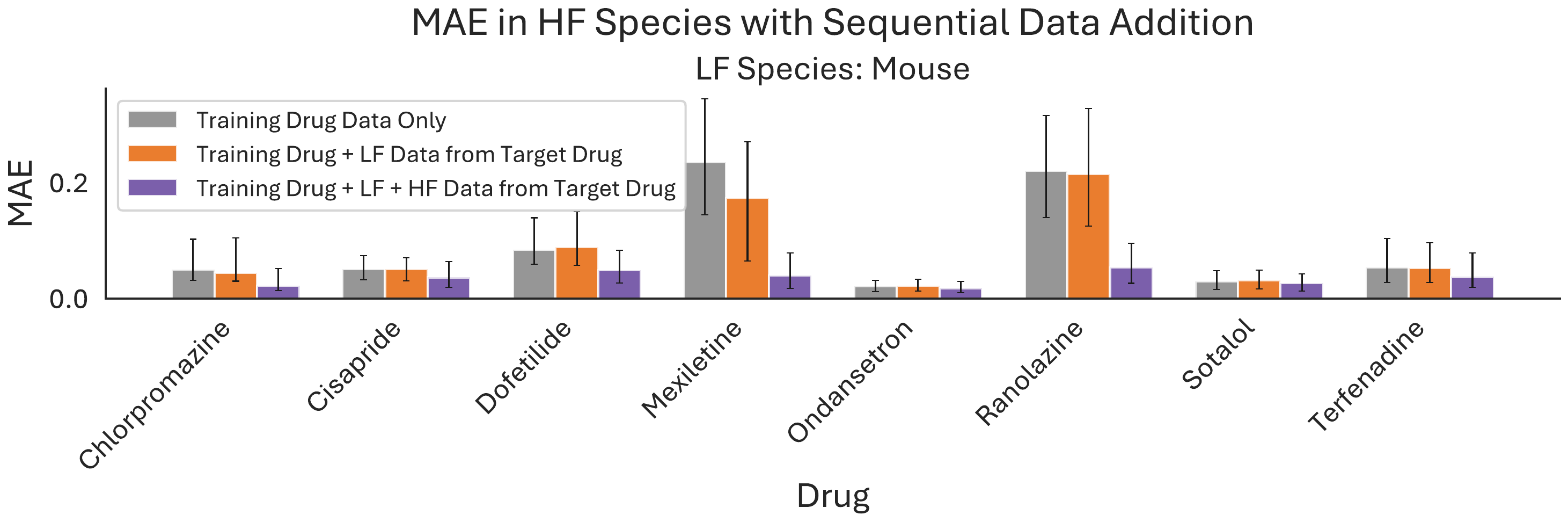}

    \vspace{1cm}

\includegraphics[width=\textwidth]{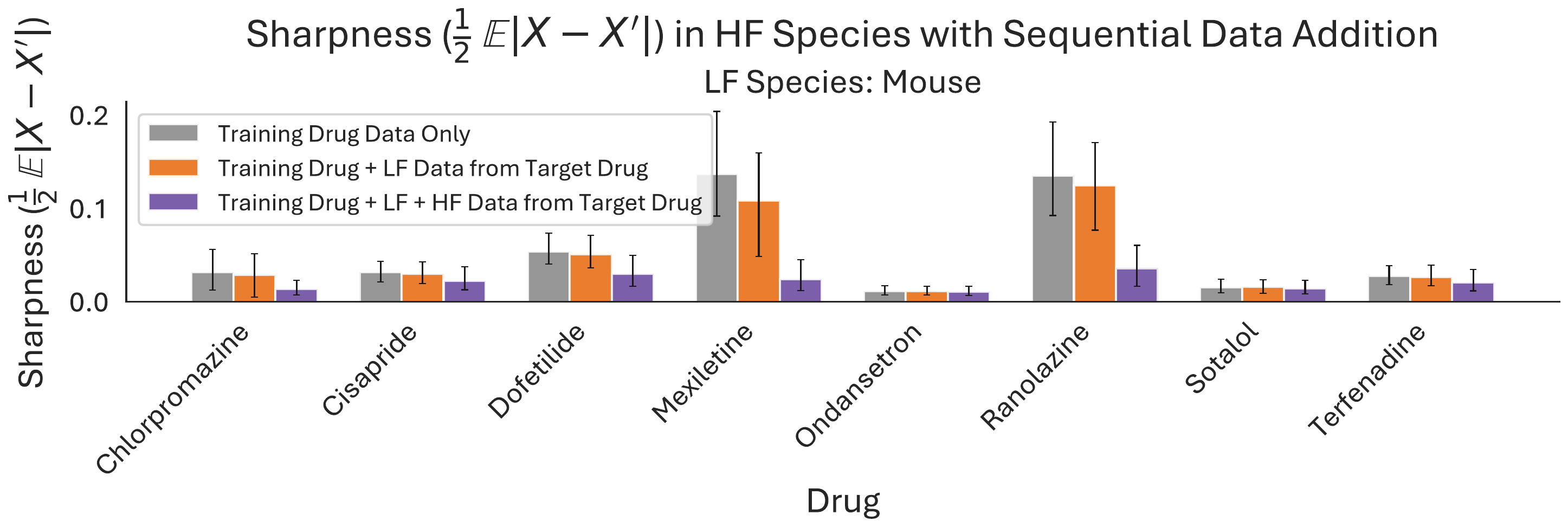}
\caption{Metrics assessing the prediction of maximum prolongation of action potential duration (APD) in humans (high-fidelity (HF) species) for eight held-out test drugs in the Comprehensive \textit{in vitro} Proarrhythmia Assay (CiPA) dataset with three-stage data inclusion. A Gaussian process (GP) is first trained on HF and low-fidelity (LF) data from four previous drugs in the CiPA dataset (grey), then including LF data for each test drug (orange), then finally further including HF data for each test drug (purple). 95\% confidence intervals (CIs) are shown across 100 randomly sampled training datasets. Mice are taken as LF species; bepridil, diltiazem, quinidine and verapamil are used as training drugs. \textbf{Top:} continuous ranked probability score (CRPS); \textbf{middle:} mean absolute error (MAE); \textbf{bottom:} sharpness. CRPS is the difference between MAE and sharpness, hence rewards predictive accuracy but penalises predictions with low uncertainty. For some drugs (e.g., ranolazine), CRPS appears to change little, since the reduction in MAE and sharpness are almost equal.}
\end{figure}

\begin{figure}
\centering
\includegraphics[width=0.8\textwidth]{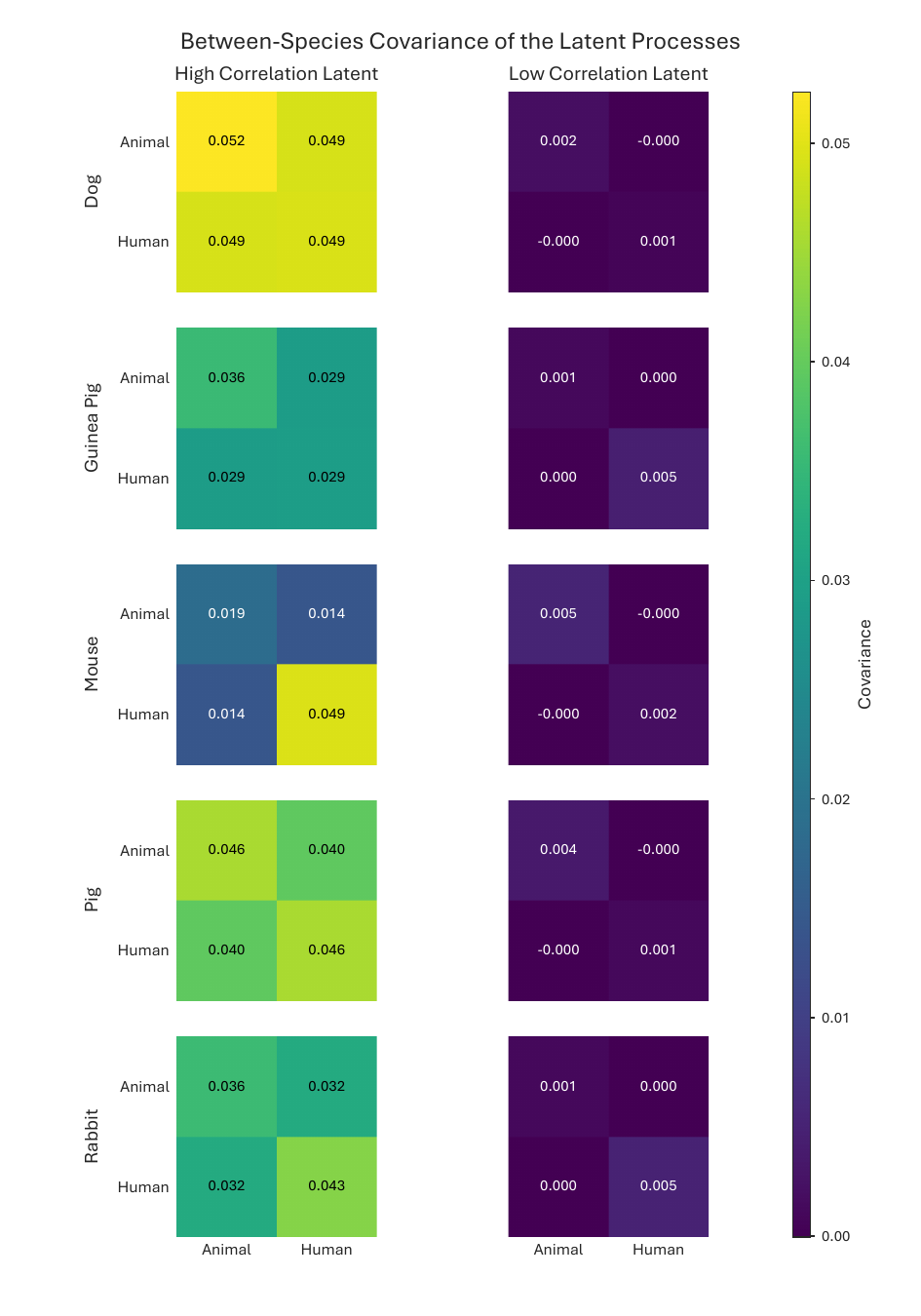}
\caption{Values of two latent between-species kernels $k_{\mathcal{S},q}(s,s')$ with $q=1,2$ (equation 2 main article) for a Gaussian process (GP) trained on human and animal data from four training drugs. Generally one latent learns a kernel with higher variance and covariance, representing larger trends present in both species, while another learns a low-variance kernel with effectively no covariance, to account for species-specific high-frequency trends. Figure \ref{fig:corr_between_species} scales these kernel matrices to correlation matrices for interpretability. For all animals apart from guinea pig the training drugs used are bepridil, diltiazem, quinidine and verapamil; dofetilide is used instead of quinidine in the guinea pig case since no quinidine data is available for guinea pigs.}
\label{fig:cov_between_species}
\end{figure}

\begin{figure}
\centering
\includegraphics[width=0.8\textwidth]{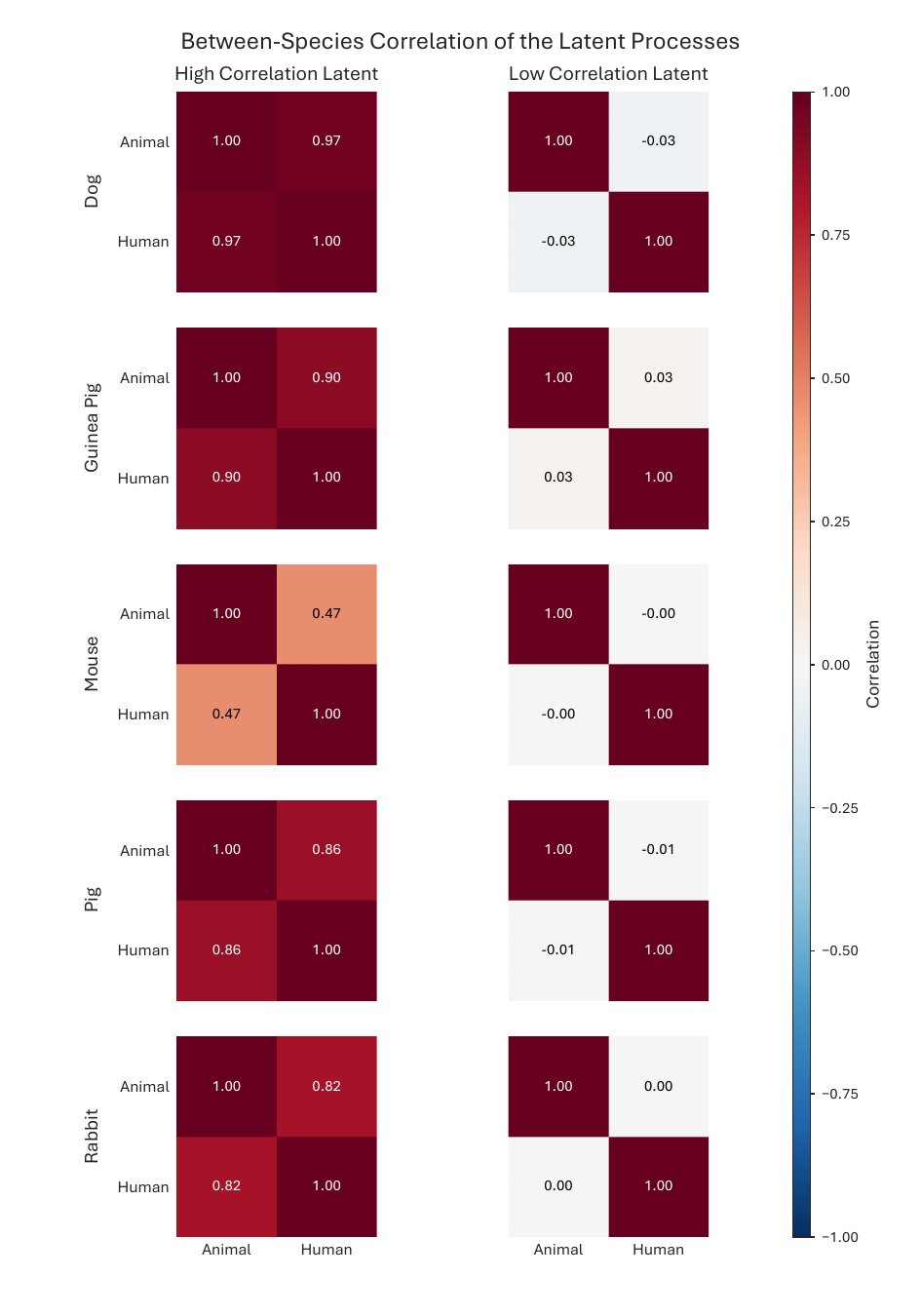}
\caption{Values of two latent between-species kernels $k_{\mathcal{S},q}(s,s')$ with $q=1,2$ (equation 2 main article) shown in figure \ref{fig:cov_between_species}, scaled to correlation matrices for interpretability, for a Gaussian process (GP) trained on human and animal data from four training drugs. Generally one latent learns a kernel with higher variance and covariance, representing larger trends present in both species, while another learns a low-variance kernel with effectively no covariance, to account for species-specific high-frequency trends. For all animals apart from guinea pig the training drugs used are bepridil, diltiazem, quinidine and verapamil; dofetilide is used instead of quinidine in the guinea pig case since no quinidine data is available for guinea pigs.}
\label{fig:corr_between_species}
\end{figure}

\begin{landscape}
\begin{table}[htbp]
\centering
\caption{Summary of clinical data for estimation of intrinsic heart rate (HR) using baseline HR. Two combinations of high-fidelity (HF) and low-fidelity (LF) species are used: human as HF with dog as LF (model 1), as well as dog as HF with mouse as LF (model 2). In model 1, the intrinsic HR is measured using an electrocardiogram (ECG) in live subjects after autonomic blockade, while in model 2 it is measured via an isolated preparation.}
\small
\renewcommand{\arraystretch}{1.3}
\begin{tabular}{|>{\raggedright\arraybackslash}p{1.6cm}|>{\raggedright\arraybackslash}p{2.2cm}|>{\raggedright\arraybackslash}p{2.8cm}|>{\raggedright\arraybackslash}p{5.0cm}|>{\raggedright\arraybackslash}p{3.6cm}|>{\raggedright\arraybackslash}p{2.6cm}|>{\raggedright\arraybackslash}p{3.4cm}|}
\hline
\textbf{Species} & \textbf{n (sedentary / athletic)} & \textbf{Baseline HR, model input} & \textbf{Intrinsic HR, model output} & \textbf{Blocking doses} & \textbf{Role} & \textbf{Source} \\
\hline
Human & 15 (7 / 8) & 3-lead ECG, conscious & Autonomic block, 3-lead ECG & Atropine 0.04 mg/kg, propranolol 0.2 mg/kg & Model 1 HF (n = 15) & D'Souza et al., Circ Res 2017;121:1058 \cite{dsouza2017targeting} \\
\hline
Dog & 14 (7 / 7) & 3-lead ECG, conscious & Model 1: autonomic block, 3-lead ECG \newline Model 2: isolated right atrial preparation, extracellular bipolar & Atropine 0.04 mg/kg, propranolol 0.2 mg/kg & Model 1 LF (n = 13) \newline Model 2 HF (n = 14) & Soattin et al., Eur Heart J 2026, ehag358 \cite{soattin2026endurance}\\
\hline
Mouse & 13 (6 / 7) & 3-lead ECG, conscious & Isolated preparation, extracellular bipolar & Atropine 0.5 mg/kg, propranolol 1 mg/kg, recorded but not used & Model 2 LF (n = 13) & D'Souza et al., Nat Comms 2014;5:3775 \cite{dsouza2017targeting}\\
\hline
\end{tabular}
\label{tab:heart-rate-models}
\end{table}
\end{landscape}

\begin{figure}
\centering
\includegraphics[width=\textwidth]{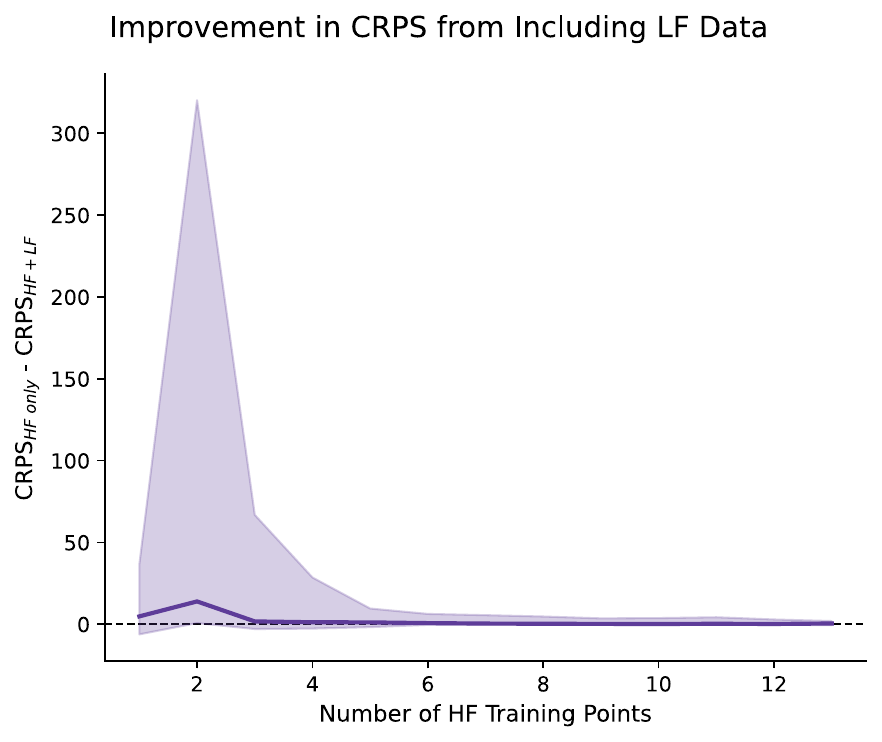}
\caption{Difference in continuous ranked probability score (CRPS) on held-out test data for prediction of drug-induced change in heart rate between a Gaussian process (GP) trained only on high-fidelity (HF) data versus a GP trained on both HF and low-fidelity (LF) data. Dogs are taken as the HF species and mice as LF. For each number of HF training points used, all remaining HF data points from the total set of 14 are taken as held-out test data. Median and 95\% confidence intervals for the average difference across all held-out data points are shown across 100 repeats of randomly sampled training data.}
\end{figure}

\begin{figure}
\centering
\includegraphics[width=\textwidth]{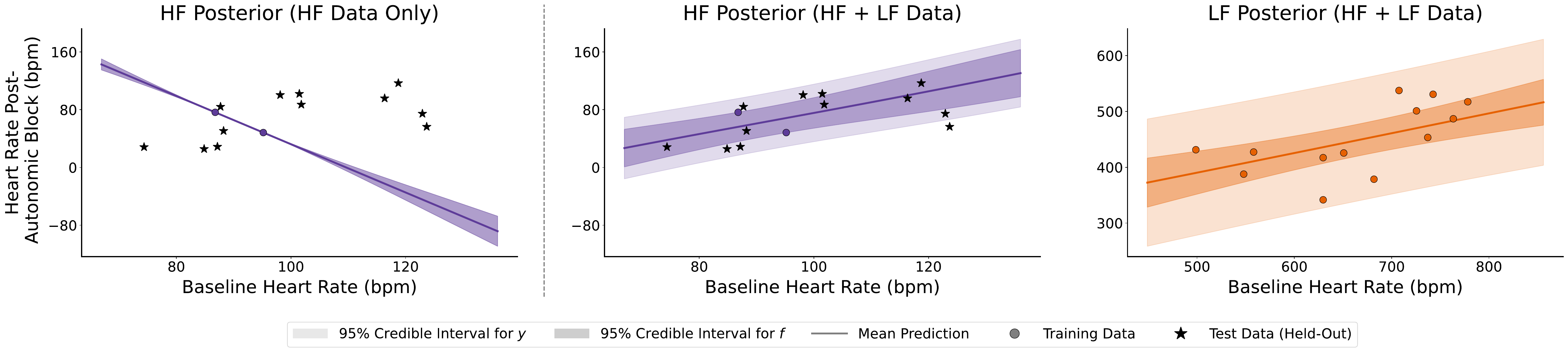}

    \vspace{1cm}

\includegraphics[width=\textwidth]{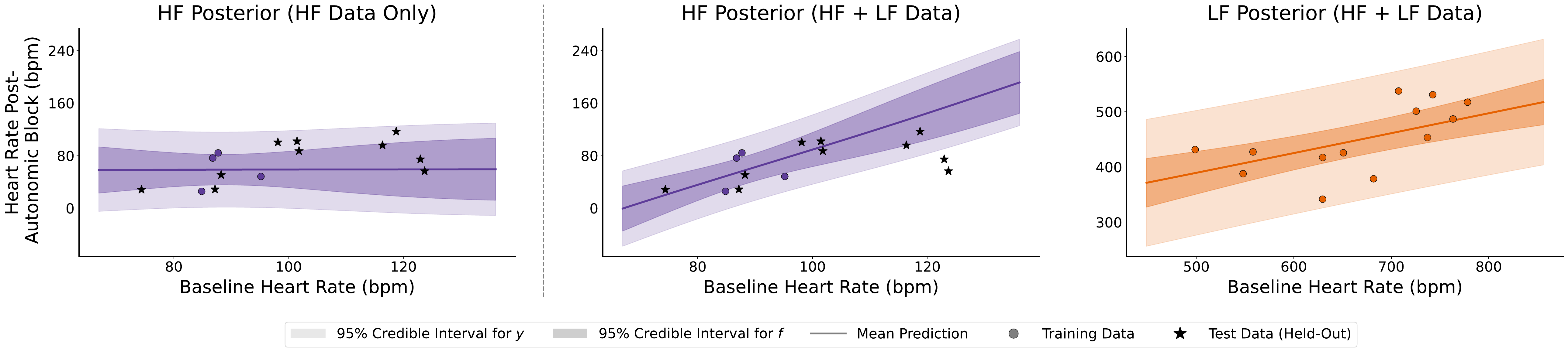}

    \vspace{1cm}

\includegraphics[width=\textwidth]{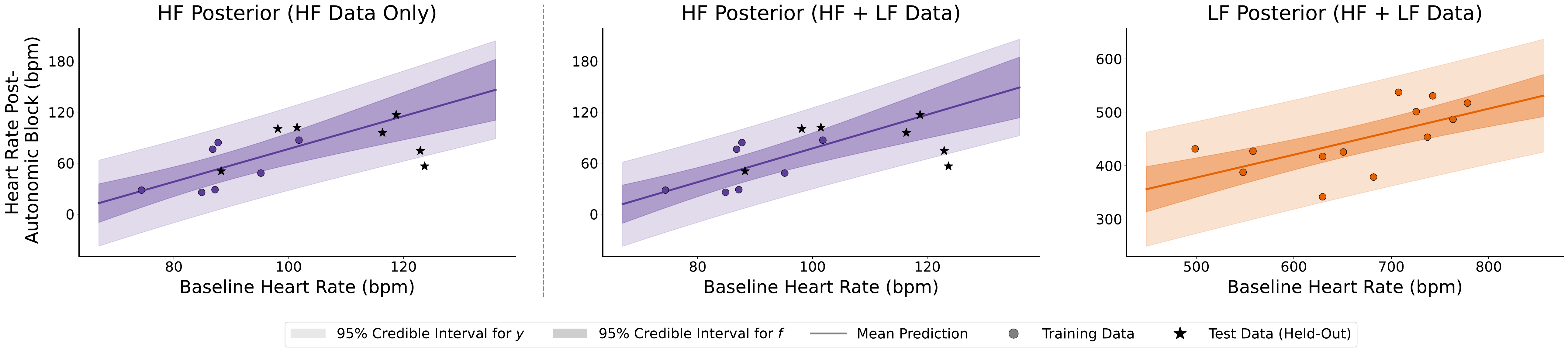}

    \vspace{1cm}

\includegraphics[width=\textwidth]{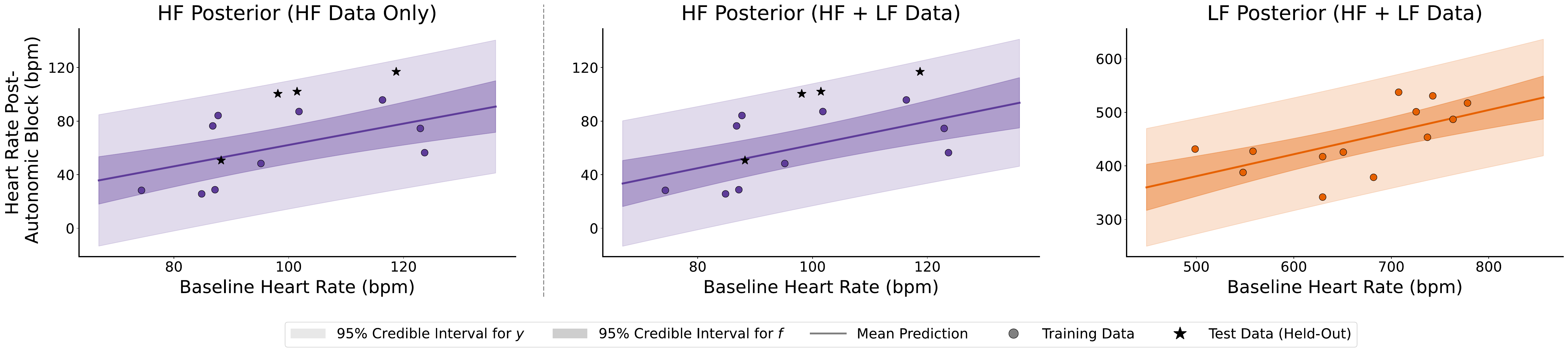}
\caption{Gaussian process (GP) posterior predictive plots for heart rate after autonomic block in humans (high-fidelity (HF)) and dogs (low-fidelity (LF)) for the following number of randomly-sampled HF training datapoints: 2 (\textbf{top}), 4, 7, 10 (\textbf{bottom}). In all cases, the addition of LF auxiliary data (right) improves estimation of HF trends (centre) when compared to using just the HF data alone (left). This improvement plateaus as the number of HF datapoints increases, as HF trends become easier to determine using just HF data.}
\end{figure}

\begin{figure}
\centering
\includegraphics[width=\textwidth]{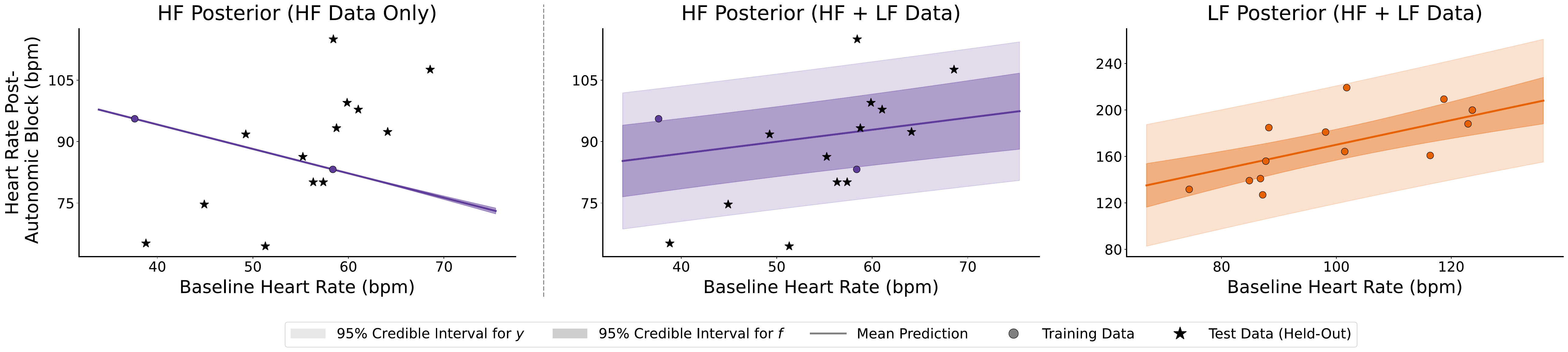}

    \vspace{1cm}

\includegraphics[width=\textwidth]{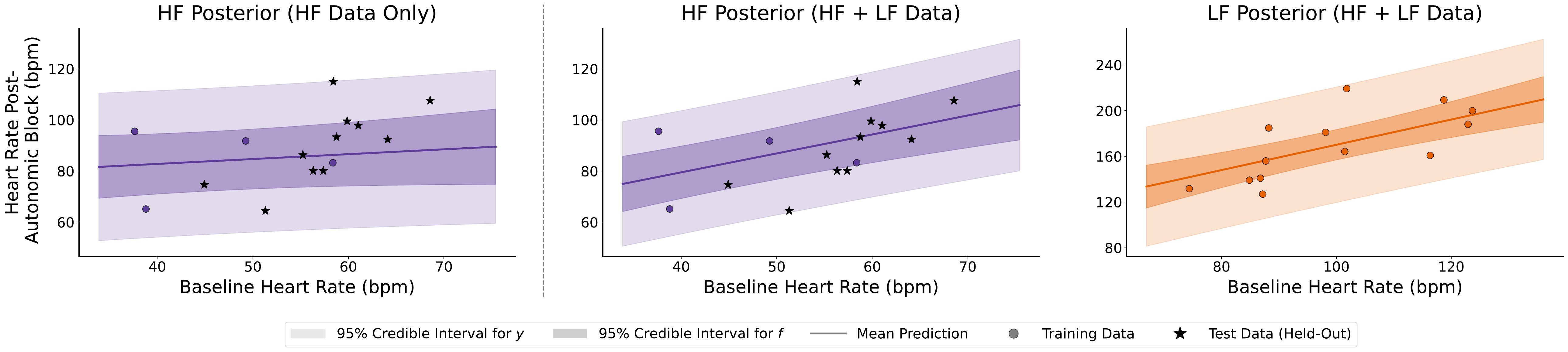}

    \vspace{1cm}

\includegraphics[width=\textwidth]{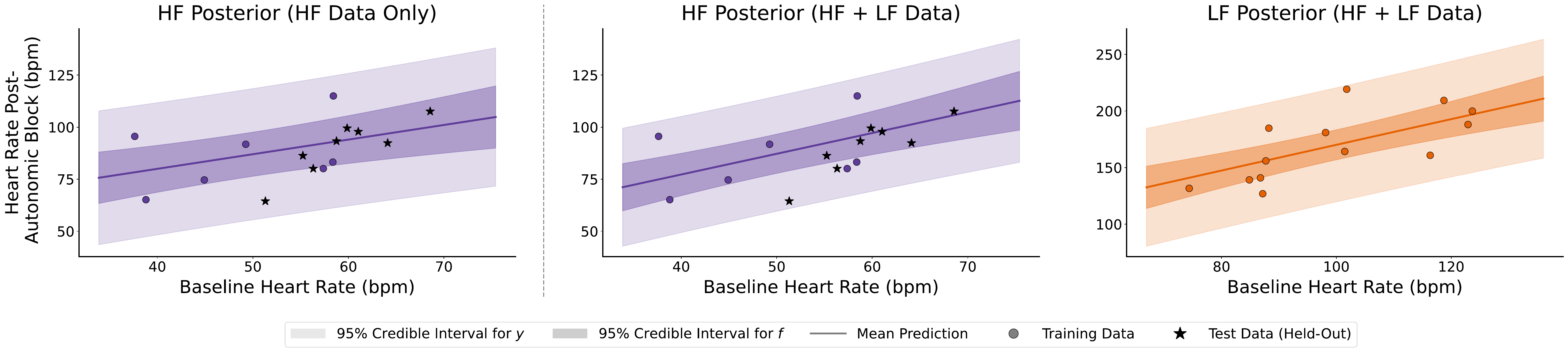}

    \vspace{1cm}

\includegraphics[width=\textwidth]{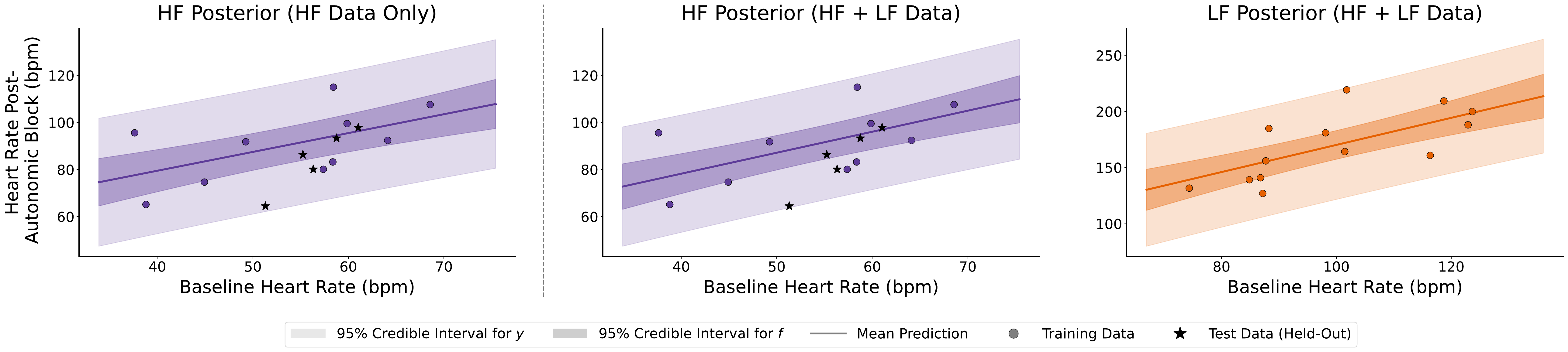}
\caption{Gaussian process (GP) posterior predictive plots for heart rate after autonomic block in dogs (high-fidelity (HF)) and mice (low-fidelity (LF)) for the following number of randomly-sampled HF training datapoints: 2 (\textbf{top}), 4, 7, 10 (\textbf{bottom}). In all cases, the addition of LF auxiliary data (right) improves estimation of HF trends (centre) when compared to using just the HF data alone (left). This improvement plateaus as the number of HF datapoints increases, as HF trends become easier to determine using just HF data.}
\end{figure}

\begin{figure}
\centering
\includegraphics[width=0.7\textwidth]{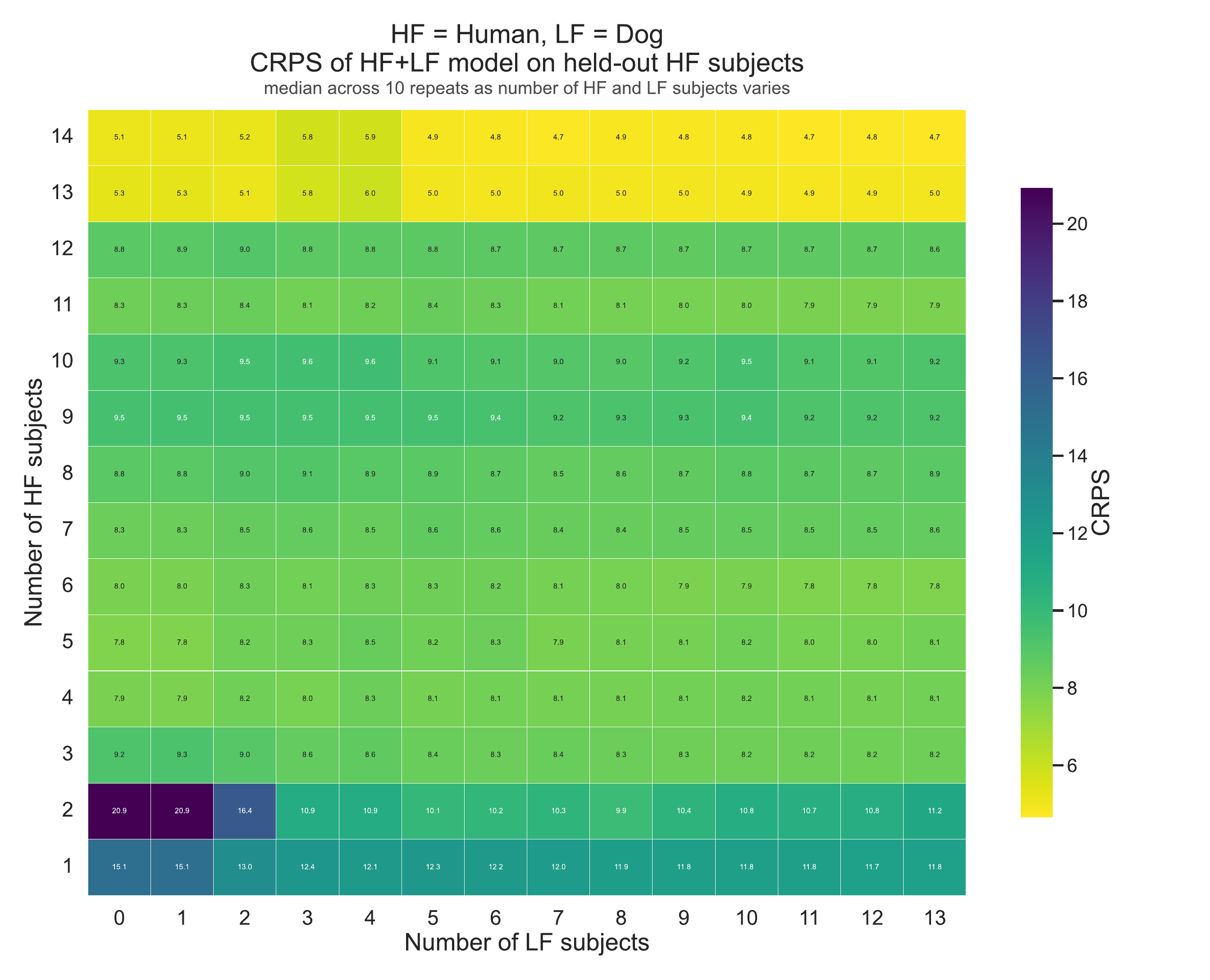}
   
    \vspace{1cm}

\includegraphics[width=0.7\textwidth]{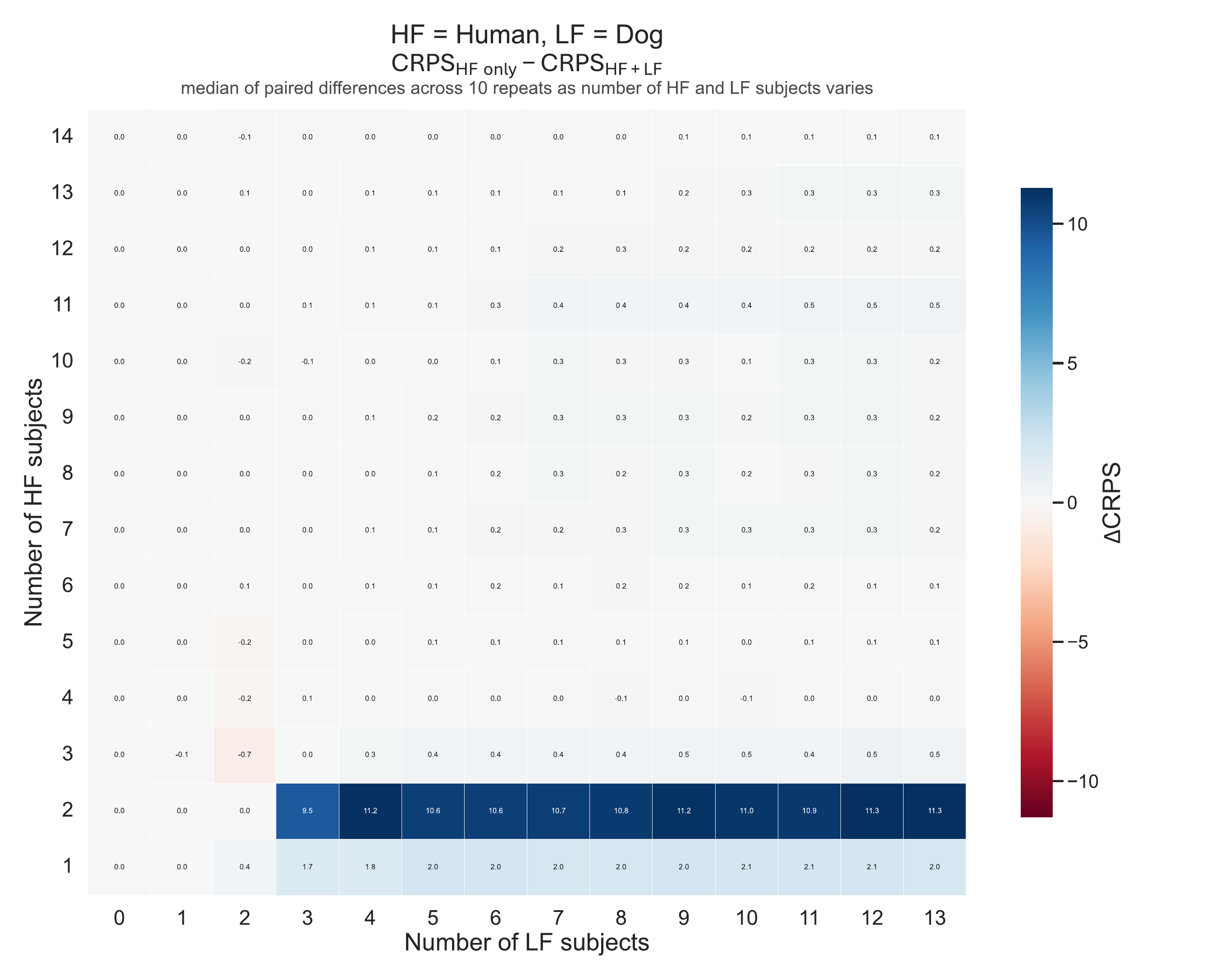}
\caption{\textbf{Above}: Continuous ranked probability score (CRPS) on held-out test data for prediction of drug-induced change in heart rate in humans (high-fidelity (HF) species) as the ratio of dog subjects (low-fidelity (LF)) to human subjects varies. \textbf{Below}: paired difference in CRPS in the HF-only vs HF+LF case for each fixed number of HF subjects. For each combination of number of HF and LF training points, the training data is randomly sampled, used to fit GP hyperparameters and obtain a posterior predictive distribution, while all remaining HF data points from the total set of 15 are taken as held-out test data. The median CRPS across 10 experimental repeats of randomly sampled training data is shown. For any fixed number of HF subjects, the addition of further LF data generally improves CRPS. This improvement plateaus as the number of HF subjects increases: LF data becomes less influential as HF data becomes richer.}
\end{figure}

\begin{figure}
\centering
\includegraphics[width=0.7\textwidth]{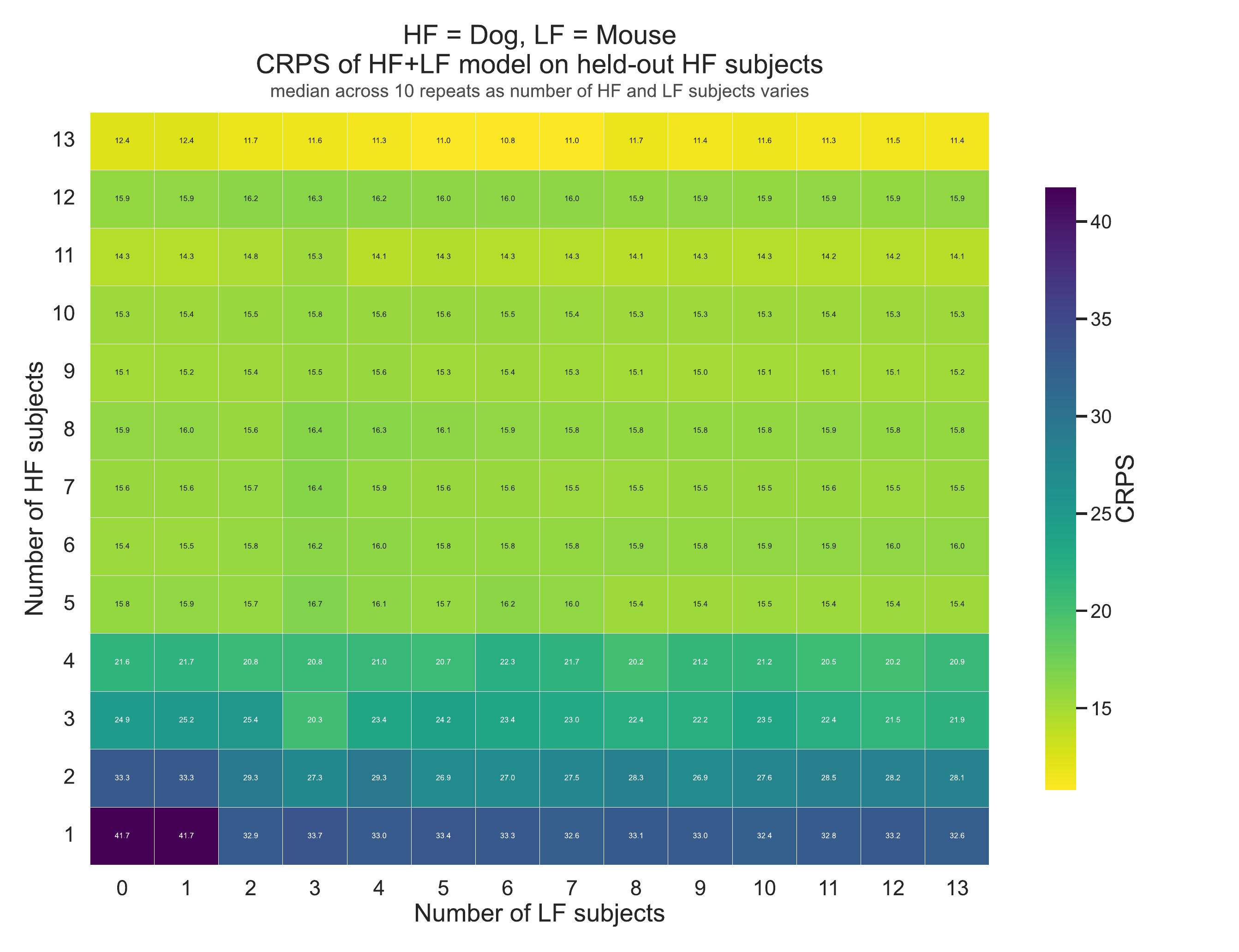}
   
    \vspace{1cm}

\includegraphics[width=0.7\textwidth]{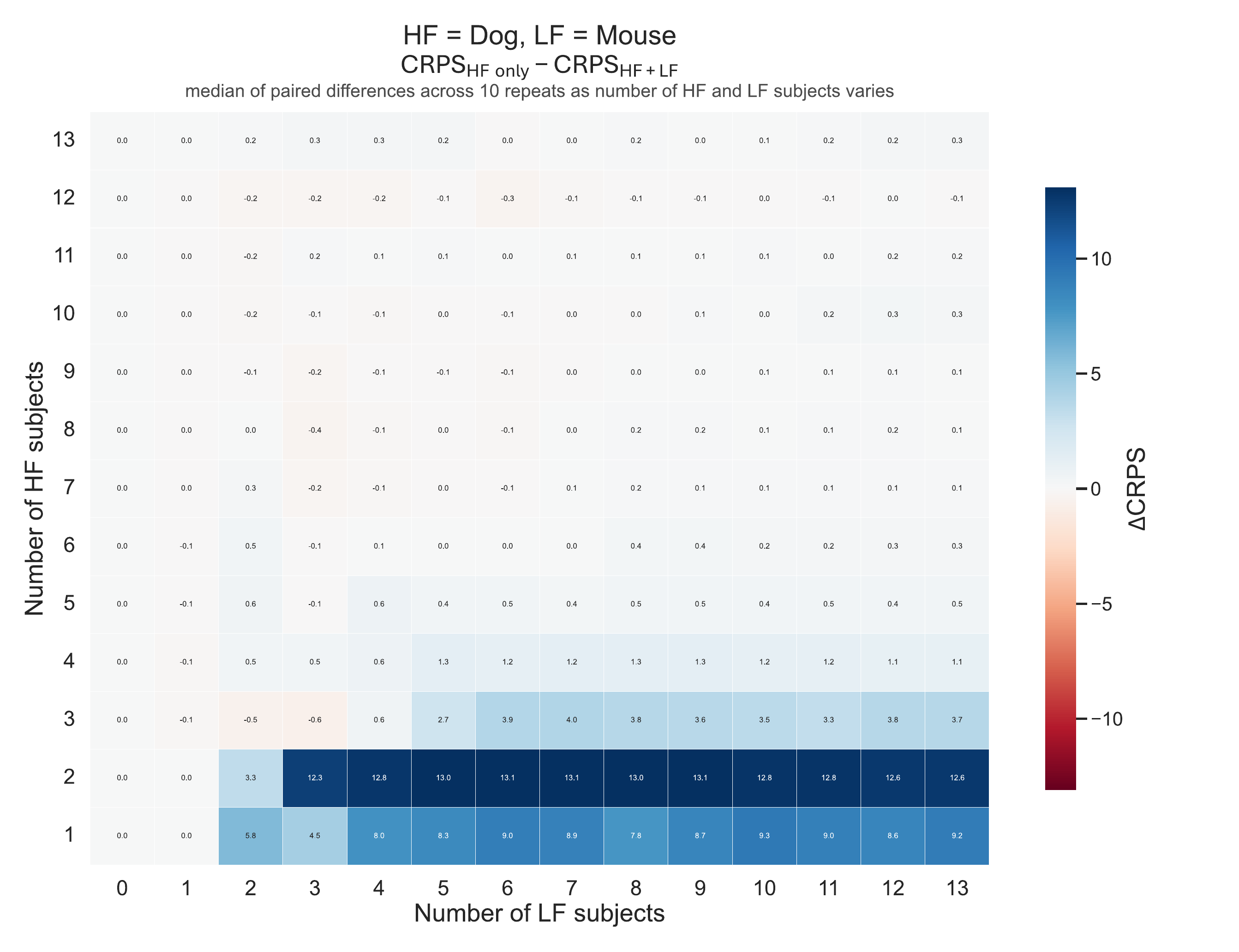}
\caption{\textbf{Above}: Continuous ranked probability score (CRPS) on held-out test data for prediction of drug-induced change in heart rate in dogs (high-fidelity (HF) species) as the ratio of mouse subjects (low-fidelity (LF)) to dog subjects varies. \textbf{Below}: paired difference in CRPS in the HF-only vs HF+LF case for each fixed number of HF subjects. For each combination of number of HF and LF training points, the training data is randomly sampled, used to fit GP hyperparameters and obtain a posterior predictive distribution, while all remaining HF data points from the total set of 15 are taken as held-out test data. The median CRPS across 10 experimental repeats of randomly sampled training data is shown. For any fixed number of HF subjects, the addition of further LF data generally improves CRPS. This improvement plateaus as the number of HF subjects increases: LF data becomes less influential as HF data becomes richer.}
\end{figure}

\begin{figure}
\centering
\includegraphics[width=\textwidth]{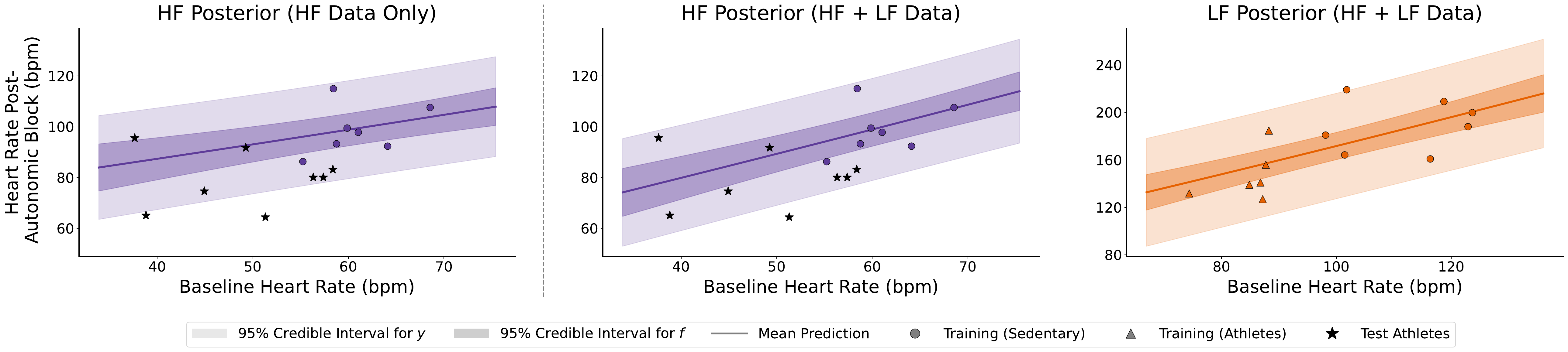}
\caption{Posterior predictive distributions for heart rate after autonomic blockade in humans (high fidelity (HF) species) and dogs (low fidelity (LF)) when HF athletic subjects are taken as held-out test data. The auxiliary LF data (right) improves estimation of HF trends (centre) when compared to using just the HF data alone (left).}
\end{figure}

% \subsection{Subhead}
% Type or paste text here. This should be additional explanatory text such as an extended technical description of results, full details of mathematical models, etc. 

% \subsection{Materials}
% Add a materials subsection if you need to.

% \subsection{Methods}
% Add a methods subsection if you need to.

%%% Each figure should be on its own page
% \begin{figure}
% \centering
% \includegraphics[width=\textwidth]{frog}
% \caption{Second figure}
% \end{figure}

% \begin{table}\centering
% \caption{This is a table}

% \begin{tabular}{lrrr}
% Species & CBS & CV & G3 \\
% \midrule
% 1. Acetaldehyde & 0.0 & 0.0 & 0.0 \\
% 2. Vinyl alcohol & 9.1 & 9.6 & 13.5 \\
% 3. Hydroxyethylidene & 50.8 & 51.2 & 54.0\\
% \bottomrule
% \end{tabular}
% \end{table}

%%% Add this line AFTER all your figures and tables
\FloatBarrier
\fi

\end{document}